\documentclass[aip,pof,graphicx]{revtex4-1} 
\pdfoutput=1
\usepackage[latin1]{inputenc}
\usepackage[T1]{fontenc}
\usepackage{lmodern}
\usepackage{graphicx}
\usepackage{amsmath}
\usepackage{amsfonts}
\usepackage{amssymb}
\usepackage{natbib}
\usepackage{graphicx}
\usepackage{nomencl}
\usepackage[svgnames*,table,dvipsnames]{xcolor} 

\newlength\epaisLigne
\newcommand{\Ghline}{\noalign{\global\epaisLigne\arrayrulewidth\global\arrayrulewidth 1.5pt}\hline \noalign{\global\arrayrulewidth\epaisLigne}}
\newcolumntype{I}{!{\vrule width 1.5pt}}

\begin{document}
\title{Transition scenario of the round jet in crossflow topology at low velocity ratios}
\author {Tristan CAMBONIE}\email[]{tristan.cambonie@espci.fr}
\author {Jean-Luc AIDER}\email[]{aider@pmmh.espci.fr}
\affiliation{Laboratoire PMMH (UMR7636 CNRS - ESPCI - UPMC Univ. Paris 6 - UPD Univ. Paris 7).
             \'Ecole Sup\'erieure de Physique et de Chimie Industrielles, 10 rue Vauquelin, 75231 Paris Cedex 5, FRANCE. Tel:~+33-1-4079~5852. Fax:~+33-1-4079~4523\\}

\begin{abstract}
We study experimentally a round Jet In CrossFlow (JICF) at low values of the jet-to-crossflow velocity ratio $R$ using instantaneous and time-averaged three-dimensions three-components (3D3C) velocimetry. The difference between instantaneous and time-averaged swirling structures of the JICF is emphasized. Through the analysis of spatial distribution of instantaneous transverse and longitudinal vortices the main transitions of the JICF are characterized for  0.15 $<$ $R$ $<$ 2.2. A new transition at very low velocity ratio is found ($R<0.3$). 
When $R$ is large enough ($R>1.25$), the classic JICF topology is recovered. In between, a deformation of the classical JICF topology is observed consisting in a progressive disappearance of the leading-edge vortices, a bending of the jet trajectory  and thus a strengthened interaction with the boundary layer. toward the wall.\\
Thanks to a state-of-the-art review on the JICF topology and using visualizations of the flow structures extracted from our experimental volumetric velocimetry measurements, this article provides a complete transition scenario of the JICF topology from the high velocity ratios to the lowest ones, and gives the topological transition threshold associated with each kind of vortex.
\end{abstract}
\maketitle 

\section{Introduction}
Jets In CrossFlow (JICF) are used in many industrial processes such as film cooling, fuel injection and flow control.
This flow has been studied for several decades and continues to be an active subject of research for many experimental or numerical research teams.  Reviews on the subject can be found in \citet{Margason1993,Karagozian2010}. 

Because the jet penetrates the crossflow, and then becomes an obstacle for it, the JICF topology is mainly governed by the competition between the jet momentum and the crossflow momentum. When studying a JICF one of the key parameters is then the momentum flux ratio, defined as $J=\rho_j V_j^2/\rho_{\infty} V_{\infty}^2$ where $\rho_j$ and $V_j$ are respectively the jet density and mean velocity while $\rho_{\infty}$ and $V_{\infty}$ are the free stream density and crossflow bulk velocity. If the jet and free stream fluid densities are equal ($\rho_{\infty}=\rho_j$), the momentum flux ratio becomes a velocity ratio $R=\sqrt(J)=V_j/V_{\infty}$. A recent study\cite{Cambonie2013} has proposed a more accurate definition of the momentum ratio and proved its relevance to study the jet counter-rotating vortex pair trajectories.

\begin{figure}[htbp!]
	\begin{center}
		\begin{tabular}{c}
\includegraphics[height=0.5\textwidth]{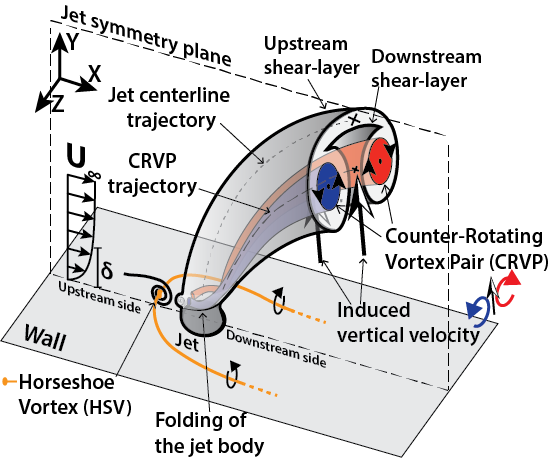}
\\a)\\
\includegraphics[height=0.5\textwidth]{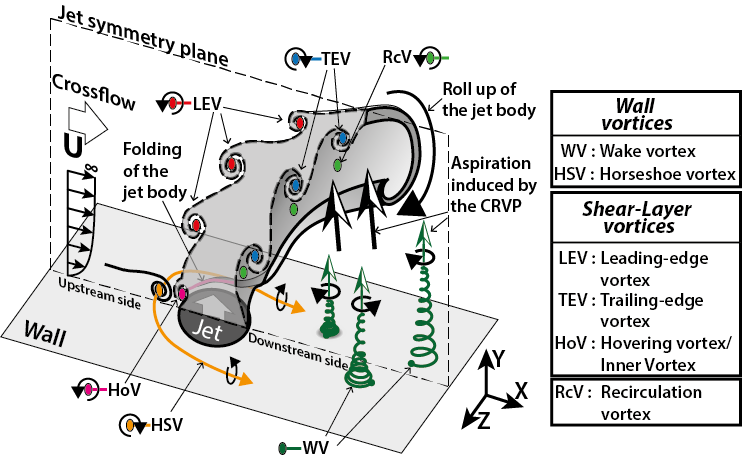}
\\b)\\
\end{tabular}
	\caption{a) Classical time-averaged topology of the high-velocity-ratio round JICF. b) Instantaneous topology of the high-velocity-ratio round JICF}
	\label{fig:JICF_Topologies}
	\end{center}
\end{figure}

The interaction between the jet and the crossflow leads to a complex three-dimensional and highly unsteady flow made of multiple shear-layers and numerous intricated vortical structures in close interaction with each other.
As a result, the study of the JICF topologies (time-averaged in Fig.~\ref{fig:JICF_Topologies}a and instantaneous in Fig.~\ref{fig:JICF_Topologies}b) and their transition thresholds is still an active research field. A lot of studies focused on the high velocity ratios, considered as the classical JICF topology. In this case the interactions between the jet and the boundary layer are weak and therefore in most cases negligible. The main
vortical structures involved in this type of flow configuration are the
counter-rotating vortex pair, the leading-edge and trailing-edge vortices, the horseshoe vortex, the hovering vortex or inner vortex and the wake vortices.

The Counter-Rotating Vortex Pair or CRVP (Fig.~\ref{fig:JICF_Topologies}a) is usually considered to be the main vortical feature of the time-averaged JICF topology and considerable attention has been devoted to its study\cite{Kelso1996,Smith1998}, and in particular its formation\cite{Kelso1996,Cortelezzi2001,Lim2001,Muppidi2006}. 

To the author's knowledge there are no recorded cases of a circular JICF configuration in which the CRVP is not present. Moreover the CRVP is the only structure of the mean field remaining far from the injection site, sometimes persisting as far as a thousand jet diameters as shown by \citet{KEFFER1963}. 
It originates from a folding of the jet body and shear-layers near the base of the jet as it is shown in Fig.~\ref{fig:JICF_Topologies}a and ~\ref{fig:JICF_Topologies}b.
This folding of the jet upstream shear-layer gets vorticity from the arms of the hovering vortex. 

The Hovering Vortex (HoV) is a junction flow vortex present in both topologies (Fig.~\ref{fig:JICF_Topologies}) which is wrapped around around the jet base. At high R, this vortex is located between the horseshoe vortex and the jet base upstream the jet\cite{Kelso1996,KELSO1995,Megerian2007a}. At low R, the hovering vortex is pushed back inside the jet pipe and for this reason has been called Inner Vortex by \citet{Bidan2013}.

The CRVP has been extensively studied : mixing rate \cite{KAMOTANI1972,BROADWELL1984,Muppidi2008,SU2004}, circulation decay \cite{BROADWELL1984,Hasselbrink2001,SU2004}, vortex cores trajectory \cite{Cambonie2013}, jet spreading\cite{Davitian2010}. The auto-induction between the CRVP branches (Fig.~\ref{fig:JICF_Topologies}a) have an important impact on the jet trajectory straightening. It also creates a vertical aspiration which  lifts up the boundary layer\cite{Kelso1996,Muppidi2005,New2006} as well as the quasi-streamwise wall vortices forming the wake vortices\cite{Kelso1996,FRIC1994} (Fig.~\ref{fig:JICF_Topologies}b). This entrainement strongly contributes to the mixing of matter and energy in the transversal planes \cite{Smith1998,Cortelezzi2001,Muppidi2007,Ziefle2009}. This is the reason why the JICF is of great practical interest for industrial applications where it can be used, for instance, to increase the mixing  or to force the transition in a boundary layer. Although the CRVP has been historically defined has a vortex of the time-averaged velocity field\cite{KAMOTANI1972,BROADWELL1984,COELHO1989,ANDREOPOULOS1984,FRIC1994,Cortelezzi2001} it is considered also by some studies \cite{Kelso1996,Camussi2002} as an instantaneous vortical structure which breaks down a few diameter after the jet injection\cite{Kelso1996,New2006}. 

The HorseShoe Vortex (HSV) is present in both time-averaged and instantaneous topologies (Fig.~\ref{fig:JICF_Topologies}a,b). It comes from the roll-up of the boundary layer and wraps around the base of the jet while being stretched by the cross-flow. It is therefore intrinsically linked to the junction flow\cite{Simpson2001} created just upstream the jet\cite{BAKER1979,KROTHAPALLI1990,KELSO1995}. Among the JICF vortical structures, it can be put in the category of the wall vortices, which originates from the wall boundary layer vorticity and are generated due to the jet presence.

The wake vortices\cite{MOUSSA1977,FRIC1994,Smith1998} also called upright vortices (Fig.~\ref{fig:JICF_Topologies}b) are also wall vortices. Despite the similitude with a von K\'arm\'an alley,  it has been proven by \citet{FRIC1994} that these tornado-like structures are not related to a von K\'arm\'an instability. They originate from foldings of the boundary layer that are lifted up by the aspiration induced by the CRVP\cite{FRIC1994,Bidan2013}. Therefore, the wake vortices only exist in a limited velocity ratio range\cite{Kelso1996,FRIC1994} ($2<R<10$), when the CRVP trajectory is altogether high enough for these structures to develop and close enough from the boundary layer for them to be lifted up by the CRVP.

The shear-layer vortices are the fundamental vortical structures of the instantaneous JICF topology.  They consists of a system of intricate loop vortices located along the shear layers on both sides of the jet\cite{Lim2001,New2006}. Their legs and arms are entangled together in the region which becomes the CRVP in the time-averaged velocity field. \\
The Leading-Edge Vortices (LEV) are created from the upstream shear-layer while the Trailing-Edge Vortices (TEV) (also called Lee-side vortices by \citet{New2006}) develop from the downstream shear-layer. At high R, the LEVs and TEVs are created by the Kelvin-Helmholtz instability. 
For a long time, their topology was assumed to be ring-like, similar to the ring shear-layer vortices of the free jet or the ring structures of the pulsed round jet.
This  misinterpretation of the JICF topology, still very common nowadays, have been spread by many reference articles \cite{COELHO1989,FRIC1994,Kelso1996,Cortelezzi2001,Camussi2002} and consolidated by decades of 2D measurements and visualizations in the symmetry plane where such ring-like topology is plausible. Nevertheless, it has been since proved that pulsed and steady jet topologies as well as free jet and steady JICF topologies are fundamentally different. \citet{Lim2001} have shown that the steady JICF topology at high velocity ratios is not organized in ring shear-layer vortices but in loop vortices located on both sides of the jet: the LEVs and TEVs (Fig.~\ref{fig:JICF_Topologies}b and \ref{Fig:Topoinsta_vortexcores}). It explains why the LEVs and TEVs can be asynchrone as in the case of intermediate or low velocity ratios when the TEV\rq{}s swirling strength stays strong while the LEVs are weak or non existent. \citet{Lim2001} have also shown an experimental example where the downstream shear layer does not develop a Kelvin-Helmholtz instability and where only LEVs are present. Numerical simulations by \citet{Marzouk2007} have provided an insightful explanation of the LEVs and TEVs formation. It explains how rings of fluid particles (which is not the same than vorticity rings) are deformed to adopt LEV or TEV shapes, offering an interesting link with the more classical and misinterpreted ring-like topology.

The Recirculation Vortices (RcV) develop on the shear layer on the downstream side of the low velocity area located just behind the jet, considered as the recirculation area. They form a few diameters after the jet exit and are there convected downstream by the jet entrainment. 

Surprisingly, most of the studies dedicated to the round steady JICF with a straight injection pipe focused on the high velocity ratios. Only a few articles\cite{ANDREOPOULOS1984, Peterson2004, Gopalan2004, Megerian2007a, Ilak2012, Bidan2013} show results for a round straight (jet axis perpendicular to the wall) JICF for $R<1.5$. Except two articles\cite{Ilak2012,Bidan2013}, only one or two different velocity ratios below $R=2$ are investigated in each study which is insufficient to really evaluate the influence of the velocity ratio on the JICF topology. To our knowledge, until now, there has been no systematic and extensive experimental study for low velocity ratios performed with the same experimental setup. 
Based on a numerical study and the stability theory, \citet{Ilak2012} proposed a transition scenario of the JICF topology at low velocity ratio. For the lowest velocity ratios, they observe a steady counter rotating vortex pair whose destabilization at higher R leads to the formation of the hairpin. Recently, \citet{Bidan2013} have used numerical simulations and experiments to propose another scenario. It is based on the horseshoe vortex and inner vortex destabilization and shedding to qualify the flow state as attached, transitional or fully detached. Both scenarii are discussed and compared with the present study in the final section of this article.

A few transitions of the jet and of its vortices have already been documented in the literature. 
A transition of the jet shear-layer from convectively unstable to absolutely unstable has been shown to exist\cite{Megerian2007a,Davitian2010} around $R\approx 3\sim3.5$.
Using a different approach, \citet{Camussi2002} also have detected around $R\approx3$ a transition of the properties of the jet shear-layer vortices.
At $R=2.5$, the wake vortices starts appearing in the numerical simulations of \citet{Bagheri2009}. At $R=2.25$, an elliptic instability of the CRVP is also observed on their simulations.
Between $0.65<R<0.675$, the numerical simulations of \citet{Ilak2012} show an Hopf bifurcation transition between a steady state and an unstable state. This result will be thoroughly discussed in the following.
In the study of \citet{Bidan2013} other transitions of the low velocity ratio JICF have been found.
At $R=0.225$, the horseshoe vortex (HSV) starts to shed. At $R=0.275$, the shedding of the inner vortex (low velocity ratio hovering vortex) is the marker of the end of the attached jet regime.
At $R=0.6$, \citet{Bidan2013} define the beginning of the fully detached JICF regime by the stabilization of the HSV at the jet base.
Between the attached and fully detached regime, an inversion of preeminence of the wall vortices and the shear layer vortices is observed at $R=0.425$. The transition scenario proposed by \citet{Bidan2013} will be more detailed and discussed later in this article.

In view of a state-of-the-art review on the JICF topology and using visualizations of the vortices extracted from experimental volumetric velocimetry measurements, this study aims at providing a complete transition scenario of the JICF topology from the high velocity ratios to the lowest ones. After detailing the experimental setup and methods  in section 2, time-averaged and instantaneous isosurfaces are shown in section 3 and used for describing the transition of the JICF topologies. In the fourth part of this article, the statistical spatial distributions of instantaneous vortices in the symmetry plane and in the transverse planes are used as another way to evaluate the topology transition based on instantaneous vortices. Both parts confirm the existence of a very low velocity ratio transition of the JICF topology which is exposed in section 5 and compared to the more recent studies found in the literature. Finally, a complete transition scenario of the JICF topology is presented in the last part of this article. It integrates both our results and other studies results and gives the topological transition threshold associated with each kind of vortical structures (LEV, HSV, HoV).

\section{Experimental facilities and volumetric velocimetry technique}
\subsection{Hydrodynamic channel}

\begin{figure}[htbp!]
	\begin{center}
			\begin{tabular}{c}
		\includegraphics[width=0.7\textwidth]{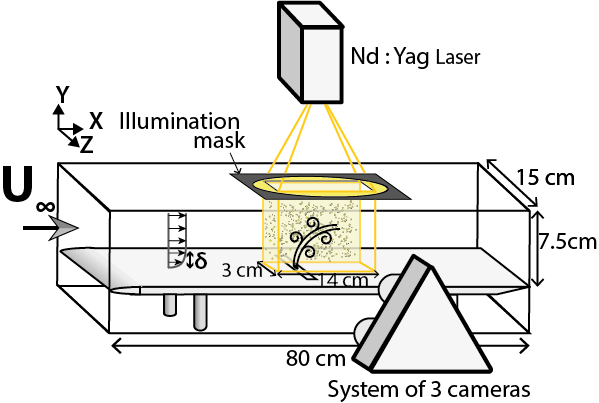}
			\end{tabular}
	\caption{Sketch of the experimental setup}
	\label{fig:plate}
	\end{center}
\end{figure}

	The experiments were carried out in a low-speed hydrodynamic channel in which the flow is driven by gravity. A divergent part, two honeycombs and a convergent section reduce the free-stream turbulence and suppress undesired large structures. The test section is 80 cm long with a rectangular cross section 15 cm wide and 10 cm high (Fig. \ref{fig:plate}). Altuglas walls allow for easy optical access from any direction.  A custom made plate with a NACA0020 leading-edge profile is used to start the crossflow boundary layer. For our range of crossflow velocities $U_\infty$, the boundary layer thickness $\delta$ varies from 1 to 2.5 cm (Table \ref{Tbl:VRdtable}). 

\subsection{Geometrical and physical parameters}
The jet supply system was custom made to allow for an easy change of jet nozzle geometry. Water enters a plenum and goes through a volume of glass beads designed to homogenize the incoming flow. The flow then goes through a removable plate in which one can design the desired jet nozzle geometry. In the following, we focus on cylindrical nozzles with different diameters $d$. 
The jet axis is normal to the flow.  The jet exit is located 42 cm downstream the leading edge of the flat plate. By convention, the crossflow always goes from left to right in the 2D representations and in the $x>0$ direction in the 3D representations.

The mean vertical jet velocity $\bar{V_{j}}$ ranges between $1.17$ to 8.8~cm.s$^{-1}$, the crossflow velocity $U_{\infty}$ between 0.9 and 17.7~cm.s$^{-1}$. It leads to velocity ratios $R = \bar{V_{j}}/ U_{\infty}$ ranging between $0.16$ to $2.13$. An extensive experimental study over more than 20 different experiments ranging from R~=~0.16 to R~=~2.13 has been carried out. We also define the crossflow Reynolds number $Re_{\infty}=U_{\infty}\cdot\delta/\nu$ and the jet Reynolds number $Re_{jet}=\bar{V_{j}}\cdot d/\nu$ which range respectively between $307<Re_{\infty}<1820$ and $94<Re_{jet}<704$. 

1000 instantaneous velocity fields are recorded for each configuration to retrieve the statistical properties of the instantaneous velocity fields and to ensure statistical convergence of the time-averaged velocity fields. Table~\ref{Tbl:VRdtable} presents all the parameters (velocity ratios, Reynolds numbers, diameters, boundary layer thicknesses and transition parameters) of the experiments shown in Fig.~\ref{fig:NbHV-HSV_vsVR},~\ref{fig:Frise},~\ref{fig:NbLEV-StLEV}. 
For the particular experiments shown in the Fig.~\ref{fig:isoLci_varVR_mean}, \ref{fig:isoVormag_varVR_mean}, \ref{fig:isoLci_varVR_insta}, \ref{FigTopoInstaJetRond:topoblob_varVR}, \ref{FigTopoJetRond:TrajCRVPinsta_backgrounds}, \ref{FigTopoJetRond:TrajCRVPinsta_XZ}, \ref{FigTopoJetRond:TrajCRVPinsta_highVR}, the free-stream velocity $U_{\infty}$ has been fixed to $U_{\infty}=6.55\pm0.05$~cm.s$^{-1}$ for the lower velocity ratios (R=\{0.16, 0.34, 0.55, 1.16, 1.34\}), where the interaction between the jet and the boundary layer is important. The other physical parameters of interest can be found in Table \ref{Tbl:VRdtable}. The higher velocity ratio cases (R=\{1.71, 2.13\}) come  from another set of experiments and correspond respectively to $U_{\infty}=$~2.56 and 1.65~cm.s$^{-1}$. In these cases, the jet has a very low interaction with the boundary layer.

\begin{table}[htb!]
\begin{center}
\begin{footnotesize}
\begin{tabular}{{IcI*{11}{c|}cI}}
\Ghline 
\textbf{Exp }&1&2&3&4&5&6&\cellcolor{Green}\textcolor{white}{\textbf{7}}&\cellcolor{Green}\textcolor{white}{\textbf{8}}&\cellcolor{Green}\textcolor{white}{\textbf{9}}&\cellcolor{Green}\textcolor{white}{\textbf{10}}&\cellcolor{Green}\textcolor{white}{\textbf{11}}&\cellcolor{Green}\textcolor{white}{\textbf{12}}\\\hline
\textbf{d} (mm)&8&8&8&8&8&8&8&8&8&8&8&8\\\hline
$\mathbf{\delta}$ (mm)&10.3&12.35&15.02&16.35&19.02&22.83&13.98&13.69&13.61&12.15&14.69&13.28\\\hline
$\mathbf{Re_\infty}$&1820&1160&840&675&504&422&921&901&890&792&961&843\\\hline
$\mathbf{Re_{jet}}$&259&259&259&266&274&292&94&182&290&471&608&704\\\hline
$\mathbf{R}$&0.18& 0.39& 0.58& 0.81& 1.29& 1.98& 0.16& 0.34& 0.55& 0.9& 1.16& 1.39\\\hline
$\mathbf{S_J}$&0.14&0.25&0.31&0.40&0.45&0.83&0.09&0.20&0.32&0.53&0.62&0.84\\\Ghline 
\multicolumn{13}{c}{}\vspace*{-0.03\textheight}\\ 
\Ghline 
\textbf{Exp }&\cellcolor{Green}\textcolor{white}{\textbf{13}}&\cellcolor{Green}\textcolor{white}{\textbf{14}}&15&16&17&18&19&20&21&22&23&\\\hline
\textbf{d} (mm)&10&10&10&10&10&10&10&10&10&10&10&\\\hline
$\mathbf{\delta}$ (mm)&16.35&18.29&12.12&14.42&16.96&12.21&14.74&17.46&12.53&14.88&16.73&\\\hline
$\mathbf{Re_\infty}$&419&307&1591&962&550&1596&979&569&1645&988&540&\\\hline
$\mathbf{Re_{jet}}$&438&359&640&640&640&640&640&640&640&640&640&\\\hline
$\mathbf{R}$& 1.71& 2.13& 0.49& 0.96& 1.97& 0.49& 0.96& 1.96& 0.49& 0.96& 1.98&\\\hline
$\mathbf{S_J}$&1.04&1.16&0.40&0.67&1.16&0.40&0.65&1.12&0.39&0.64&1.19&\\\Ghline
\end{tabular}
\end{footnotesize}
\end{center}
\caption{Diameters, boundary layer thicknesses, jet and crossflow Reynolds numbers, velocity ratios, and transition parameters $S_J$ of the experiments shown on Fig.~\ref{fig:NbHV-HSV_vsVR}, \ref{fig:Frise}, \ref{fig:NbLEV-StLEV}. Like in Fig.~\ref{fig:NbHV-HSV_vsVR}, \ref{fig:Frise}, \ref{fig:NbLEV-StLEV}, the particular experiments shown in this article in the Fig.~\ref{fig:isoLci_varVR_mean}, \ref{fig:isoVormag_varVR_mean}, \ref{fig:isoLci_varVR_insta}, \ref{FigTopoInstaJetRond:topoblob_varVR}, \ref{FigTopoJetRond:TrajCRVPinsta_backgrounds}, \ref{FigTopoJetRond:TrajCRVPinsta_XZ}, \ref{FigTopoJetRond:TrajCRVPinsta_highVR} are marked in green.}\label{Tbl:VRdtable}
\end{table}

\subsection{3D Particle Tracking Velocimetry measurements}
3D defocusing digital particle image velocimetry (3D-DDPIV) measurements have been performed using a system designed by TSI (Volumetric 3-components Velocimetry system, V3V) on the basis of the work of \citet{Pereira2000,Pereira2002,Pereira2006}. In a first step, the intensity peaks corresponding to each particles are detected in each camera frame for each time step. Then, using a spatial calibration, the triplets of 2D particle coordinates are used to reconstruct for each time step a 3D field of particle positions. A particle tracking step, between t and t+dt, leads to the instantaneous raw velocity field. Finally, a last step interpolates this raw velocity field on a grid in order to be able to use classical visualization tools and more generally to post-process the data. More details can be found in \citet{Pereira2002,CaAi14}.
The set-up was designed and the physical parameters were chosen to optimize the quality of the instantaneous velocity fields, using previous work\cite{CaAi12,CAMBONIEthesis,CaAi14}. 
The flow is seeded with 50 $\mu$m particles, with a visual concentration of $5.10^{-2}$ particles per pixel\cite{CaAi14}. The flow is illuminated through an illumination mask located on the upper wall (Fig. \ref{fig:plate}) and the particles are tracked in the volume using three cameras facing the side wall. The three cameras of this system are 4~MP double-framed with a 12~bit output. Volumetric illumination is generated using a 200~mJ pulsed Nd:YAG laser and two perpendicular cylindrical lenses. Synchronization is ensured by a TSI synchronizer. The measurement volume is 14$\times$6$\times$3~cm$^3$ and is homogeneously illuminated. 4 mm voxels and a 75 \% overlap lead to interpolation grid with one velocity vector per millimeter for both the instantaneous and time-averaged velocity fields. 

\subsection{Analysis of the 3D velocity fields using the $\lambda_{Ci}$ criterion}
Instantaneous and time-averaged swirling structures of the flow are visualized using isosurfaces of $\lambda_{Ci}$ which is a detection criterion of swirling structures initially proposed by \citet{Zhou1999} and improved by \citet{Chakraborty2005,Chakraborty2007}. This criterion is in reality twofold, based on the complex eigenvalues of the gradient velocity tensor $\mathcal{D}=\overrightarrow{\nabla} \overrightarrow{u}$. 
$\mathcal{D}$ is diagonalized to retrieve the real eigenvalue $\lambda_r$ which quantify the strain value in the main strain direction, and the complex eigenvalues $\lambda_{cr} \pm \lambda_{ci}$ which quantifies the fluid motion in the main rotation plane.
The set of criteria consists in (i):
$\lambda_{ci}\geq\epsilon_1$ and (ii):
$\frac{\lambda_{cr}}{\lambda_{ci}}\leq\epsilon_2$.
The first criterion is related to the swirling strength of the vortex. $T=\frac{2\pi}{\lambda_{ci}}$ measures the period of rotation of a fluid particle around the vortex core. The second criterion measures the compactness of the spiraling orbits. $\frac{\lambda_{cr}}{\lambda_{ci}}=0$ corresponds to a closed orbit in the swirling plane, while $\frac{\lambda_{cr}}{\lambda_{ci}}>0$ indicates an outward spiral and $\frac{\lambda_{cr}}{\lambda_{ci}}<0$ corresponds to an inward spiral.\\
The parameters $\epsilon_1$ and $\epsilon_2$ have to be chosen with respect to the relevant time and length scales of the physical problem considered.
This criterion is appropriate for a JICF: it captures only the swirling strength of the rotating fluid motion and does not take into account any contributions from surrounding shear, which proves to be necessary when vortices grow inside a boundary layer.\\
The same definition is applied to the 2D gradient velocity tensors to define similar criteria $\lambda_{Ci\ X}$, $\lambda_{Ci\ Y}$ and $\lambda_{Ci\ Z}$ which only detect swirling motions respectively along the X, Y and Z directions\cite{Christensen2002}.
While the 3D criterion $\lambda_{3D\ Ci}$ can be seen as a swirling magnitude, necessarily positive or null, the 2D criteria $\lambda_{Ci\ X,Y\ or\ Z}$ have been associated with the local sign of the fluctuating vorticity in order to discriminate clockwise from anti-clockwise vortices.
\\

\section{Qualitative overview of the time-averaged and instantaneous coherent structures of the JICF for $\mathbf{0.15~<~R~<2.2}$}
\subsection{Time-averaged structures}

\begin{figure}[htb!]
      \begin{center}
            \hspace*{-0.5cm}\begin{tabular}{Ic|c|cI}
             \Ghline\vspace{-0.025\textheight}&&\\
    \includegraphics[width=0.34\textwidth]{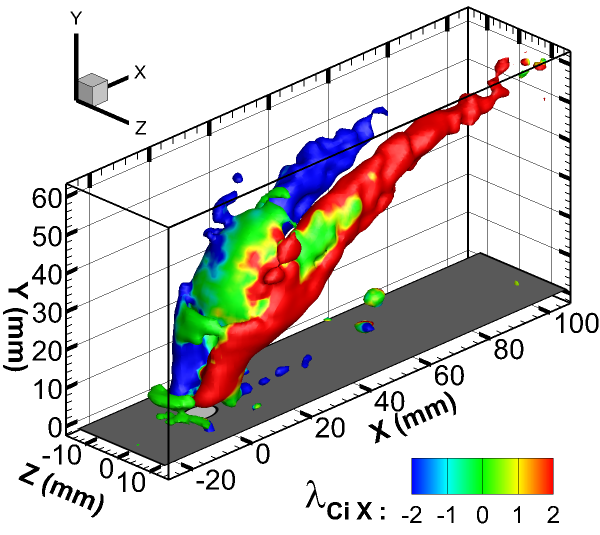}
    &    \includegraphics[width=0.34\textwidth]{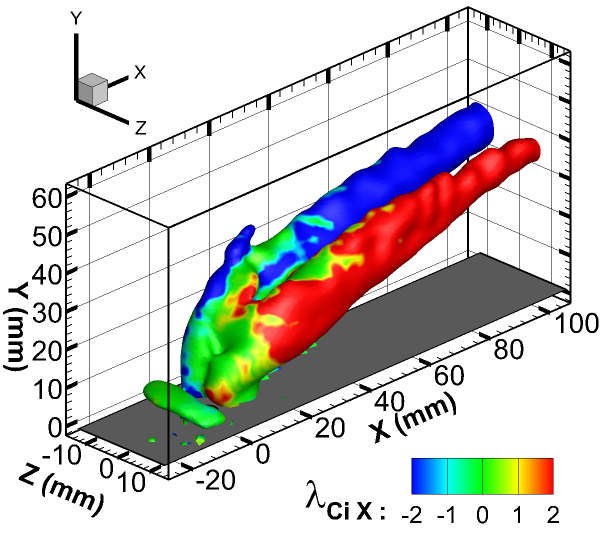}
    &    \includegraphics[width=0.34\textwidth]{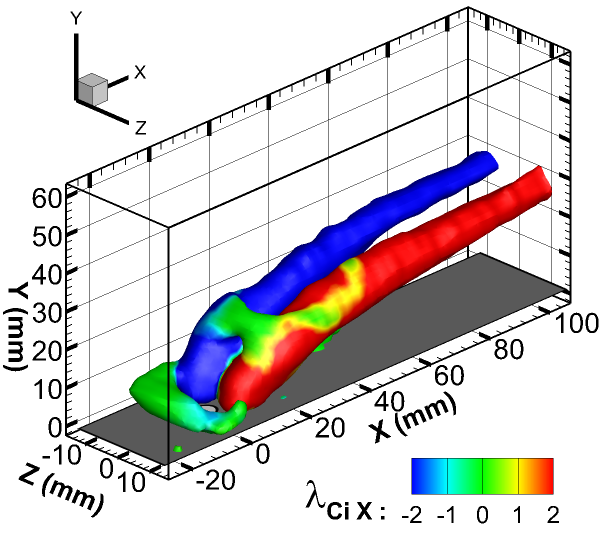}
    \\	a) R=1.71&b) R=1.39&c) R=1.16\\\hline\vspace{-0.025\textheight}&&\\
    \includegraphics[width=0.34\textwidth]{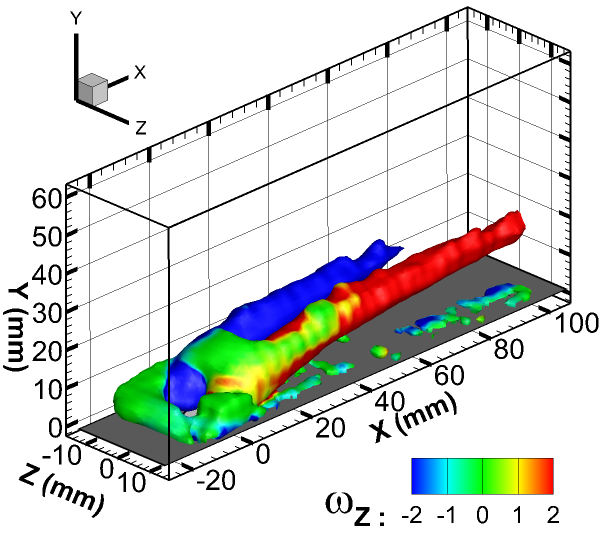}
    &    \includegraphics[width=0.34\textwidth]{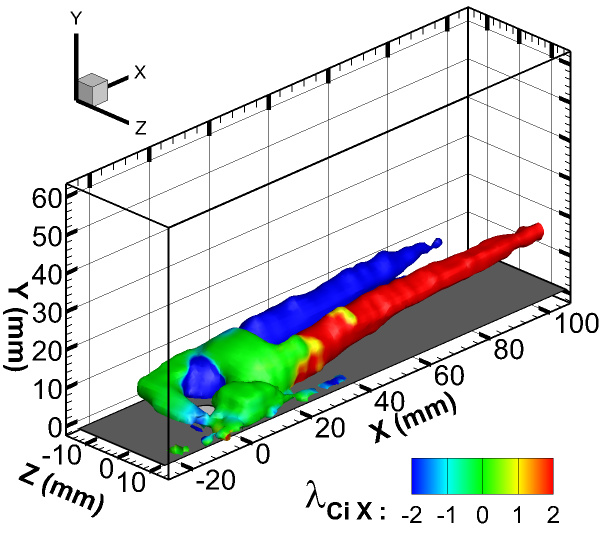}
    &    \includegraphics[width=0.34\textwidth]{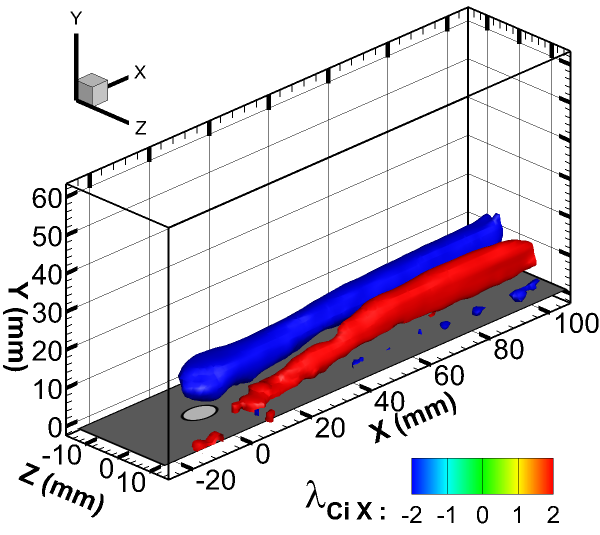}
    \\    d) R=0.90&e) R=0.55&f) R=0.16\\\Ghline
    \end{tabular}
    \caption{Visualization of the pair of  time-averaged counter-rotating streamwise vortices using  isosurfaces of $\lambda_{CI}=1.5\ s^{-1}$ colored by the longitudinal swirl $\lambda_{CI\ X}$ for six decreasing velocity ratios.}
\label{fig:isoLci_varVR_mean}
    \end{center}
\end{figure}

	Figure~\ref{fig:isoLci_varVR_mean} shows the swirling structures in the mean velocity field using isosurfaces of $\lambda_{Ci} = 1.5$~s$^{-1}$ colored by $\lambda_{Ci\ X}$, the longitudinal swirl. For each velocity ratio, the well-known Counter-Rotating Vortex Pair (CRVP) is visualized. For velocity ratios from R~$=1.71$ to R~$=0.55$, two other structures can be noticed. The first one, the horseshoe vortex \citep{Kelso1996}, wraps upstream the jet exit around the base of the jet. This vortex originates from the junction flow \citep{Simpson2001} between the jet and the crossflow, and is then intrinsically related to the obstacle made by the jet for the crossflow.
 The second one, a bridge of $\lambda_{Ci}$ connecting the positive and negative branches of the CRVP, is due to the strong swirling motion along the Z axis of the recirculation area behind the jet.
The CRVP is the only swirling structure which subsists at the lowest velocity ratio R$=0.16$. For R~$=0.16$, the horseshoe and recirculation vortices disappear and can not be found even by lowering the $\lambda_{Ci}$ isosurface value.
The CRVP is the main swirling structure of the time-averaged velocity field. It is a robust structure present for all velocity ratios. Its trajectory lowers when the velocity ratio is decreased\cite{CaAi14}. 

\begin{figure}[htb!]
      \begin{center}
            \hspace*{-0.5cm}\begin{tabular}{Ic|c|cI}
             \Ghline\vspace{-0.025\textheight}&&\\
    \includegraphics[width=0.34\textwidth]{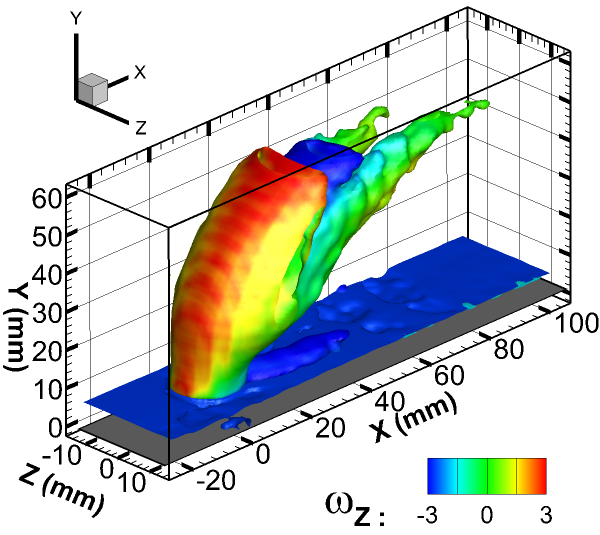}
    &    \includegraphics[width=0.34\textwidth]{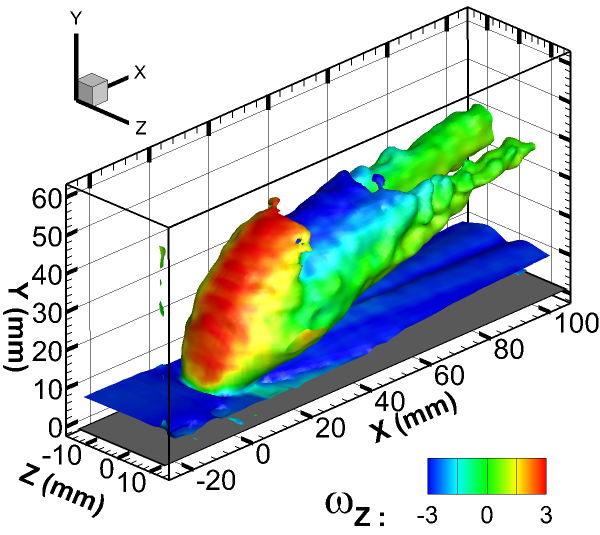}
    &    \includegraphics[width=0.34\textwidth]{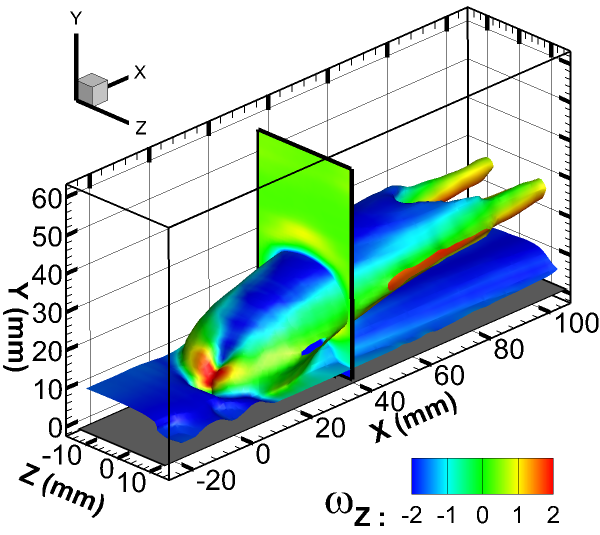}
    \\	a) R=1.71&b) R=1.39&c) R=1.16\\\hline\vspace{-0.025\textheight}&&\\
    \includegraphics[width=0.34\textwidth]{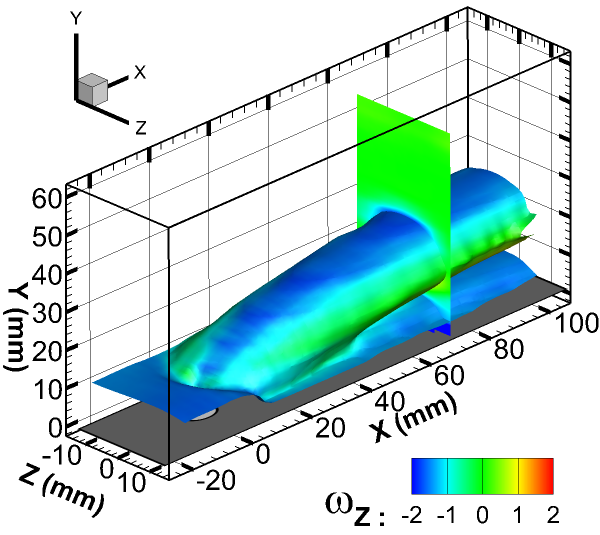}
    &    \includegraphics[width=0.34\textwidth]{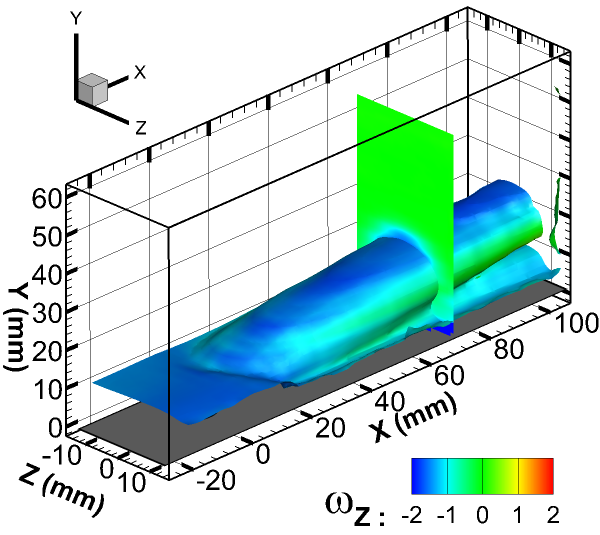}
    &    \includegraphics[width=0.34\textwidth]{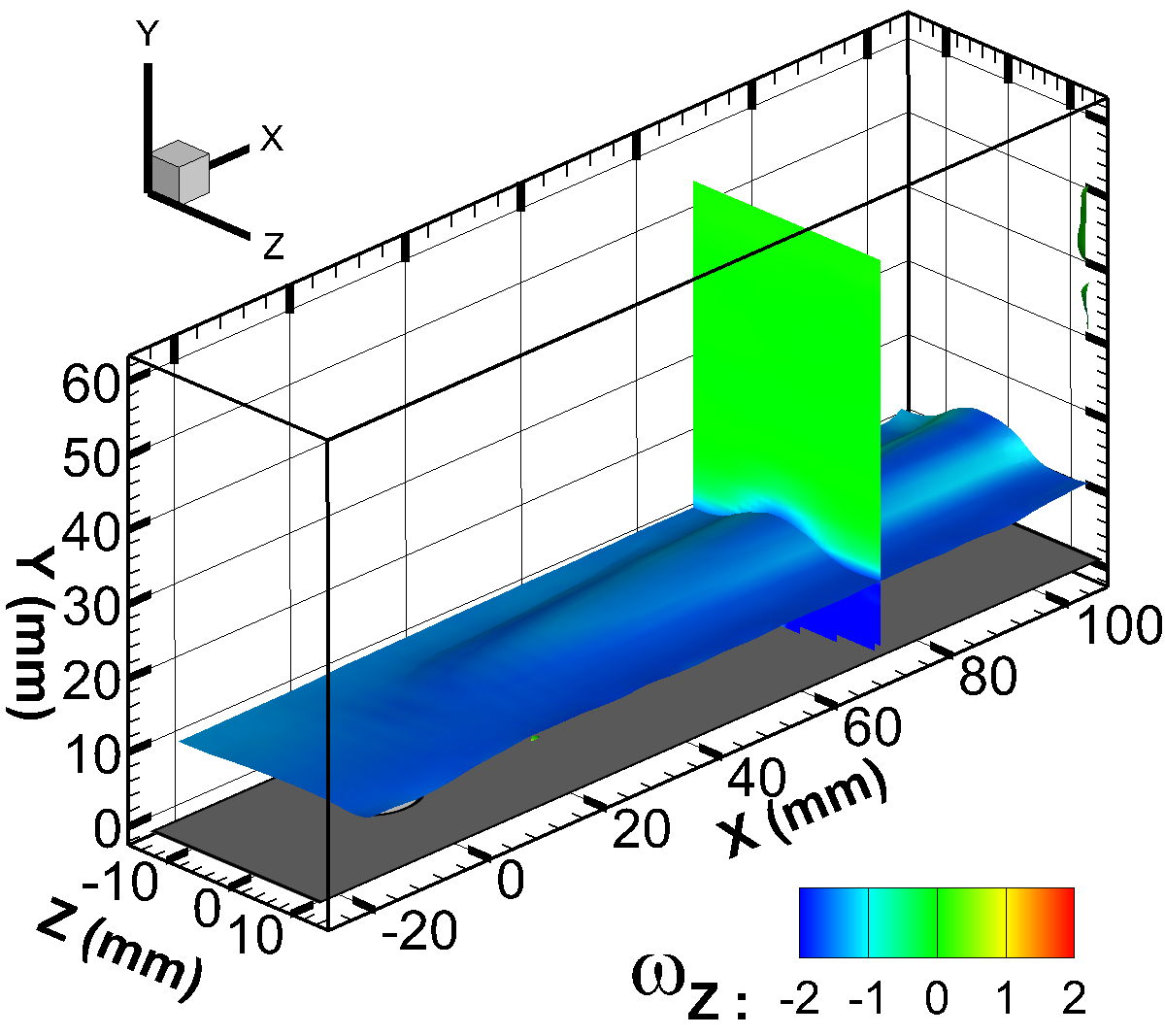}
    \\    d) R=0.90&e) R=0.55&f) R=0.16\\\Ghline
    \end{tabular}
    \caption{Time-averaged vorticity magnitude isosurfaces colored by transversal vorticity $\omega_{Z}$. $|\omega|=2\ s^{-1}$ (a, b, c). $|\omega|=1.5\ s^{-1}$ (d, e, f).}
\label{fig:isoVormag_varVR_mean}
    \end{center}
\end{figure}

On Fig.~\ref{fig:isoVormag_varVR_mean}, isosurfaces of vorticity magnitude $|\omega|$ show the whole time-averaged jet topology with both swirl and shear. Because the vorticity magnitude includes the transversal shear, the boundary layer and its deformation due to the presence of the jet can be clearly seen. At high velocity ratios, the jet deeply penetrates into the crossflow resulting into a strong shear between the jet and the crossflow. With respect to the crossflow orientation, one can define an upstream and a downstream shear-layer. The isosurfaces of Fig.~\ref{fig:isoVormag_varVR_mean} are colored by the transversal vorticity $\omega_{Z}$ to easily distinguish the upstream from the downstream jet shear-layer.\\ 
For high velocity ratios (R~$=1.71$, Fig.~\ref{fig:isoVormag_varVR_mean}a), the magnitude of the upstream and downstream shear-layers are very close. The jet crosses briefly the boundary layer without significant interaction and quickly leaves the measurement volume by the top. The jet is then in the \lq\lq{}fully detached jet regime\rq\rq{}\cite{Bidan2013}. For R~$=1.39$ (Fig.~\ref{fig:isoVormag_varVR_mean}b), the upstream shear-layer is shorter than the downstream one and therefore less intense. For a close-to-one velocity ratio (R~$=1.16$, Fig.~\ref{fig:isoVormag_varVR_mean}c), the positive shear between the jet and the crossflow velocities is considerably weaker on the upstream jet side. It can still be visualized in a $\omega_{Z}$ colored transverse plane. On the opposite, the downstream side keeps a strong shear due to the velocity difference between the jet velocity and the very low velocity area provided by the covering of the crossflow by the jet. 

For low velocity ratios (R~$=0.9$ and R~$=0.55$, Fig. \ref{fig:isoVormag_varVR_mean}d,e), the upstream shear layer disappears completely while the negative downstream shear layer stays strong. The crossflow bent over the jet, interacts with the boundary layer and starts forming in the vicinity of the jet exit a semi-cylindrical vortical structure. It is a vorticity magnitude shell connected with and embedded into the boundary layer,  similar to the one predicted by \citet{Gopalan2004}.

For each velocity ratio from R~$=1.71$ to R~$=0.55$, a small lowering upstream the jet exit of the transverse shear $\omega_{Z}$ isosurface denotes the existence of the vertical velocity induced by the horseshoe vortex.
This lowering can not be seen anymore for the lowest velocity ratio (R~$=0.16$).
At this very low velocity ratio, the jet is in an "attached regime" as suggested by \citet{Bidan2013}. The whole jet structure stays embedded in the boundary layer, forming in the time-averaged field a fully closed vorticity shell\cite{Bidan2013}.

\subsection{Instantaneous swirling structures}
\begin{figure}[p!]
      \begin{center}
            \vspace*{-1.5cm}\begin{tabular}{Ic|c|cI}
            \Ghline\vspace{-0.025\textheight}&&\\
    \includegraphics[width=0.33\textwidth]{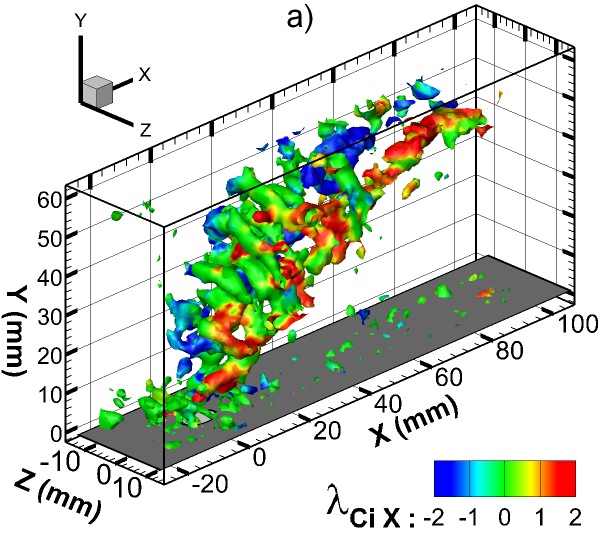}
    &
    \includegraphics[width=0.33\textwidth]{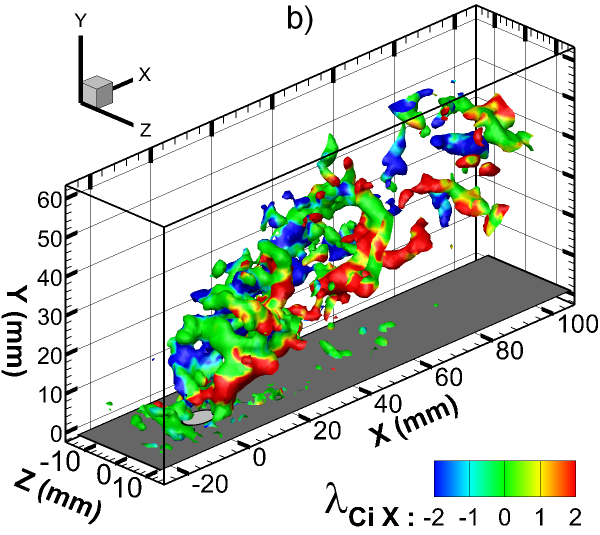}
    &
    \includegraphics[width=0.33\textwidth]{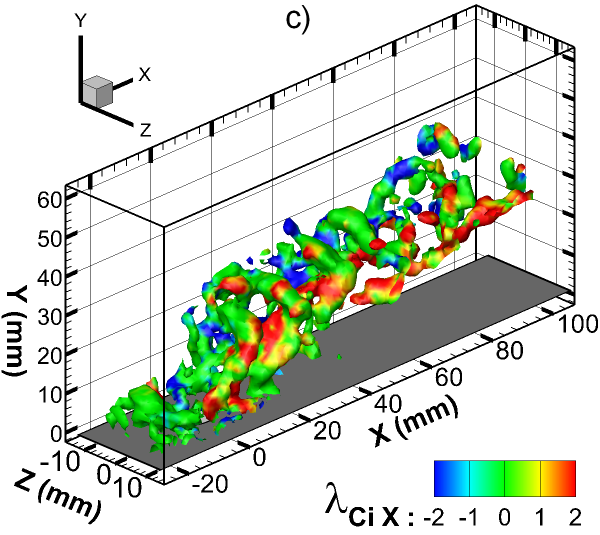}
    \\
    \vspace{-0.025\textheight}&&\\
    \includegraphics[width=0.33\textwidth]{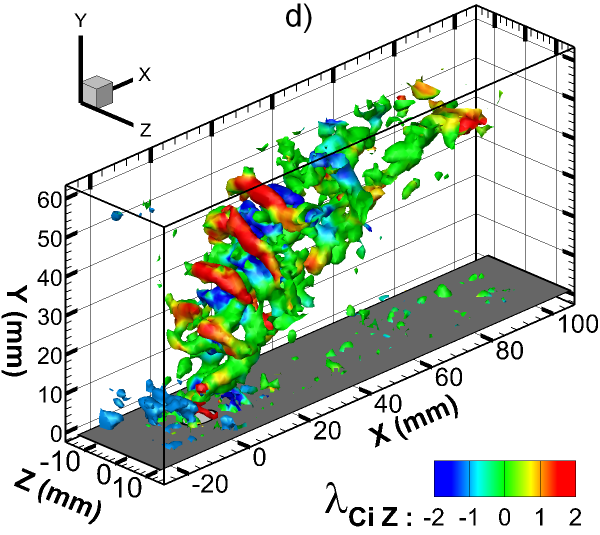}
    &
    \includegraphics[width=0.33\textwidth]{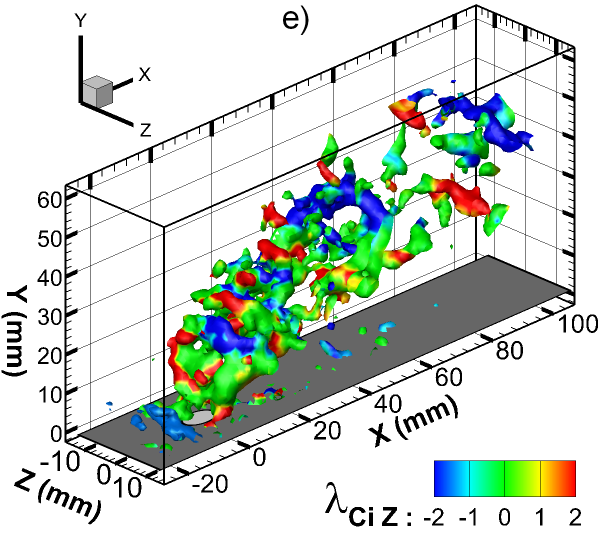}
    &
    \includegraphics[width=0.33\textwidth]{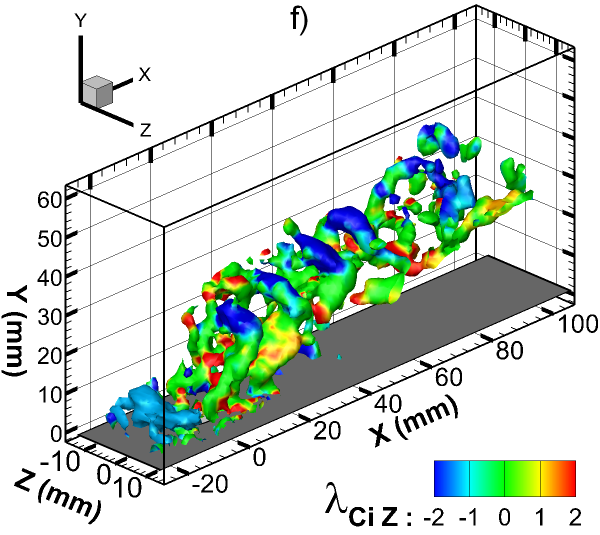}
\vspace{-0.01\textheight}\\\hline
    R=1.71&R=1.39&R=1.16\\\Ghline\multicolumn{3}{c}{}\vspace{-0.025\textheight}\\\Ghline\vspace{-0.03\textheight}&&\\
    \includegraphics[width=0.33\textwidth]{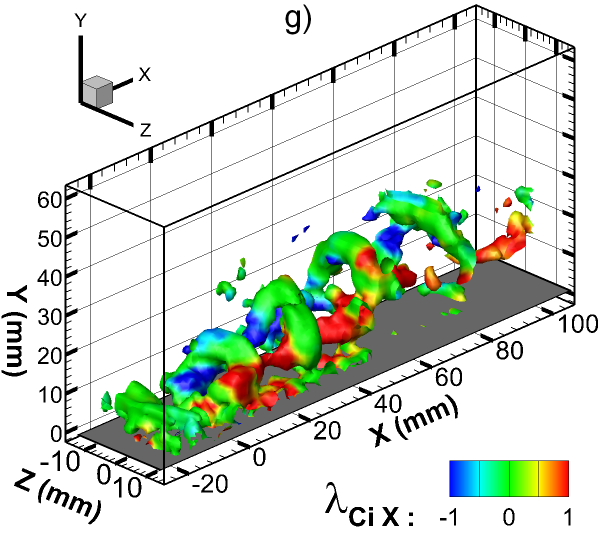}
    &
    \includegraphics[width=0.33\textwidth]{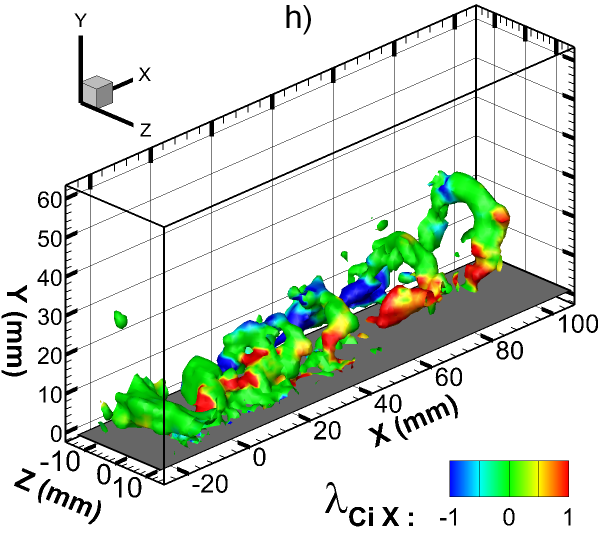}
    &
    \includegraphics[width=0.33\textwidth]{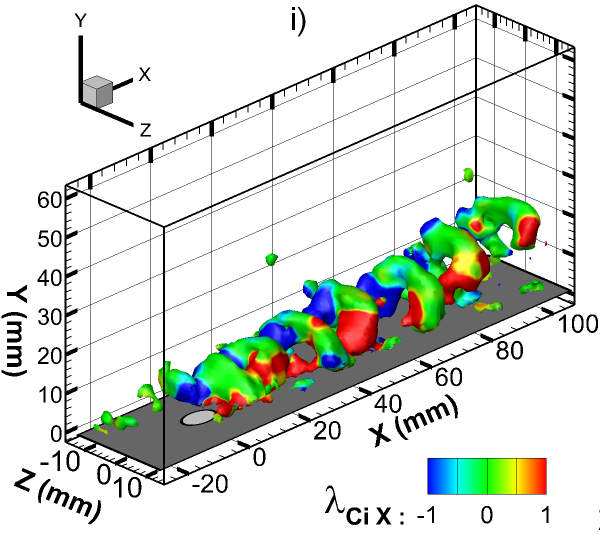}
    \\
\vspace{-0.025\textheight}&&\\
    \includegraphics[width=0.33\textwidth]{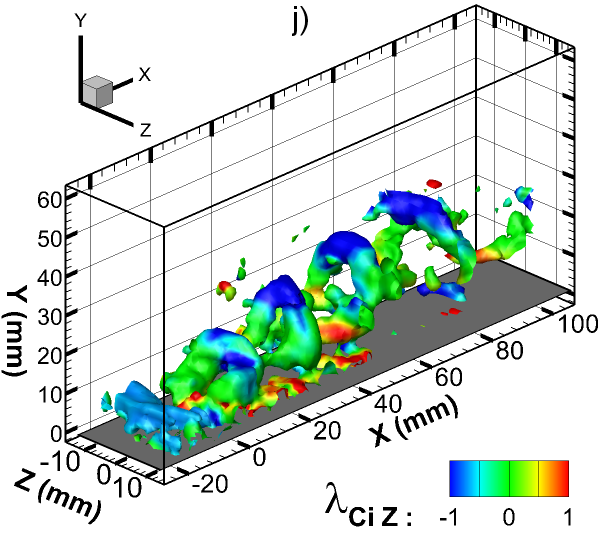}
    &
    \includegraphics[width=0.33\textwidth]{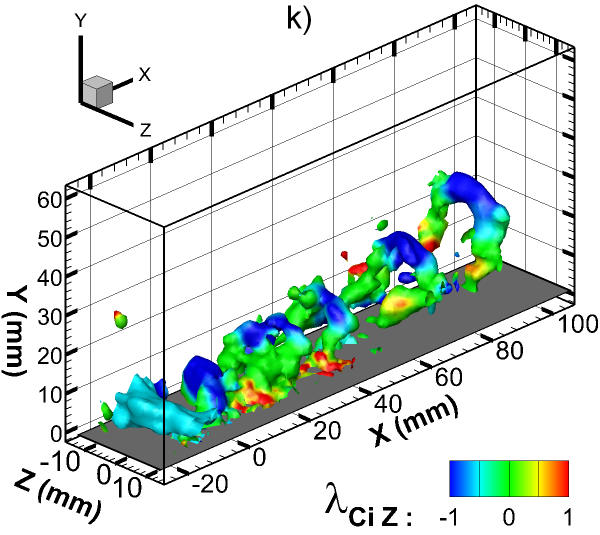}
    &
    \includegraphics[width=0.33\textwidth]{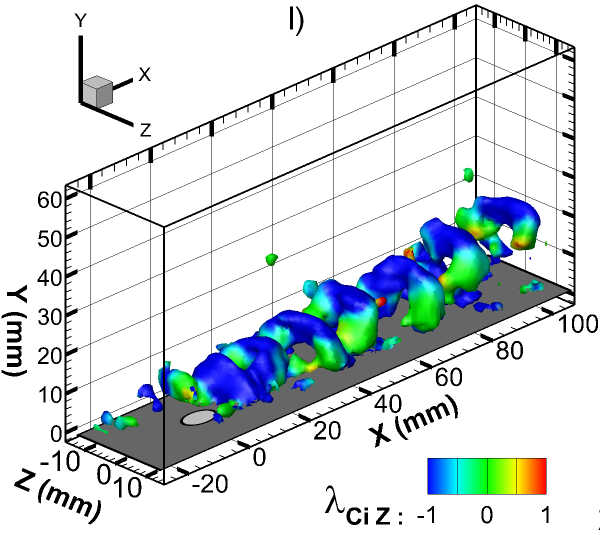}
    \vspace{-0.01\textheight}\\\hline
    R=0.90&R=0.55&R=0.16\\\Ghline
    \end{tabular}
    \caption{Visualization of the instantaneous  swirling structures using $\lambda_{CI}$ isosurfaces (movies online). For the six velocity ratios two color maps are used: the upper figure is colored by the longitudinal swirl $\lambda_{CI\ X}$ (a,b,c,g,h,i) while the lower figure is colored by the transversal swirl $\lambda_{CI\ Z}$ (d,e,f,j,k,l). $\lambda_{CI}=2.5\ s^{-1}$ (a,b,c). $\lambda_{CI}=1.5\ s^{-1}$ (d,e,f).}
\label{fig:isoLci_varVR_insta}
    \end{center}
\end{figure}

	For the same velocity ratios, Fig. \ref{fig:isoLci_varVR_insta} shows the swirling structures of the instantaneous velocity field using isosurfaces of $\lambda_{Ci}$ colored by the longitudinal and transversal swirls, respectively $\lambda_{Ci\ X}$ (Fig. \ref{fig:isoLci_varVR_insta}a,b,c,g,h,i) and $\lambda_{Ci\ Z}$ (Fig. \ref{fig:isoLci_varVR_insta}d,e,f,j,k,l). It exhibits very different vortex systems depending on the velocity ratio.
	
\begin{figure}[htb!]
      \begin{center}
            \begin{tabular}{c}
    \includegraphics[width=0.8\textwidth]{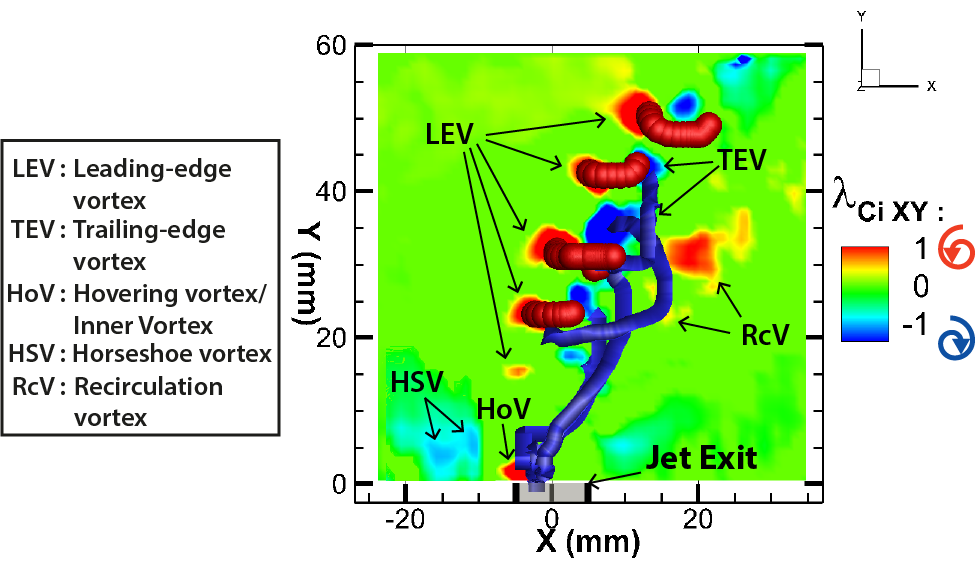}
    \\ 
a)\\
    \end{tabular}
\caption{ Visualization of the symmetry plane colored in $\lambda_{Ci\ Z}$. Vortex cores show the LEVs and TEVs arms and legs. R=1.71 case. These vortex cores are defined as streamlines of the [$\lambda_{Ci\ X}$, $\lambda_{Ci\ Y}$, $\lambda_{Ci\ Z}$] field started from $\lambda_{Ci\ Z}$ maxima in the symmetry plane, and each line is colored with the color of the maximum it is issued from.}
    \label{Fig:Topoinsta_vortexcores}
    \end{center}
\end{figure}

	For a high velocity ratio (R~$=1.71$, Fig.~\ref{fig:isoLci_varVR_insta}a,d), one can see the vortices entanglement characteristic of the jet in crossflow topology.  Figure~\ref{Fig:Topoinsta_vortexcores} shows a more detailed view of this high velocity ratio JICF topology. The symmetry plane is colored in $\lambda_{Ci\ Z}$. Vortex cores are colored with the same color that the jet shear layer vortices (LEV and TEV). They are used to illustrate the entanglement of these vortices outside of the symmetry plane. At this velocity ratio, the instantaneous observations support the JICF topology of loop vortices proposed by \citet{Lim2001} (cf Fig.~\ref{fig:JICF_Topologies}b) and supported by \citet{Marzouk2007}.  The destabilization of the upstream and downstream shear-layers by the Kelvin-Helmholtz instability leads to the formation of periodically shed vortex loops located on both sides of the jet: the shear-layer vortices. The Leading-Edge Vortices (LEV) are located along the upstream shear-layer while the Trailing-Edge Vortices (TEV) are located on the downstream side of the jet.  The heads of the LEV ($\lambda_{Ci\ Z}>0$, in red) and TEV ($\lambda_{Ci\ Z}<0$, in blue) are easily seen with the $\lambda_{Ci\ Z}$ coloring (Fig.~\ref{fig:isoLci_varVR_insta}d). The $\lambda_{Ci\ X}$ coloring shows that the swirling motions of the arms and legs of the shear-layer vortices (LEV and TEV) are the same than the swirling motion of the branches of the time-averaged CRVP (Fig.~\ref{fig:isoLci_varVR_insta}a). It shows how the entanglement and advection of the arms and legs of the instantaneous shear layer vortices lead to the CRVP when time-averaged\citep{Cortelezzi2001,Marzouk2007}. A small instantaneous horseshoe vortex can also be seen upstream of the jet exit.

Because decreasing the velocity ratio leads to a weakening of the upstream shear-layer, the shear layer vortices  weaken as well. This phenomenon is visible for R~$=1.39$ (Fig.~\ref{fig:isoLci_varVR_insta}e)  and R~$=1.16$ (Fig.~\ref{fig:isoLci_varVR_insta}f) where the $\lambda_{Ci}$ isosurface exhibits less and less red $\lambda_{Ci\ Z}>0$ on the upstream side. 

For R~$=0.9$ (Fig.~\ref{fig:isoLci_varVR_insta}g and \ref{fig:isoLci_varVR_insta}j) and for the same isosurface value, no leading-edge vortex can be seen anymore in the upstream shear layer. The disappearance of the LEVs uncovers the downstream side and makes the TEVs directly visible. They are still well defined along the strong lee-side shear-layer.

For lower velocity ratios (R~$=\{1.16, 0.90, 0.55\}$, Fig.~\ref{fig:isoLci_varVR_insta}c, \ref{fig:isoLci_varVR_insta}f, \ref{fig:isoLci_varVR_insta}g, \ref{fig:isoLci_varVR_insta}h, \ref{fig:isoLci_varVR_insta}j, \ref{fig:isoLci_varVR_insta}k), the vortex organization on the downstream side of the jet follows the topologies observed by \citet{Blanchard1999,Bidan2013}, with a succession of very well-defined hairpin-shaped trailing-edge vortices. The trajectory of the TEVs is lowered when the velocity ratio is decreased. The instantaneous HSV swirling intensity progressively becomes comparable, then larger than the TEV's swirling intensity. It illustrates the smooth transition observed by \citet{Bidan2013} from the detached jet regime, where the shear-layer vortices are preeminent,  toward the attached jet regime dominated by the influence of the near wall vortices.

For R~$=0.16$, an alley of hairpin-like vortices can still be observed, but the horseshoe vortex can not be seen anymore, unless it became a part of the hairpin vortices formation process and can no longer be clearly distinguishable from them. This observation suggests that a transition occurs at very low velocity ratio toward a flow topology very different from the classical high velocity ratio JICF flow topology. One can also notice that it does not correspond to the recent numerical result obtained by \citet{Ilak2012}. This important point will be discussed later.\\

\section{Evolution of the time cumulative spatial distributions of instantaneous vortices  for $\mathbf{0.15<R<2.2}$}
\begin{figure}[htb!]
      \begin{center}
            \begin{tabular}{c}
    \includegraphics[width=0.5\textwidth]{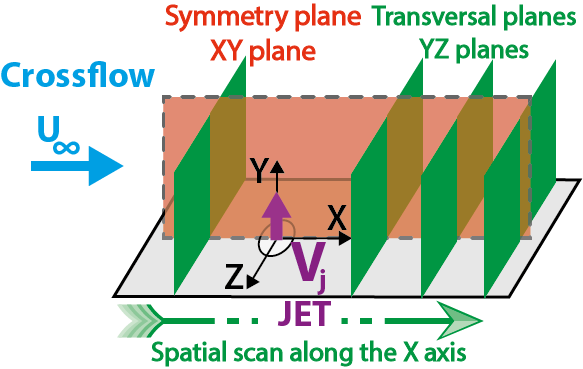}
    \end{tabular}
\caption{Reference detection planes.}
    \label{Fig:blobdetection}
    \end{center}
\end{figure}

	To highlight and clarify the salient features and transitions of the instantaneous velocity fields, a statistical approach has been adopted. We focused on the time cumulative spatial distribution of vortices in specific planes of interest: the symmetry plane and the transverse planes (Fig.~\ref{Fig:blobdetection}). One can see on Fig.~\ref{fig:JICF_Topologies}b that the symmetry plane intersects most of the swirling structures at the exception of the vortices leading to the CRVP in the time-averaged velocity field. Hence after a brief description of the vortex detection procedure, we focus on the cumulative distribution of vortices in the symmetry plane and its link with the instantaneous JICF topology. In the following sub-section, a similar study in the transverse planes has been conducted to take into account the topological changes associated with the streamwise-oriented vortices. 

\subsection{Vortex subgrid detection}

The first step is to define a rigorous quantitative criterion to automatically search and identify the main swirling structures present in the successive instantaneous 3D velocity fields. For each vortex, the position of the peak of the intensity swirl is detected using the local extremum of the 2D $\lambda_{Ci}$ showing the component of rotation in the plane of interest ($\lambda_{Ci\ Z}$ in the symmetry plane,  $\lambda_{Ci\ X}$ in the transversal planes). 
Starting from the rough position of each maximum on the interpolation grid, a sub-grid location is determined using the swirling intensity of the nodes in the close-neighborhood of each maxima.

\subsection{Evolution of the cumulative distribution of swirling structures in the symmetry plane for decreasing velocity ratios}

\begin{figure}[htb!]
      \begin{center}
a)    \includegraphics[width=0.7\textwidth]{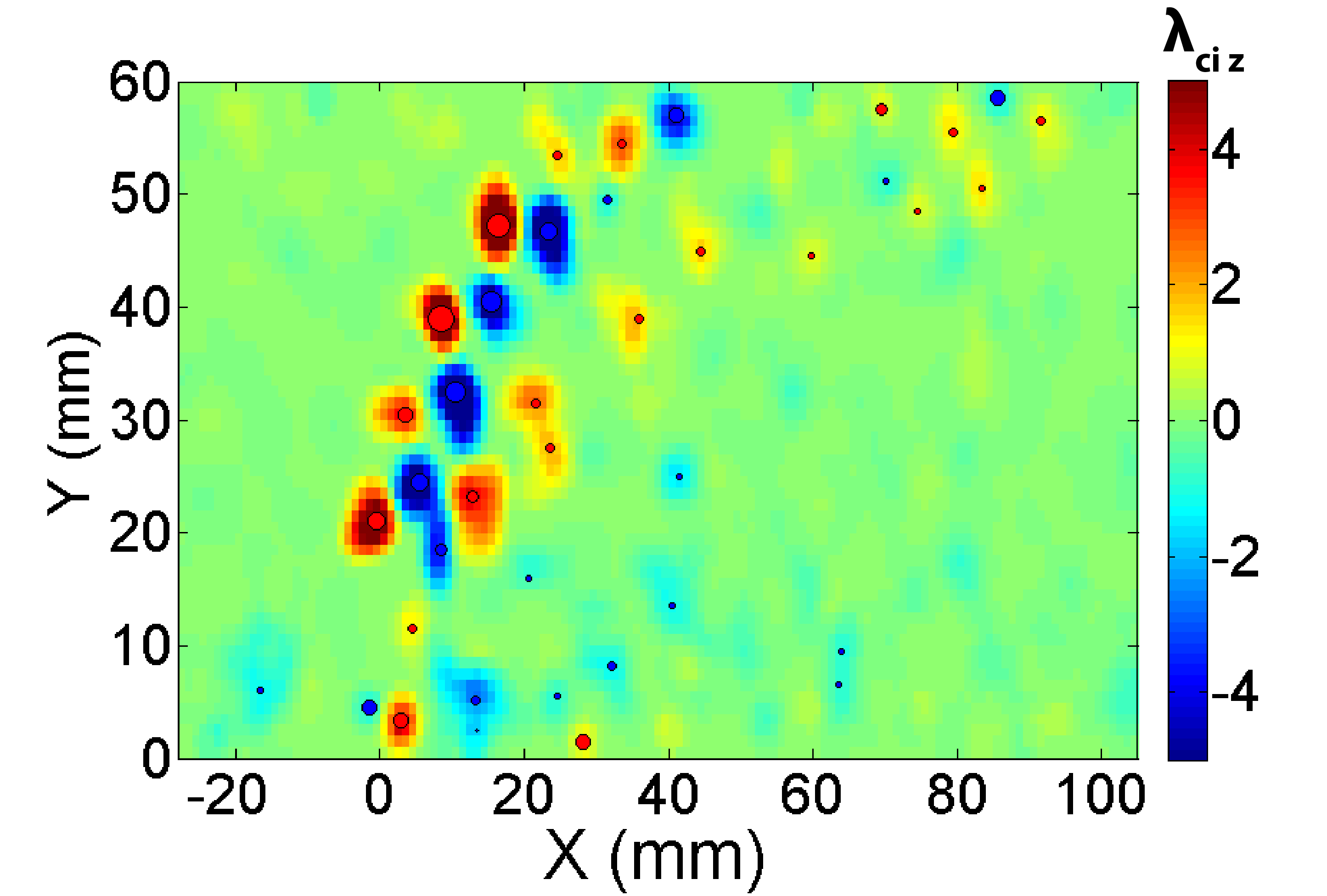}
\\b)    \includegraphics[width=0.7\textwidth]{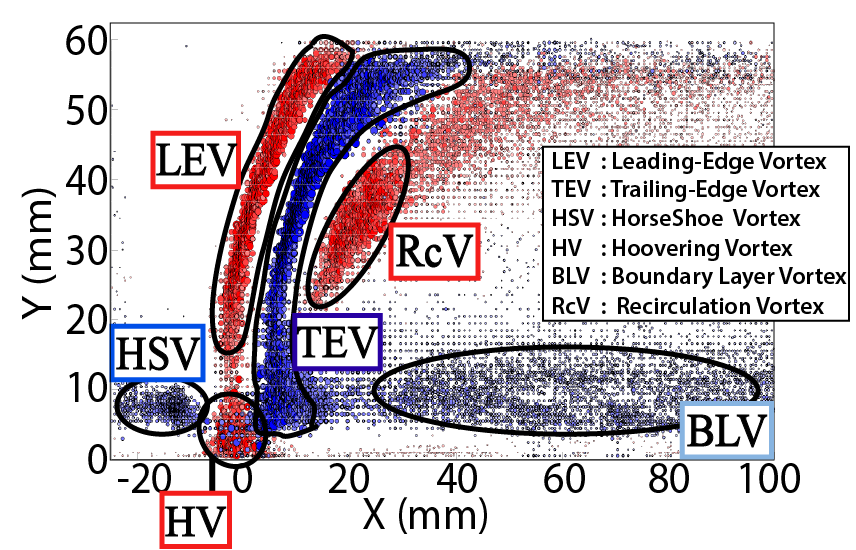}
\caption{a) Positions of the detected instantaneous vortices in the symmetry plane at a given time step. An online video shows this detection process on a 50 fields time-serie. b) R=2.13. Cumulative distribution over 1000 time steps of all the positions of the detected instantaneous vortices in the symmetry plane. An online video shows this cumulative process on the firsts 50 fields. The size and coloration of each marker are proportional to the swirling intensity peak value of each vortex.}
    \label{Fig:CumulativeDistribution}
    \end{center}
\end{figure}

To quantitatively study the topology of the instantaneous 3D velocity fields, a statistical approach has been adopted.	
On the jet symmetry plane of each instantaneous velocity fields, the positions of the twenty first positive and negative swirling structures are found using the local maximum detection algorithm discussed in the previous section. Keeping twenty maxima for each time step is enough to ensure that all relevant vortical structures of the symmetry plane have been detected. 

Figure~\ref{Fig:CumulativeDistribution}a shows an example of instantaneous positions of the positive (clockwise) and negative (anti-clockwise) extrema of the $\lambda_{Ci\ Z}$ field in the symmetry plane of the jet. 
Figure \ref{Fig:CumulativeDistribution}b shows the superposition for 1000 timesteps of all the positions of the detected instantaneous positive and negative extrema of the $\lambda_{Ci\ Z}$ field in the symmetry plane of the jet. The size and coloration of each marker are proportional to its swirling intensity peak to graphically emphasize the stronger vortices. This representation allows for an easy visualization of the spatial distribution of the main spanwise swirling structures.

\begin{figure}[p!]
      \begin{center}
            \vspace*{-1cm}\begin{tabular}{Ic|cI}
\Ghline\multicolumn{2}{IcI}{\vspace*{-0.03\textheight}}\\
\multicolumn{2}{IcI}{\includegraphics[width=0.5\textwidth]{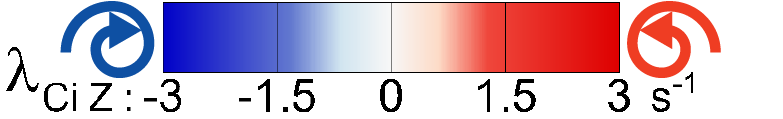}}
\vspace*{-0.01\textheight}\\\Ghline&\vspace*{-0.03\textheight}\\
    \includegraphics[width=0.45\textwidth]{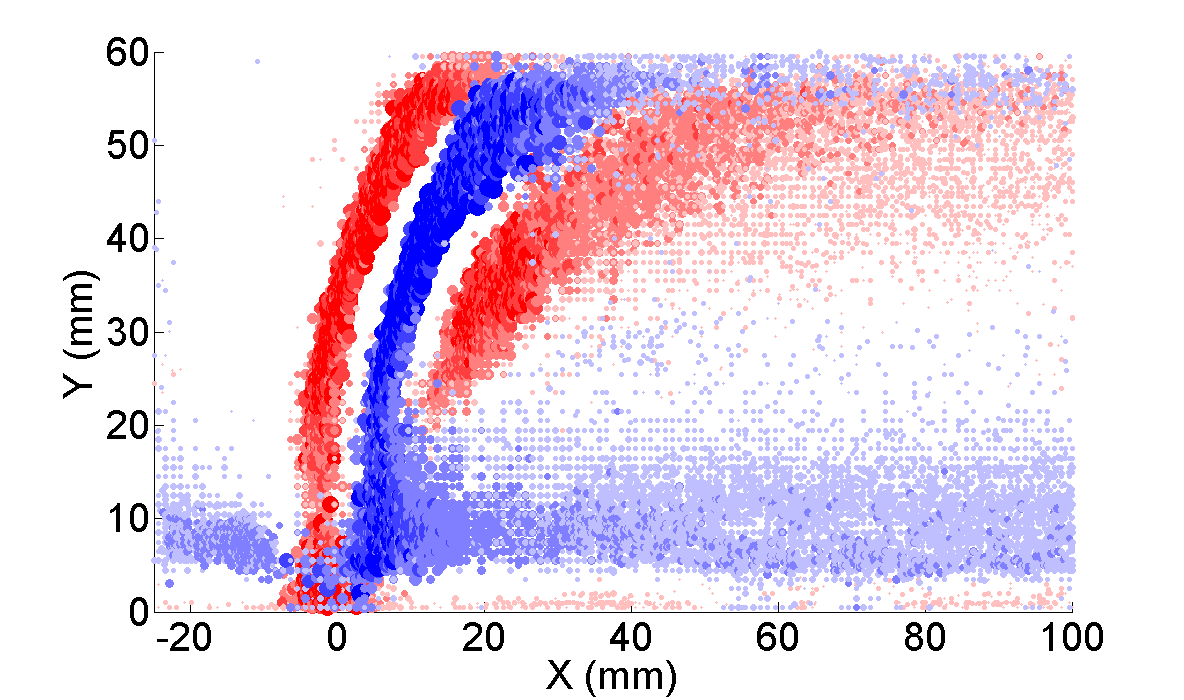}
    &    \includegraphics[width=0.45\textwidth]{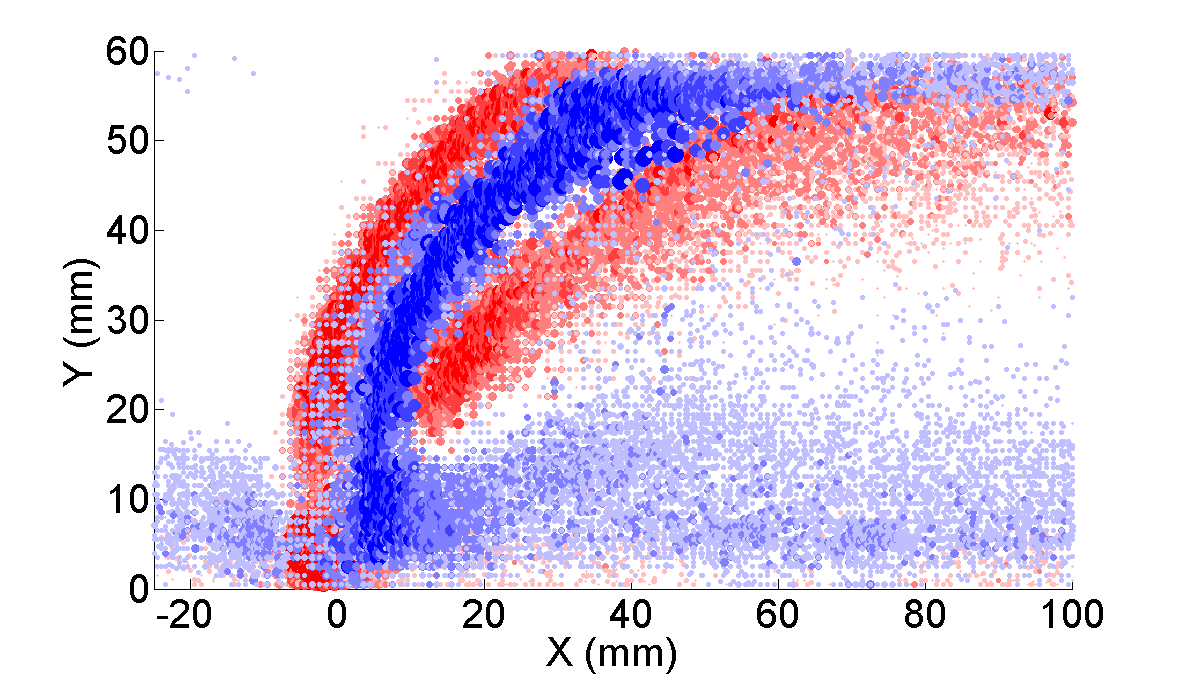}
    \vspace*{-0.01\textheight}\\ 
a) R=2.13&b) R=1.71\\\hline&\vspace*{-0.03\textheight}\\
    \includegraphics[width=0.45\textwidth]{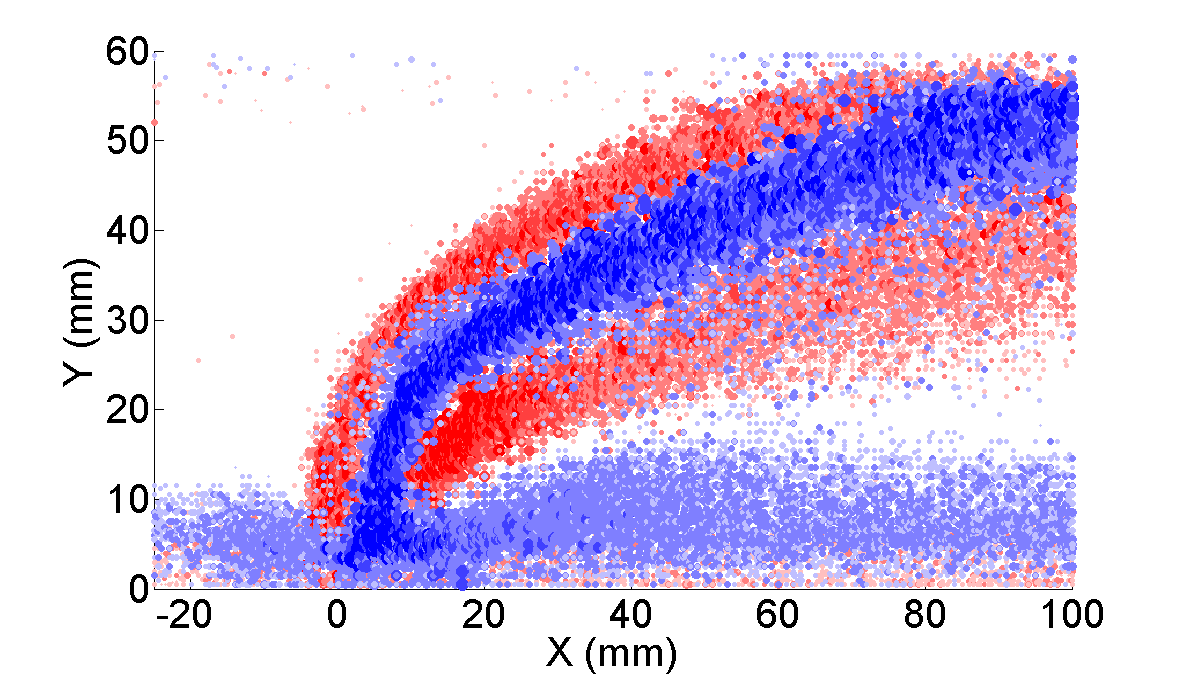}
    &    \includegraphics[width=0.45\textwidth]{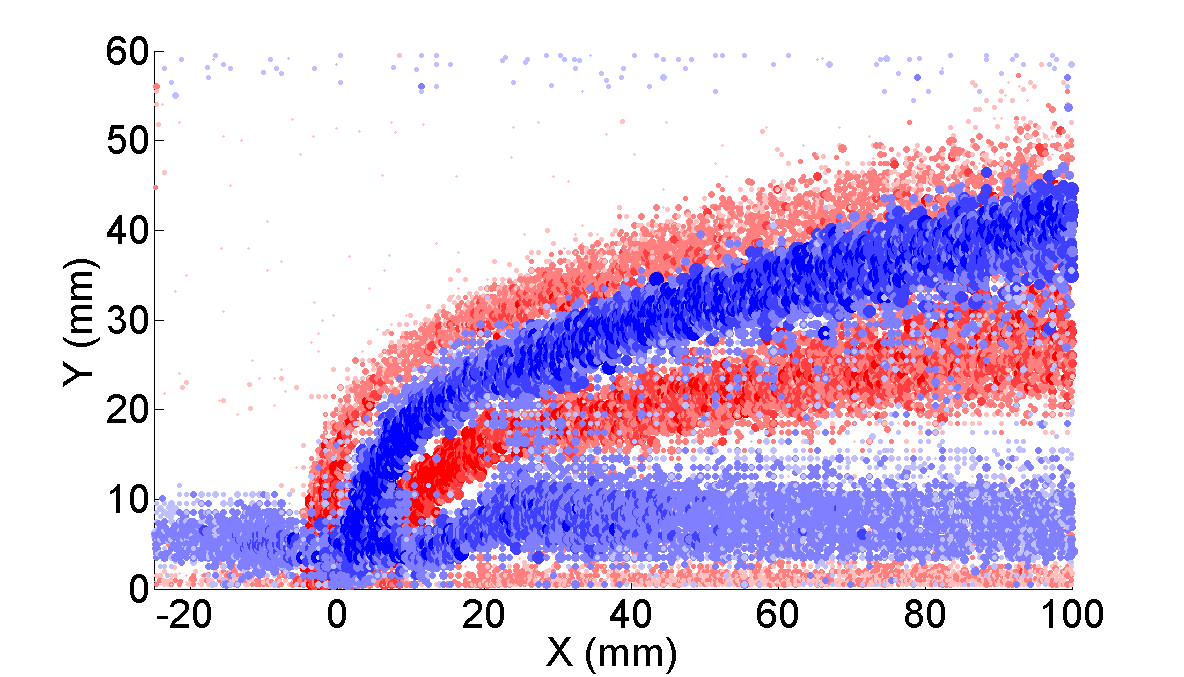}
    \vspace*{-0.01\textheight}\\ 
    c) R=1.39&d) R=1.16\\  \hline&\vspace*{-0.03\textheight}\\
    \includegraphics[width=0.45\textwidth]{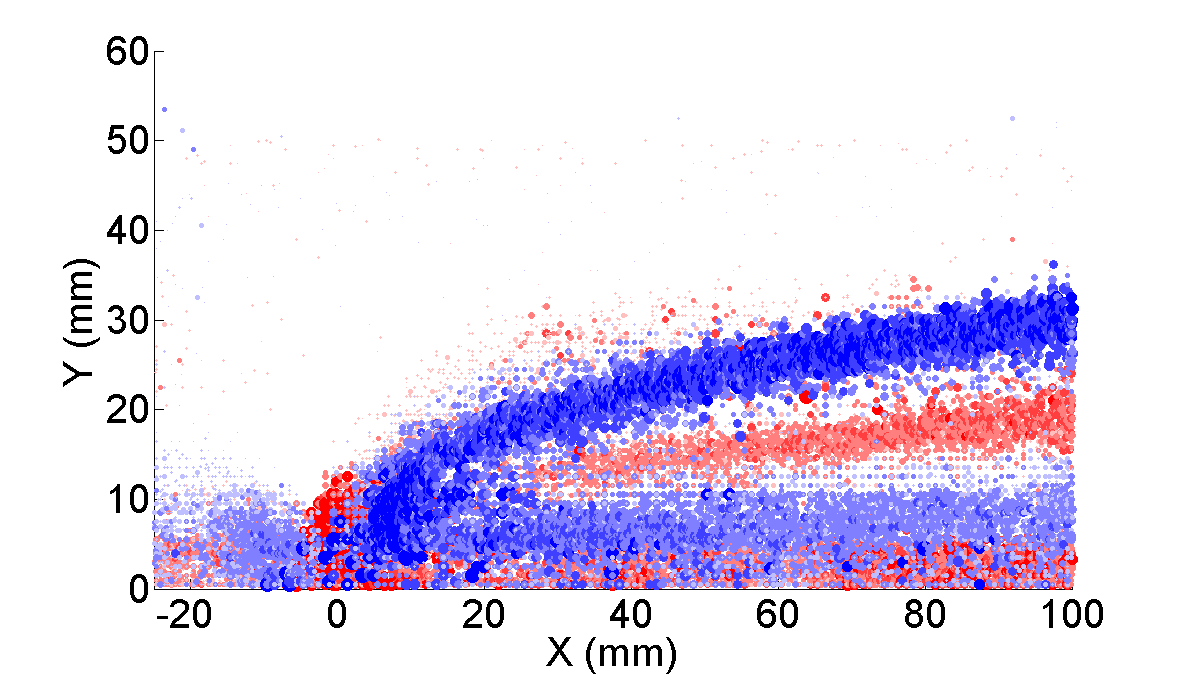}
    &    \includegraphics[width=0.45\textwidth]{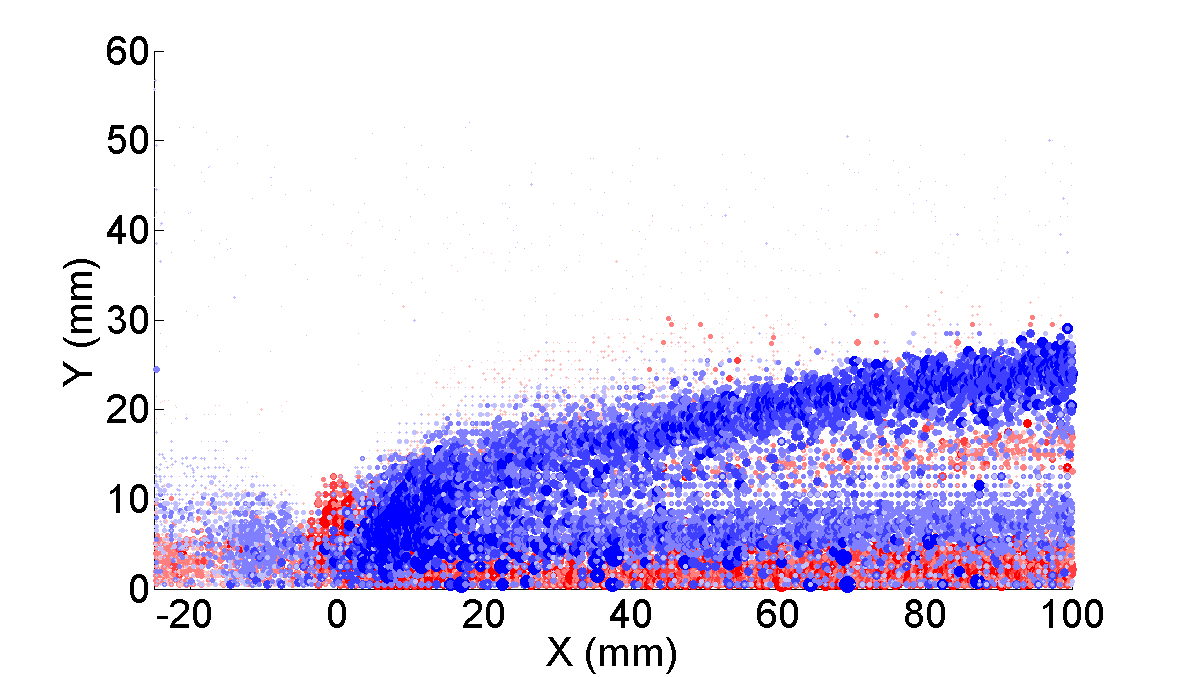}
    \vspace*{-0.01\textheight}\\ 
      e) R=0.90&f) R=0.55 \\ 
\hline&\vspace*{-0.03\textheight}\\  
    \includegraphics[width=0.45\textwidth]{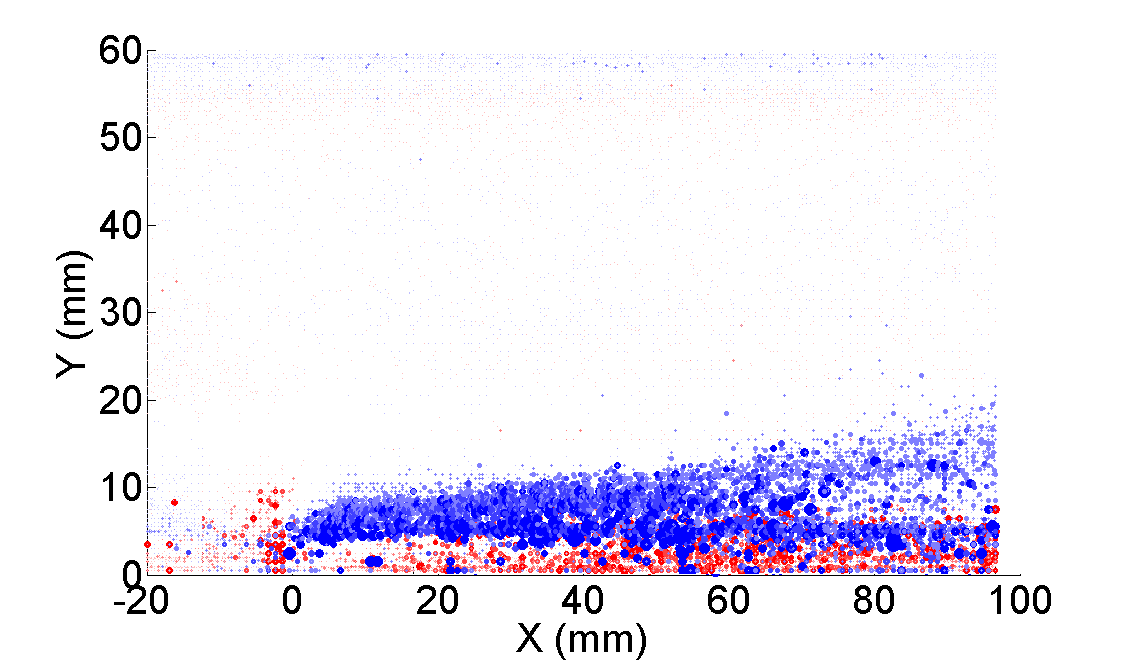}
    &    \includegraphics[width=0.45\textwidth]{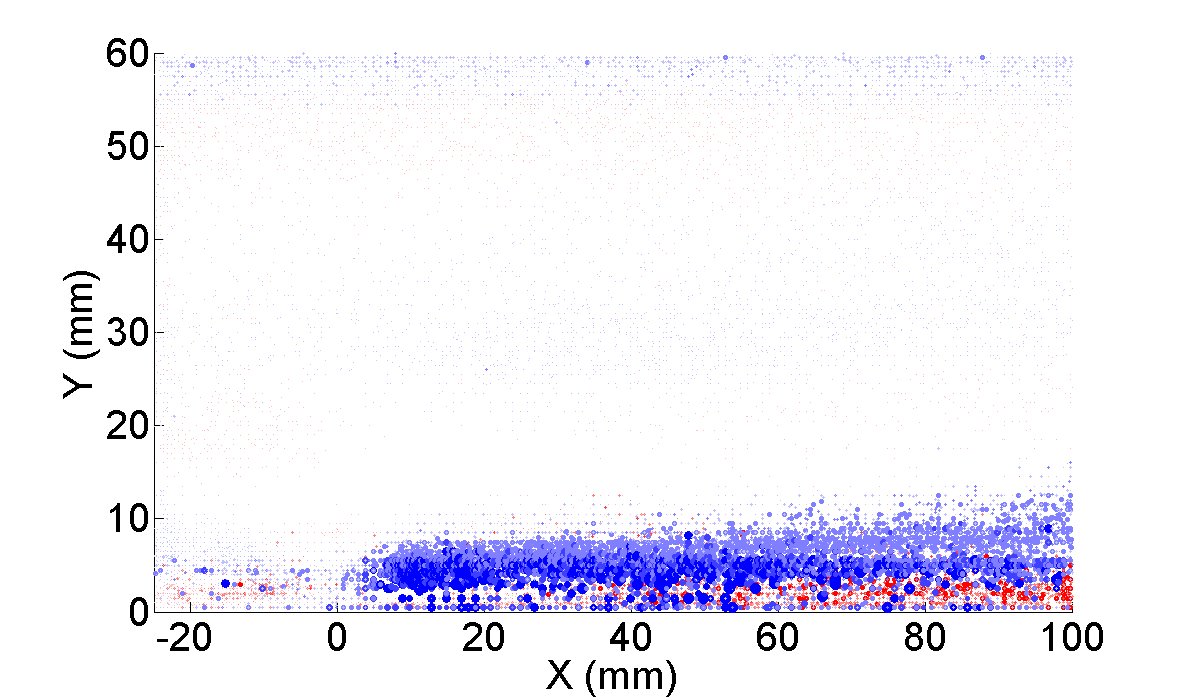}
\vspace*{-0.01\textheight}\\ 
     g) R=0.34&h) R=0.16\\\Ghline
    \end{tabular}
\caption{Cumulative distribution over 1000 time steps of all the positions of the detected instantaneous vortices in the symmetry plane for decreasing velocity ratios. The size and coloration of each marker are proportional to the swirling intensity peak value of each vortex.}
\label{FigTopoInstaJetRond:topoblob_varVR}
    \end{center}
\end{figure}

Different regions can be distinguished around the jet. With this representation, the jet trajectory is clearly visible: the convection of the LEVs and TEVs along the upstream and downstream shear layers forms two clouds of opposite sign on both sides of the jet. 
The horseshoe vortex (HSV) forms a small distribution of negative swirling structures located upstram near the jet exit. 
Another small cloud of anti-clockwise vortices ($\lambda_{Ci\ Z}>0$, in red in Fig. \ref{Fig:CumulativeDistribution}b) can be seen at the base of the jet on the upstream side. Even if this distribution may seem to be part of the upstream shear layer vortices, it has to be distinguished from the cloud created by the LEVs. Both clouds are spatially distinct, separated by a region where the vortices are less intense and where their distribution is less dense. From its spatial localization at the base of the jet, its anti-clockwise rotation and, above all, its linked behavior with the HSVs distribution (Fig. \ref{FigTopoInstaJetRond:topoblob_varVR}), this distribution is undoubtedly composed of Hovering Vortices. This vortex, like the horseshoe vortex, results from the junction flow between the jet and the crossflow and has already been observed using dye visualization\cite{KELSO1995,Kelso1996}. This vortex has also been recently observed by \citet{Bidan2013}. Because at very low R, it is located in the jet pipe near the exit, they also called it Inner Vortex.\\
On the lee-side of the jet, lies a recirculation area characterized by very low velocities. Beyond this recirculation area, the crossflow finally bypass the jet resulting in a strong positive shear layer on the downstream side of the recirculation area where Recirculation Vortices are created (RcV).
Finally, some local weak swirls of the boundary layer can be seen in the symmetry plane, in the upper part of the boundary layer. Hence, we call them Boundary Layer Vortices (BLV). 
We use this cumulative spatial distribution of swirling structures in the symmetry plane to study the evolution of the JICF topology when the velocity ratio is decreased from R~$=2.13$ to R~$=0.16$ as illustrated on Fig.~\ref{FigTopoInstaJetRond:topoblob_varVR}.

For R~$=2.13$ (Fig. \ref{FigTopoInstaJetRond:topoblob_varVR}a), the swirling strength of the LEVs and TEVs are very close. The standard flow topology shown in Fig. \ref{fig:JICF_Topologies}b  is recovered. The experiments of Fig.~\ref{FigTopoInstaJetRond:topoblob_varVR}b,c,d,e,f,g,h respectively correspond to the ones presented in Fig.~\ref{fig:isoLci_varVR_mean}(a,b,c,d,e,f), Fig. \ref{fig:isoVormag_varVR_mean}a,b,c,d,e,f and Fig~\ref{fig:isoLci_varVR_insta}a,b,c,d,e,f.

In the Fig.~\ref{FigTopoInstaJetRond:topoblob_varVR}(a,b), the velocity gradients between the jet and the outer flow create strong upstream and downstream shear layers. Destabilization of the shear layers due to Kelvin-Helmholtz instability leads to the formation of the loop vortices which are convected along the shear layers and form the big positive and negative spatial distributions which corresponds to the boundaries of the jet in the symmetry plane. 

For $1.5>$~R~$>1$ (Fig.~\ref{FigTopoInstaJetRond:topoblob_varVR}(c, d)), while the jet and crossflow velocities become closer on the upstream side of the jet, the LEVs swirling intensity decreases. In the meantime, the jet is close enough from the upper part of the boundary layer to strengthen the interaction between the RcVs and BLVs whose numbers and swirling intensities strongly increase. 

For R~$=0.9$ (Fig.~\ref{FigTopoInstaJetRond:topoblob_varVR}(e)), the LEVs cloud has almost disappeared. It starts around X~$=20$~mm and is composed of a small number of weak vortices, still  a few times stronger than the background noise. In the following, a more quantitative analysis will help analyzing this evolution. As the jet becomes more and more embedded in the boundary layer, the confinement of the RcVs against the upper part of the boundary layer decreases their number and swirling intensity.
In the Fig.~\ref{FigTopoInstaJetRond:topoblob_varVR}(f,g,h), the LEVs distributions have totally disappeared.

Another remarkable feature of this evolution lies in the horseshoe and hovering vortices distributions.
Both distributions are still made of vortices with strong swirling strengths even for very low velocity ratio (R~$=0.55$). Both distributions can still be clearly seen in the Fig.~\ref{FigTopoInstaJetRond:topoblob_varVR}f.  The weak dependency of the HSVs swirling intensity with the velocity ratio is not surprising. Indeed, it has already been observed in the case of a junction flow in front of a cylindric obstacle\cite{BAKER1979,BAKER1980}: the swirling strength of the horseshoe vortex depends more on the diameter of the obstacle than on its height. In the present case, the lowering of the jet trajectory with decreasing velocity ratios matters less than the diameter of the jet nozzle which is constant in the present experiments.\\

\begin{figure}[htb!]
      \begin{center}\begin{tabular}{cc}
\includegraphics[height=0.3\textwidth]{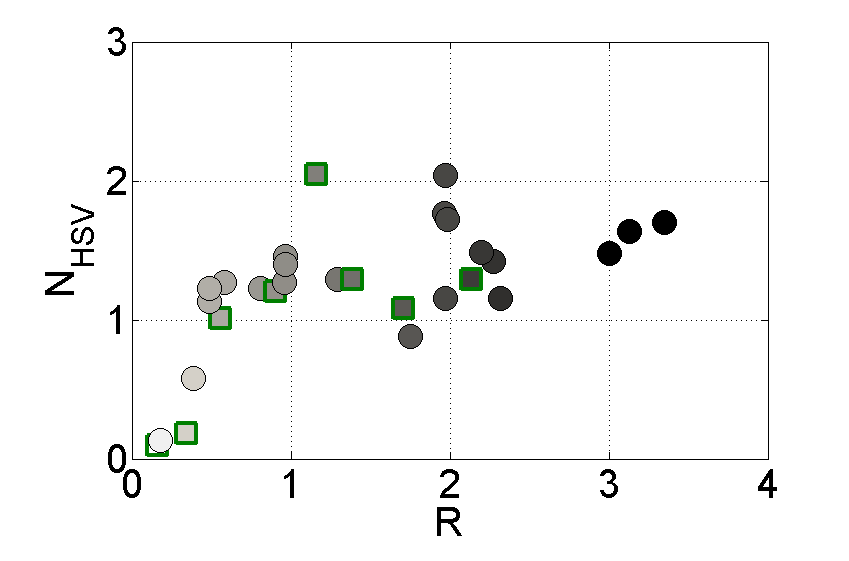}
&\includegraphics[height=0.3\textwidth]{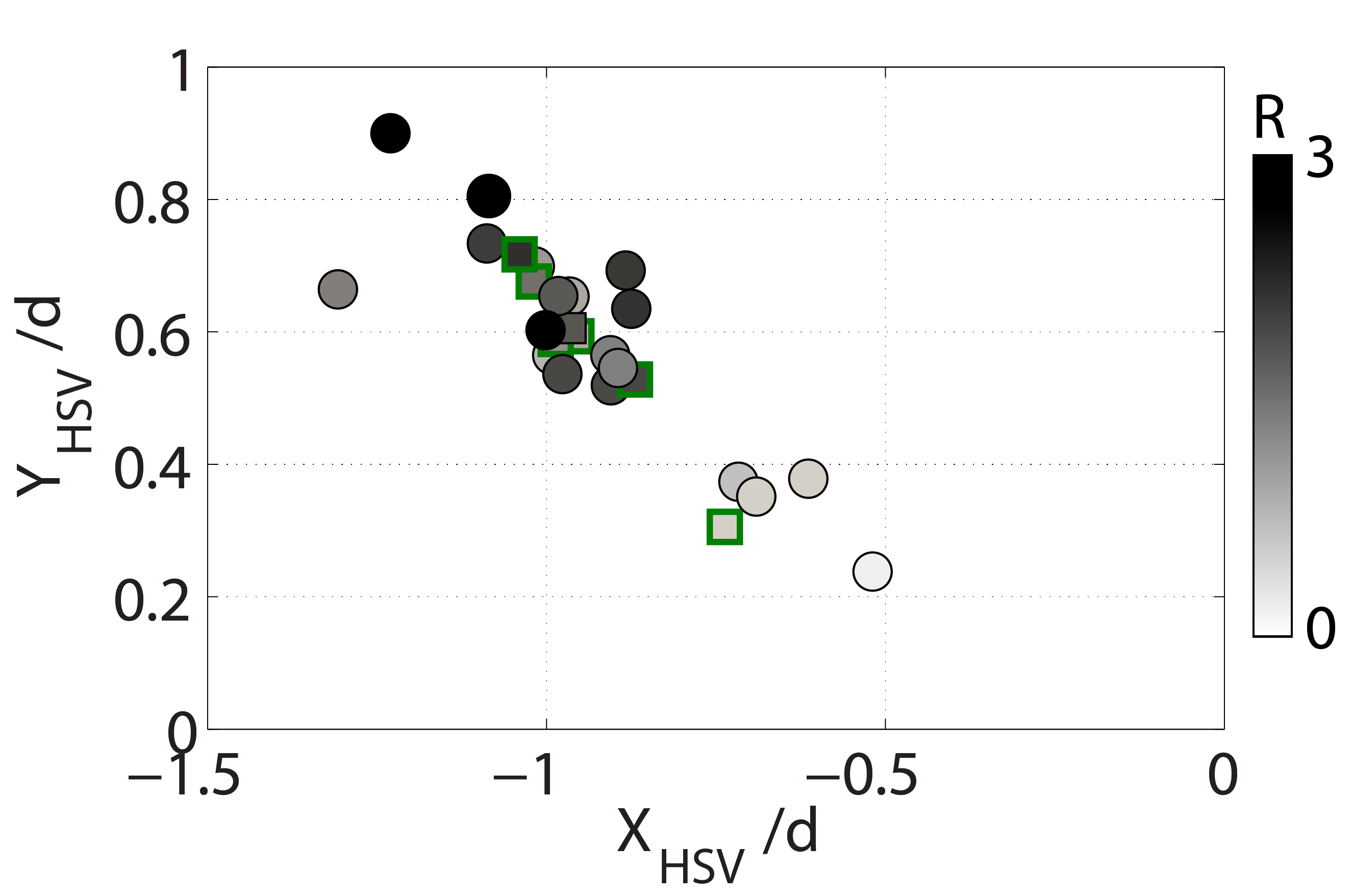}
\\
a)&b)
\end{tabular}
    \caption{a) Number of horseshoe vortex per instantaneous velocity field $N_{HSV}$ as a function of the velocity ratio. b) Spatially averaged position of the HSV computed as the center of mass of the HSV cloud. The particular experiments shown in this article in the Fig.~\ref{fig:isoLci_varVR_mean}, \ref{fig:isoVormag_varVR_mean}, \ref{fig:isoLci_varVR_insta}, \ref{FigTopoInstaJetRond:topoblob_varVR}, \ref{FigTopoJetRond:TrajCRVPinsta_backgrounds}, \ref{FigTopoJetRond:TrajCRVPinsta_XZ}, \ref{FigTopoJetRond:TrajCRVPinsta_highVR} are circled in green square markers. The markers are colored with respect to R to emphasize its effect on the HSV average position.}
    \label{fig:NbHV-HSV_vsVR}
    \end{center}
\end{figure}

Finally, for R~$=0.34$ and 0.16 (Fig.~\ref{FigTopoInstaJetRond:topoblob_varVR}g and \ref{FigTopoInstaJetRond:topoblob_varVR}h), both junction flow (HSV+HoV) clouds can not be seen anymore. 
A statistical approach is used to determine more quantitatively this phenomenon. From the spatial distribution of vortices, one can evaluate the averaged number of vortices present in each regions at each time steps as a function of R or $S_j$. For instance, the averaged number of horseshoe vortices per instantaneous velocity field $N_{HSV}$ is defined as $N_{HSV}=\frac{n_{HSV}}{n}$, 
where n~=~1000 is the number of instantaneous velocity field used for a given configuration, and $n_{HSV}$ is the total number of negative vortices detected in the HSV clouds of the vortices distribution in the symmetry plane (Fig.~\ref{FigTopoInstaJetRond:topoblob_varVR}). The same definition is used for the number of hovering vortices per instantaneous velocity field $ N_{HoV}$. 

Figure \ref{fig:NbHV-HSV_vsVR}a shows $N_{HSV}$ as a function of the velocity ratio R. Unlike the progressive disappearance of the leading edge vortices, the strong junction flow vortices disappear sharply without any noticeable weakening of their swirling strength nor diminution of their number.
This sharp disappearance of the HSV cloud is probably related to the limitation of resolution of the velocimetry technique. Indeed, owing to the lateral positioning of the cameras, the first 2 mm above the wall are less prone to precisely detect rotational motions due to the way boundaries are handled by the interpolation algorithm. Figure.~\ref{fig:NbHV-HSV_vsVR}b shows the positions of the center of mass of the HSV cloud using markers colored with respect to the experiment\rq{}s velocity ratio. The average HSV position gets closer to the wall when R is decreased. For $R<0.38$, the average HSV position lies below Y=2 mm, i.e Y/d=0.25, which explains the sudden drop in the detection of HSV.

A few strong HoV can be seen in front of the jet exit for R~$=0.34$ (Fig.~\ref{FigTopoInstaJetRond:topoblob_varVR}g). The sparse visual aspect of this spatial distribution and the actual count of strong vortices in this area show that these HoV are no longer part of a steady vortex cloud, but rather are intermittent structures. It is coherent with an intermittent shedding of Hovering Vortex confined in the jet pipe. In this situation this vortex can be called Inner Vortex.

For R~$=0.16$ (Fig.~\ref{FigTopoInstaJetRond:topoblob_varVR}h), no sign of shedding is observed consistently with the scenario proposed by \citet{Bidan2013}.

\subsection{Evolution of the three-dimensional trajectories of the instantaneous CRVP in the transverse planes for decreasing velocity ratios}
In every cross-sections (YZ planes, cf Fig. \ref{Fig:blobdetection}), the same vortex detection algorithm has been applied on the $\lambda_{Ci\ X}$ field (longitudinal swirl). First, the main positive and negative swirling structures are recovered using a classic local maxima detection algorithm. Ten positive and ten negative swirling structures have been kept for each time step.

\begin{figure}[htb!]
      \begin{center}
            \begin{tabular}{cc}
    \includegraphics[width=0.55\textwidth]{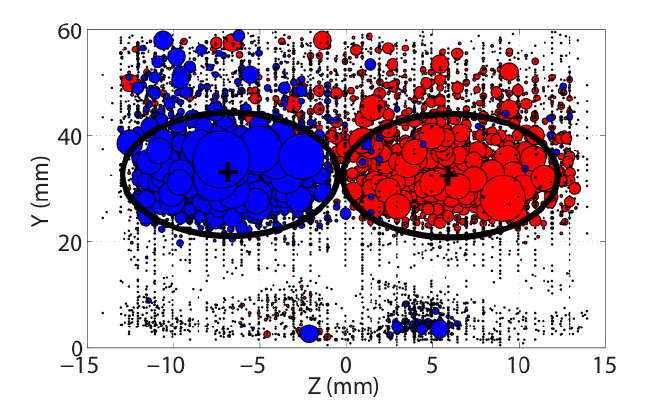}
    &    \includegraphics[width=0.45\textwidth]{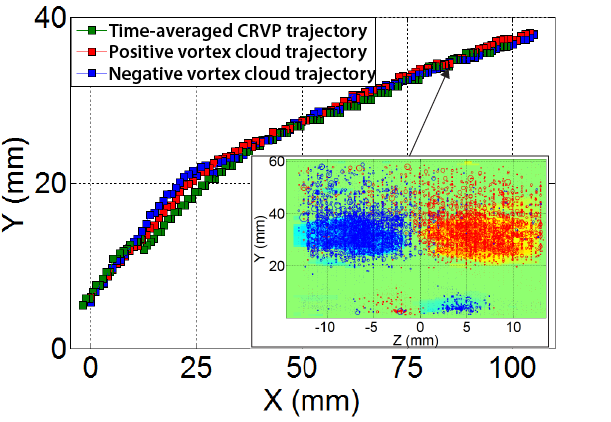}
    \\
    a)&b)
    \end{tabular}
    \caption{a) Positive and negative swirling structures detected in the X/d$=$10 plane for the R~$=1.39$ case. b) Superposition of the time-averaged CRVP trajectory with the instantaneous positive and negative vortex clouds trajectories. Inner window : Superposition of the time-averaged $\lambda_{Ci\ X}$ field with the instantaneous $\lambda_{Ci\ X}$ extrema in the X/d$=$10 plane.}
    \label{fig:Blobdetection_YZplanes}
    \end{center}
\end{figure}

Figure~\ref{fig:Blobdetection_YZplanes}a shows an example of the cumulative distribution over 1000 time steps of the positions of the positive and negative $\lambda_{Ci\ X}$ extrema
obtained for the R~$=1.39$ case at X/d$=$10. Two approximately elliptical shaped areas have very high concentrations of vortices with intense longitudinal swirl. 
 The superposition of the time-averaged $\lambda_{Ci\ X}$ field with these instantaneous $\lambda_{Ci\ X}$ extrema (Fig.~\ref{fig:Blobdetection_YZplanes}b) reveals that these counter-rotating distributions of instantaneous vortex perfectly fits the CRVP position in the time-averaged field. The trajectories of the CRVP (taken from \citet{Cambonie2013}) and of the center of mass of the instantaneous positive and negative vortex clouds are very close in the jet near-field, and perfectly coincide in the jet far-field.
Therefore, these instantaneous streamwise vortices strongly contribute to the CRVP formation as well as its sustaining.\\
In Fig.~\ref{fig:Blobdetection_YZplanes}a, two other small regions composed of counter-rotating vortices can also be seen close to the wall. 
The spatially homogeneous distribution of very weak vortices in the cross-section corresponds to a background noise of the $\lambda_{Ci\ X}$ field.
The noise distribution have been used to evaluate the signal-to-noise ratio and to define a filtering criterion.\\
The small counter-rotating structures close to the wall are neglected to focus on the main counter-rotating vortices area. Even if it would be more accurate to speak of Counter-Rotating Vortex Areas, this region will be called in the following  "instantaneous CRVP". Indeed, Fig.~\ref{fig:Blobdetection_YZplanes}b shows that the standard time-averaged CRVP position perfectly fits with these counter rotating vortex clouds and is therefore mostly the result of the time-averaging of  these instantaneous vortices. As a result, we call these instantaneous cloud vortices "instantaneous CRVP". They can be seen as the spatially-averaged influence of the swirling of each instantaneous vortices at the origin of the time-averaged CRVP.

\begin{figure}[htb!]
      \begin{center}
            \begin{tabular}{I>{\centering\vspace{-0.175\textheight}}m{2 cm}Ic|cI}
\Ghline&&\vspace*{-0.03\textheight}\\Side view\newline Z/d=0&\includegraphics[width=0.42\textwidth]{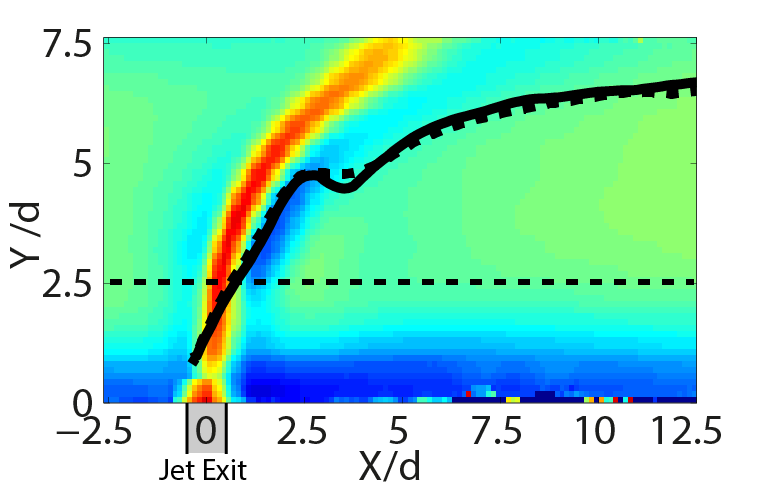}
&    \includegraphics[width=0.42\textwidth]{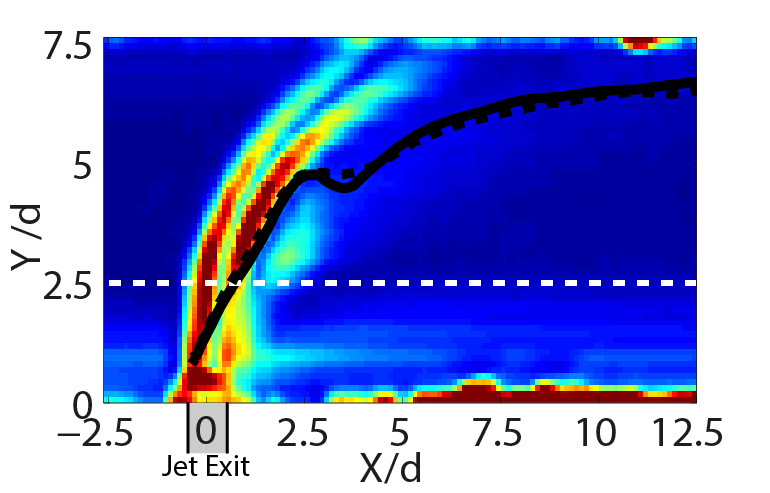}
    \vspace*{-0.01\textheight}\\ \hline&\vspace*{-0.03\textheight}\\
Top view\newline Y/d=2.5&\includegraphics[width=0.42\textwidth]{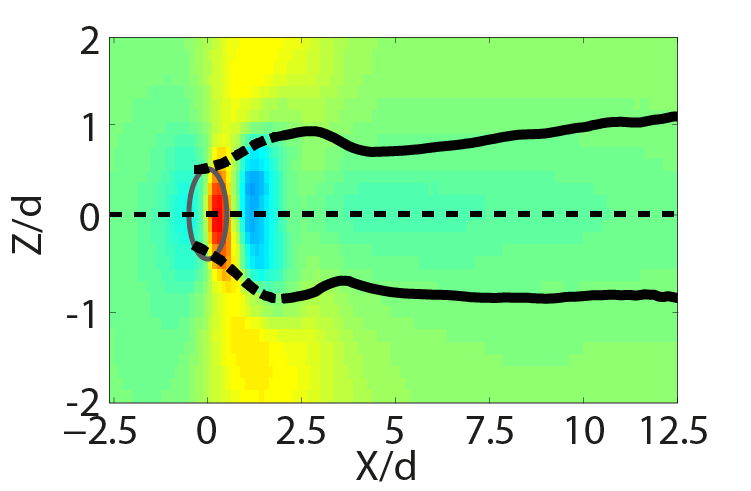}
&    \includegraphics[width=0.42\textwidth]{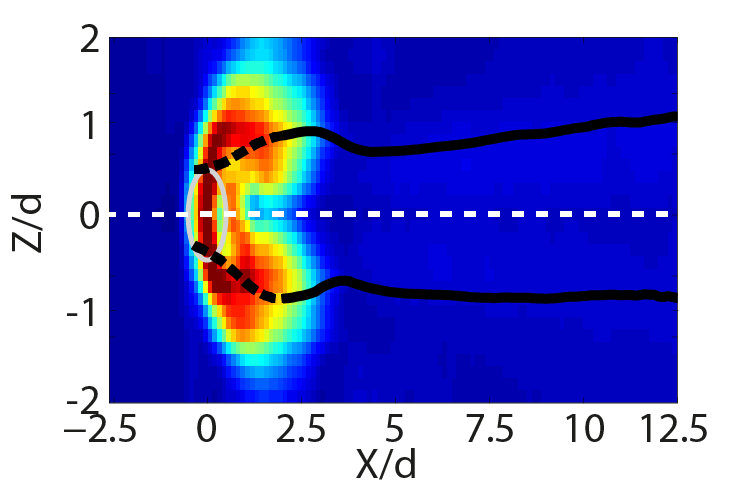}
    \vspace*{-0.01\textheight}\\ \hline&\vspace*{-0.03\textheight}\\
 &\includegraphics[width=0.37\textwidth]{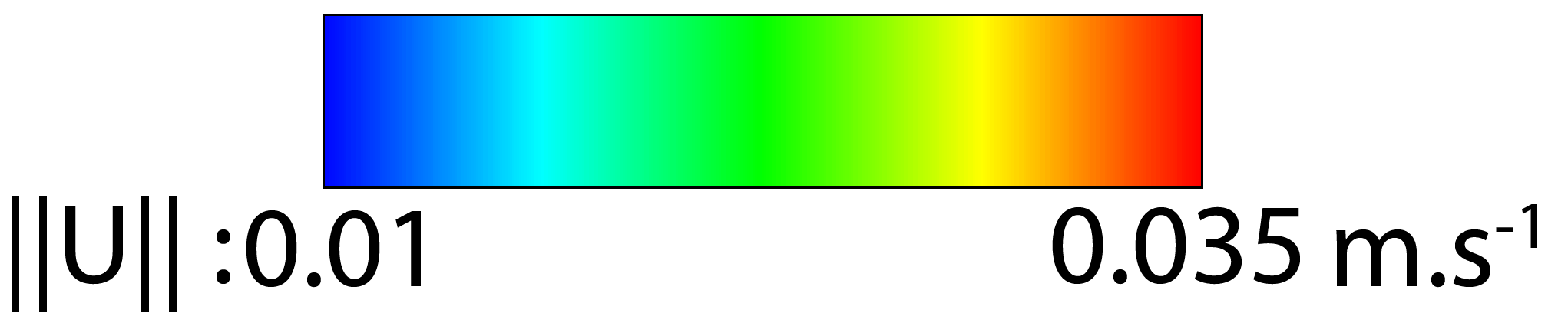}
 &\includegraphics[width=0.37\textwidth]{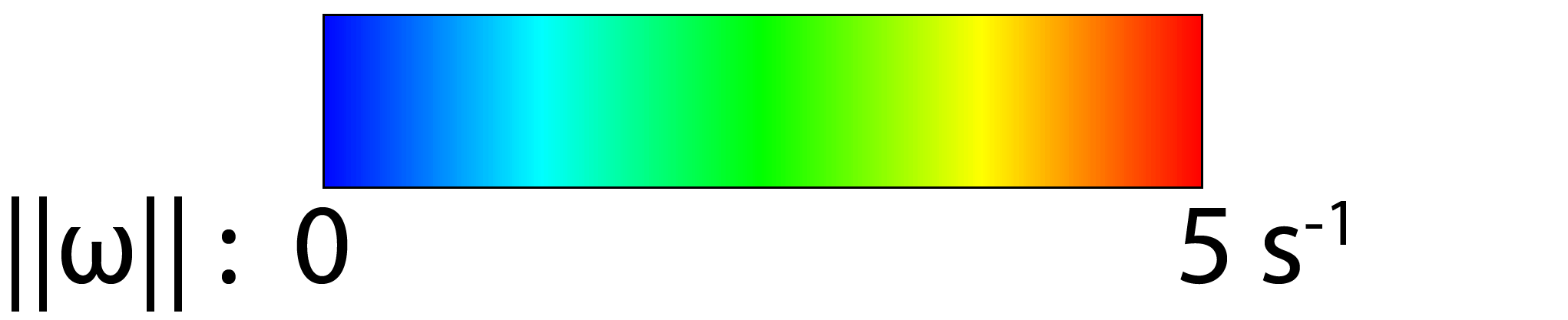}
 \vspace*{-0.02\textheight}\\
 &Velocity Magnitude&Vorticity Magnitude\\
\Ghline
    \end{tabular}
    \caption{Side and top views of the instantaneous CRVP for the R~$=1.71$ case. The background planes are colored with the velocity magnitude $||V||$ and the vorticity magnitude $||\omega||$. Dotted part of the instantaneous CRVP trajectories are located behind the planes of interest. 
Straight dotted lines help locating each plane with respect to each other. The jet exit is sketched on each image. Its elliptical shape is due to the axes scaling.}
    \label{FigTopoJetRond:TrajCRVPinsta_backgrounds}
    \end{center}
 \end{figure}   
 
The spatial positions of the instantaneous CRVP branches are defined as the "centers of mass" based on the swirling intensity of the positive and negative distributions (Fig.~\ref{fig:Blobdetection_YZplanes}~b). These positions are retrieved for each longitudinal position  (Fig.~\ref{Fig:blobdetection}), which allows for the definition of a 3-dimensional trajectory of the instantaneous CRVP branches.

At high velocity ratio, for the R~$=1.71$ case, Fig.~\ref{FigTopoJetRond:TrajCRVPinsta_backgrounds} shows the projection of this trajectory in the Z/d~$=0$ plane (side view) and Y/d~$=2$ plane (top view). 
For each orientation, dotted straight lines show the position of the other plane of interest. Dotted part of the instantaneous CRVP trajectories also show which parts are located behind the planes of interest.
For each orientation, two different backgrounds show the vertical velocity field and the vorticity magnitude field to locate the positions of each branch relative to the jet.  

The 3D instantaneous CRVP trajectories are lower than the centerline jet trajectory (side views) as has already been shown in previous numerical \cite{Salewski2008} or experimental\cite{Cambonie2013} studies.
They start approximately at the upstream limit and a few millimeters above the jet exit, a position which coincides with the arms of the first LEVs generated on the upstream shear layer. For X/d~$<$~0.5 - 1, they follow the body of the jet. Past this region, they continue to rise following the main shear layers. Following the curvature of the recirculation area, they wrap around the very low velocity region until the upper limit of the recirculation area where the branches get closer. Then, they detach and move away from each other while the global structure of the jet continues to expand transversally.

Figure~\ref{FigTopoJetRond:TrajCRVPinsta_backgrounds} shows how the instantaneous CRVP 3D trajectories follow the jet shear layer and by-pass the low velocity region confined against the downstream shear-layer. 

\begin{figure}[htb!]
      \begin{center}
            \begin{tabular}{cc}
    \includegraphics[width=0.45\textwidth]{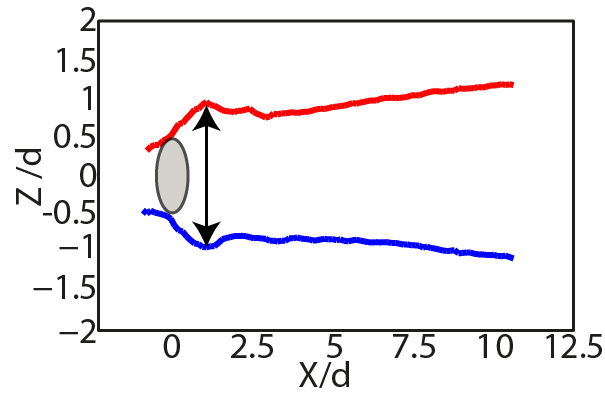}
    &    \includegraphics[width=0.45\textwidth]{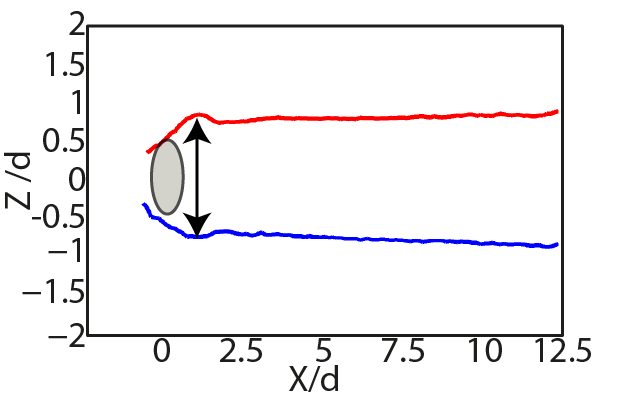}
    \vspace*{-0.01\textheight}
    \\ a) R = 1.71&b) R = 1.16\\
    \includegraphics[width=0.45\textwidth]{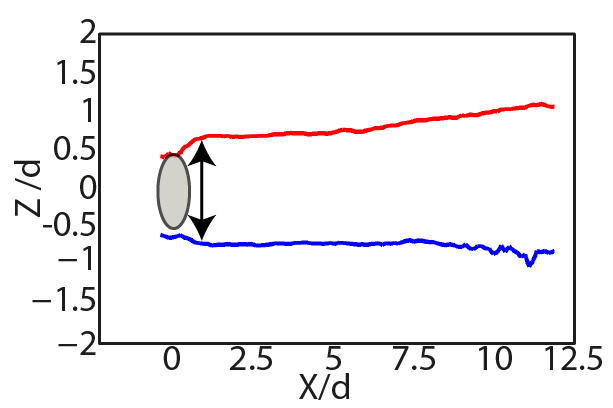}
    &    \includegraphics[width=0.45\textwidth]{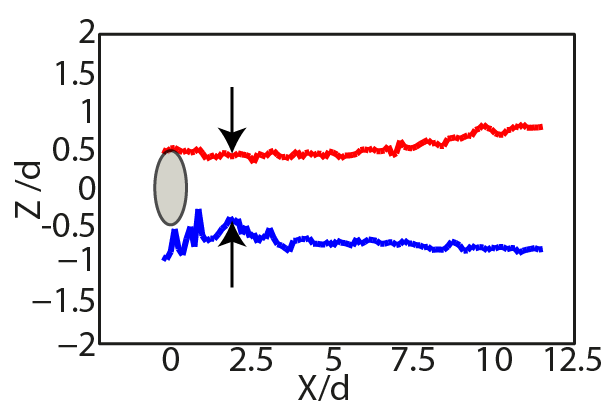}
    \vspace*{-0.01\textheight}
    \\ c) R = 0.55&d) R = 0.16\\
    \end{tabular}
    \caption{Top view of the instantanous CRVP trajectories for different velocity ratios. The jet exit is sketched on each image. Its elliptical shape is due to the axes scaling.}
    \label{FigTopoJetRond:TrajCRVPinsta_XZ}
    \end{center}
 \end{figure}   
 
Figure \ref{FigTopoJetRond:TrajCRVPinsta_XZ} shows the instantaneous CRVP trajectories for decreasing velocity ratios. For the cases R=\{1.71, 1.16, 0.51\} (Fig.~\ref{FigTopoJetRond:TrajCRVPinsta_XZ}a, b, c), the deviation induced by the low velocity area can be observed. On the opposite for R~$=0.16$, the instantaneous CRVP branches tend to get closer showing no trace of the very low velocity area.

Therefore, the vortices detected in the transverse planes also show a different behavior at very low velocity ratio, a sign that a topological transition occur.  The topological features of this transition are an evolution of the junction flow vortex properties (disappearance of the HSV and HoV) and the disappearance of the bypass of the very low velocity region by the instantaneous CRVP. Since this latter is the direct results of the jet being an obstacle for the crossflow, it also means a transition toward a new topology for the JICF.

\section{Swept-jet Topology}

In this section, an alternative to the junction flow topology at higher velocity ratios is discussed for the very low velocity ratios: the swept-jet topology.

\subsection{Experimental observation of the swept-jet flow}\label{ssec:sweptjet}

\begin{figure}[htb!]
      \begin{center}
            \begin{tabular}{cc}
            \multicolumn{2}{c}{\includegraphics[width=0.9\textwidth]{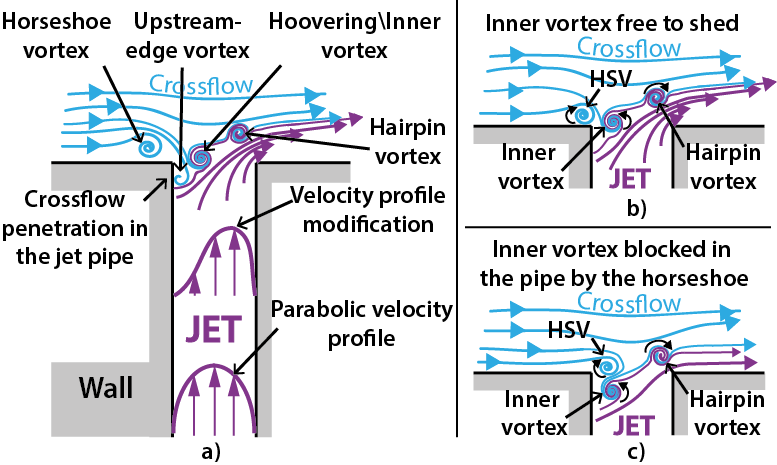}
            }\\
    \end{tabular}
    \caption{a) JICF vortices in the symmetry plane at low velocity ratios and modification of the velocity profile in the jet pipe. b) $R>0.35$ Inner vortex free to shed. c) $R<0.35$ Blockage of the inner vortex in the pipe.}\label{fig:Topo_soufflage_jet_miseenplace}
    \end{center}
\end{figure}

As shown in the previous section, for the very low velocity ratios, the jet momentum is no longer strong enough to provide a steady obstacle to the cross-flow. At low velocity ratios, the experimental and numerical studies of \citet{ANDREOPOULOS1984} and \citet{Muppidi2005} show a modification of the jet exit velocity due to the penetration of crossflow fluid inside the jet pipe as sketched in the Fig. \ref{fig:Topo_soufflage_jet_miseenplace}a. 

The formation of the HSV progressively occurs closer to the wall and to the jet exit (Fig. \ref{fig:Topo_soufflage_jet_miseenplace}b). Owing to the linear trend observed in Figure \ref{fig:NbHV-HSV_vsVR}b, we can deduce that for the experiments 1, 7 and 8 whose velocity ratio is lower than $R=0.35$, the HSV even forms above the jet exit ($-0.5<X_{HSV}$). For these very low velocity ratios, the HSV partially blocks the jet exit and confines the hovering vortex (HoV) inside the jet pipe (Fig. \ref{fig:Topo_soufflage_jet_miseenplace}b).
 For R=0.34 (Fig. \ref{FigTopoInstaJetRond:topoblob_varVR}g), the competition between the jet momentum and the crossflow momentum is balanced enough for the blockage of the HoV by the HSV not to be perfect and for a few shedding events of the HoV to occur. For R=0.16 (Fig. \ref{FigTopoInstaJetRond:topoblob_varVR}h), this is no longer the case.

\begin{figure}[htb!]
      \begin{center}
            \begin{tabular}{I>{\centering\vspace{-0.15\textheight}}m{0.5cm}cI>{\centering\vspace{-0.15\textheight}}m{0.5cm}cI}
\Ghline\multicolumn{4}{IcI}{\vspace*{-0.03\textheight}}\\
\multicolumn{4}{IcI}{\includegraphics[width=0.5\textwidth]{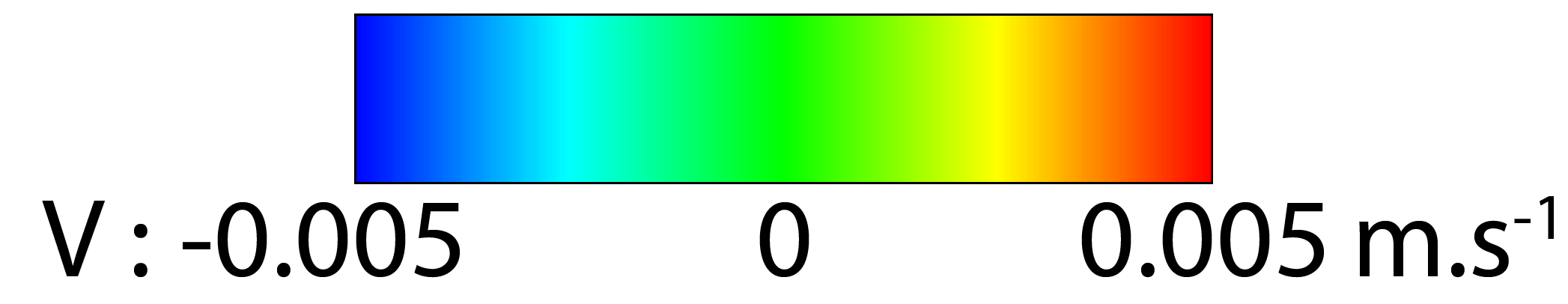}
}\vspace*{-0.01\textheight}\\\Ghline&&&\vspace*{-0.03\textheight}\\
    a)&\includegraphics[width=0.45\textwidth]{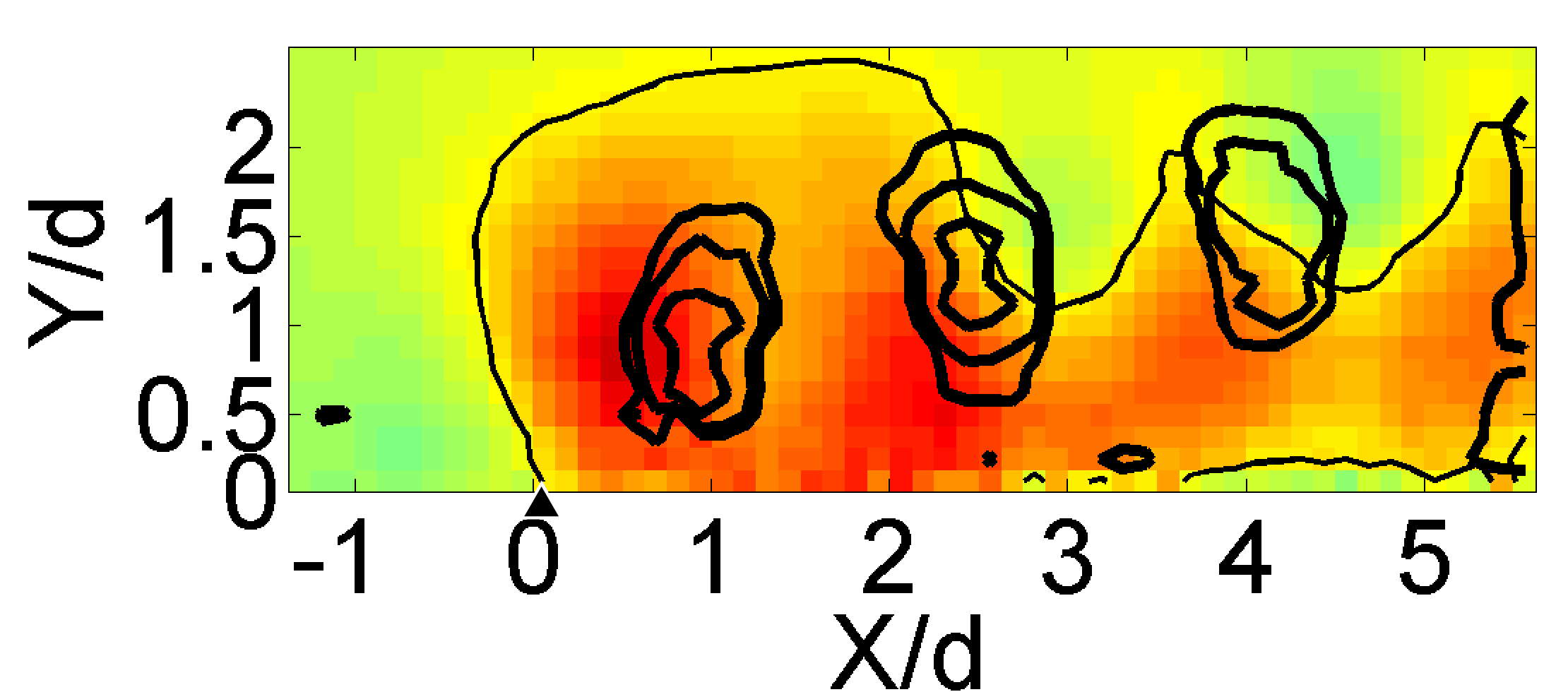}
    &    e)&\includegraphics[width=0.45\textwidth]{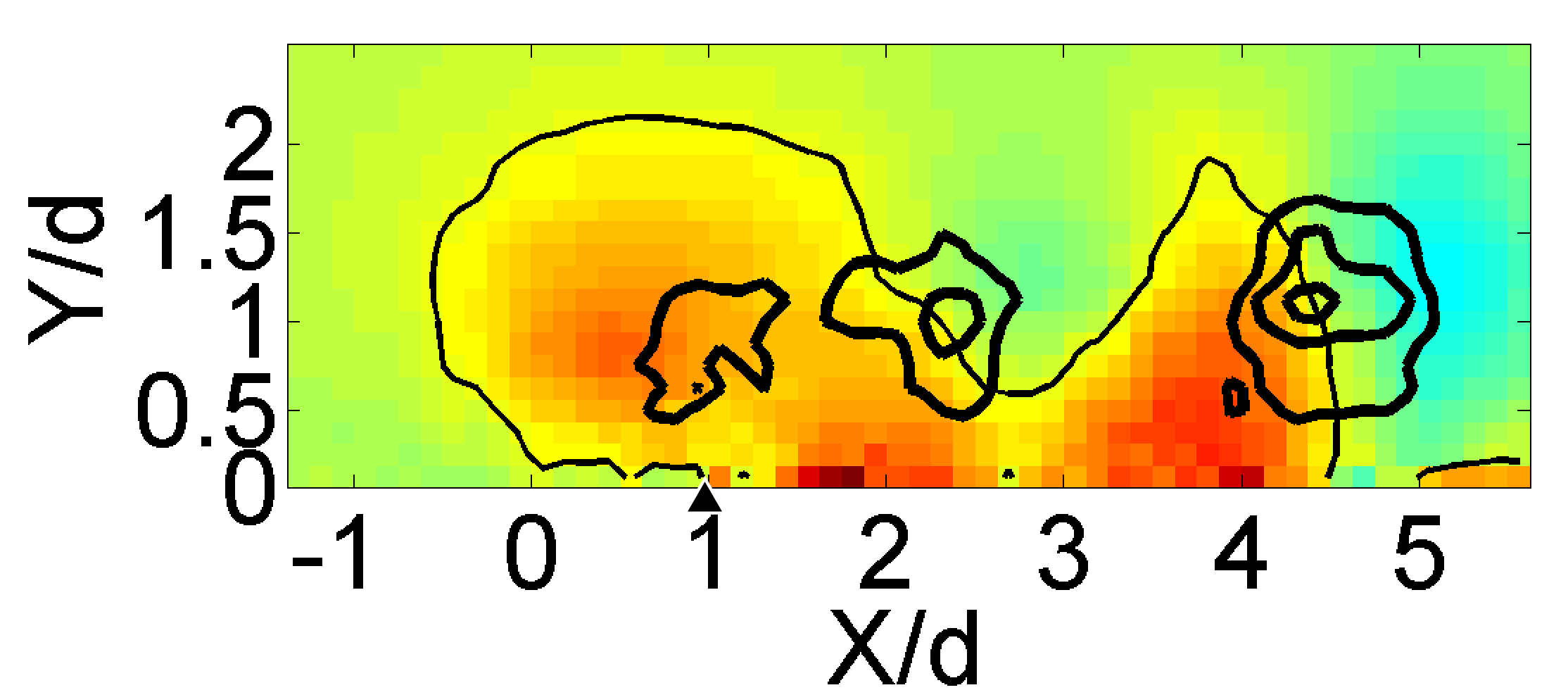}
    \\   b)&\includegraphics[width=0.45\textwidth]{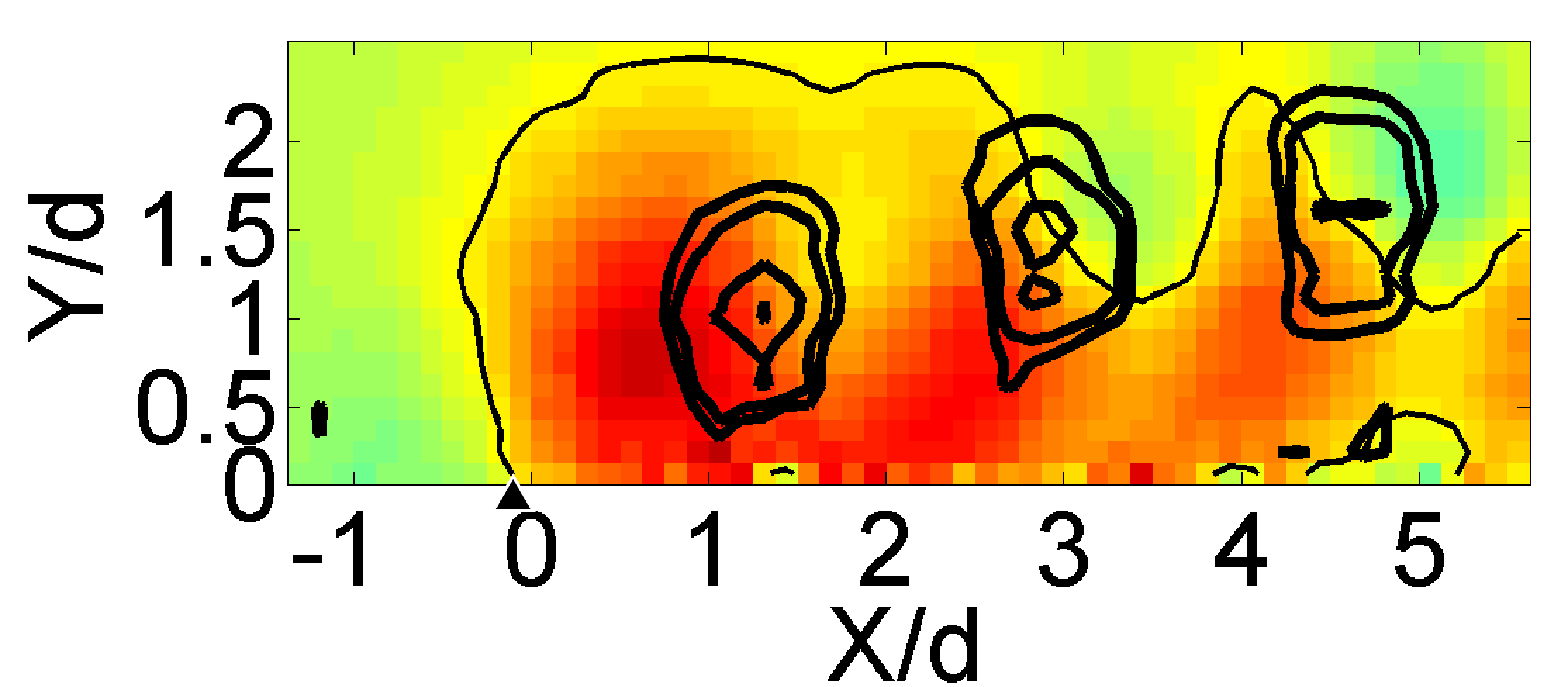}
    &    f)&\includegraphics[width=0.45\textwidth]{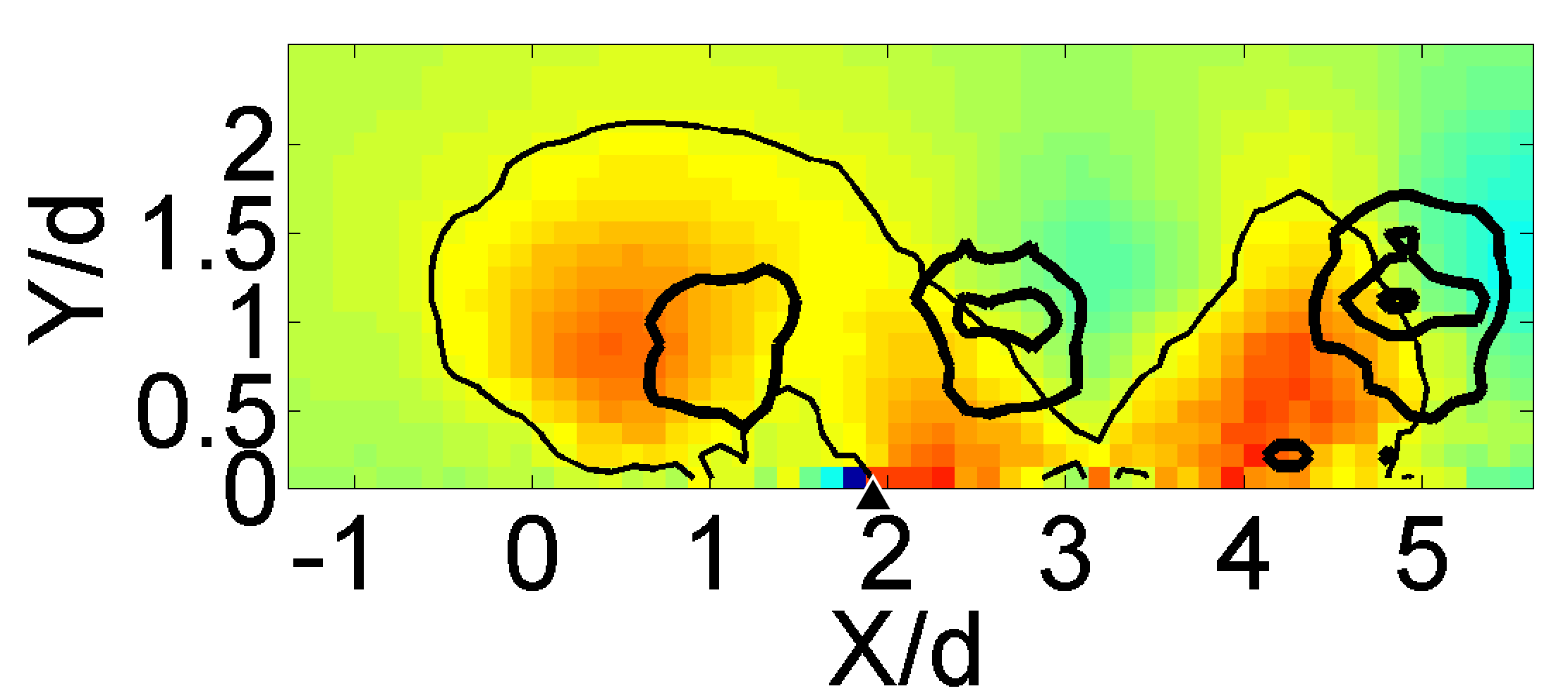}
    \\    c)&\includegraphics[width=0.45\textwidth]{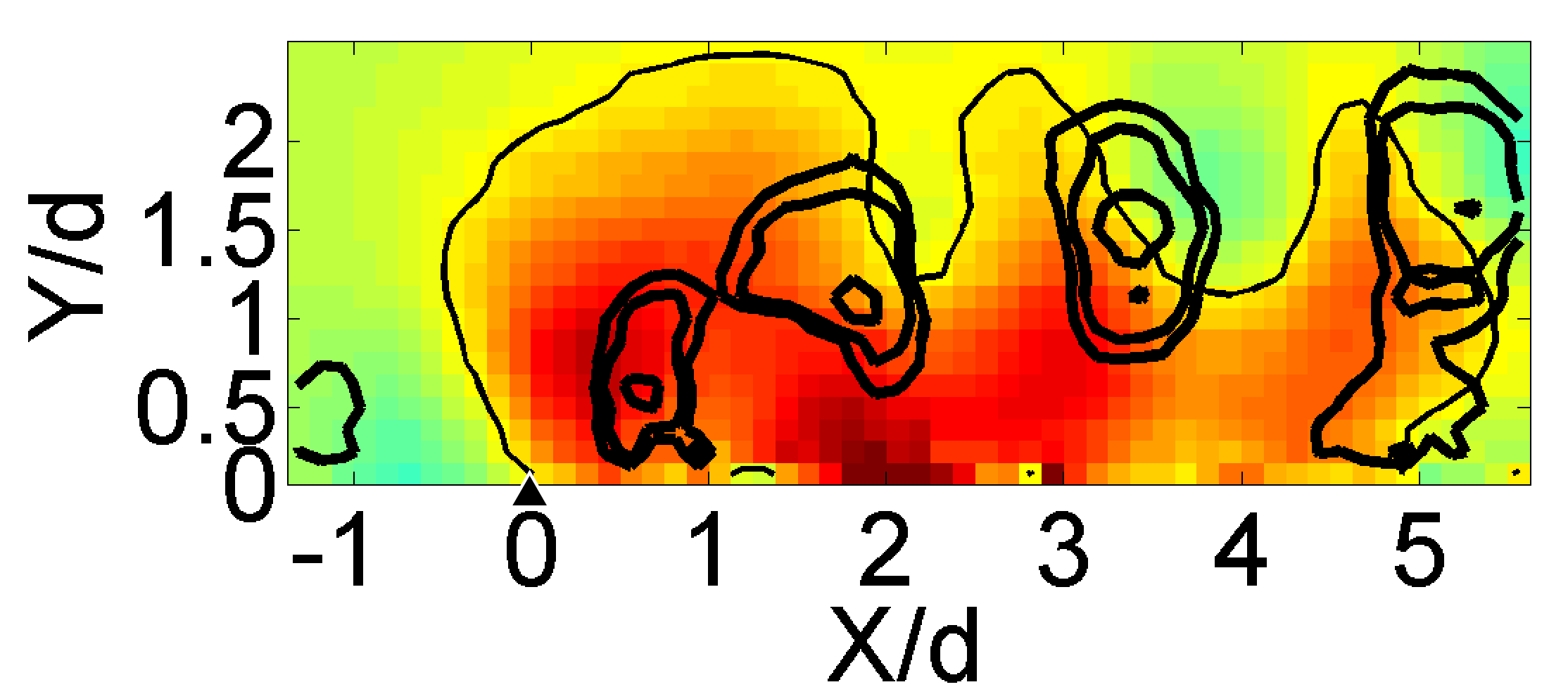}
    &    g)&\includegraphics[width=0.45\textwidth]{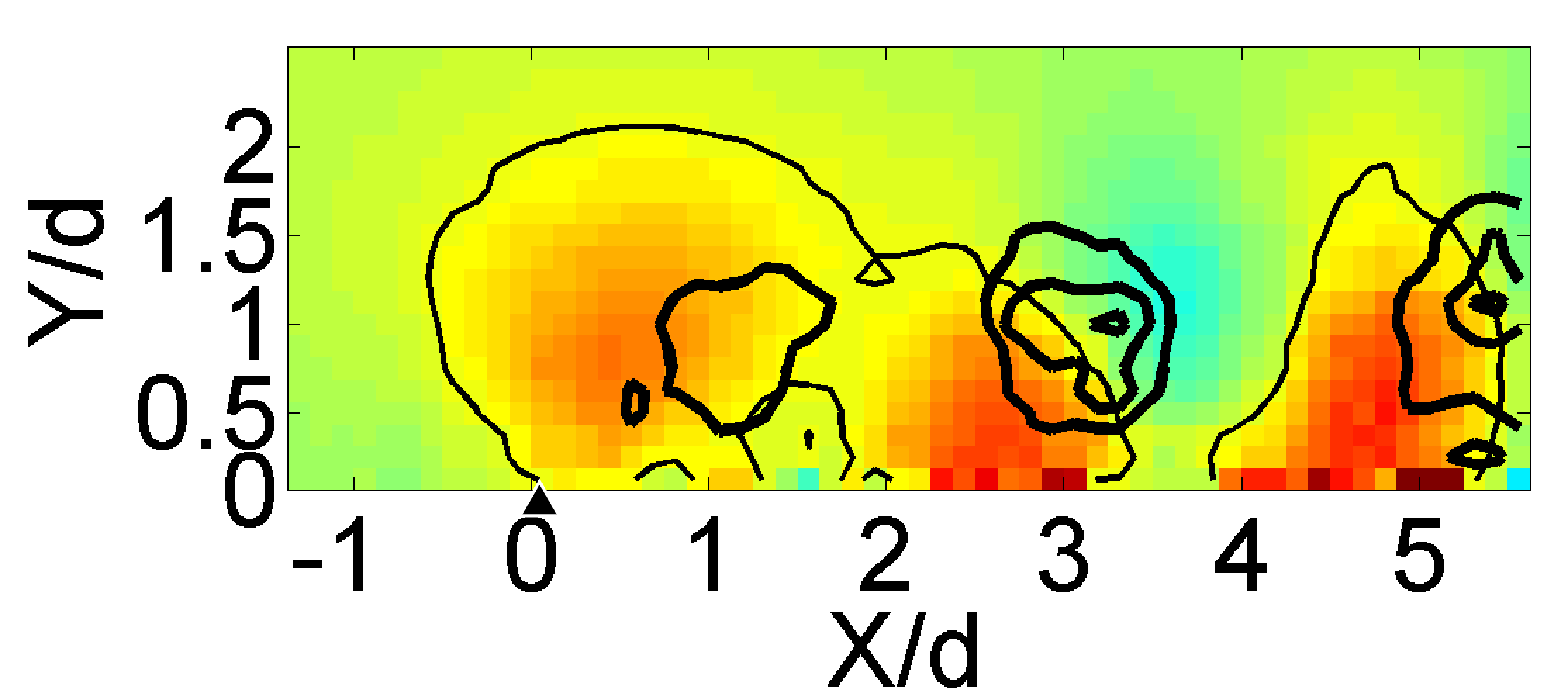}
    \\    d)&\includegraphics[width=0.45\textwidth]{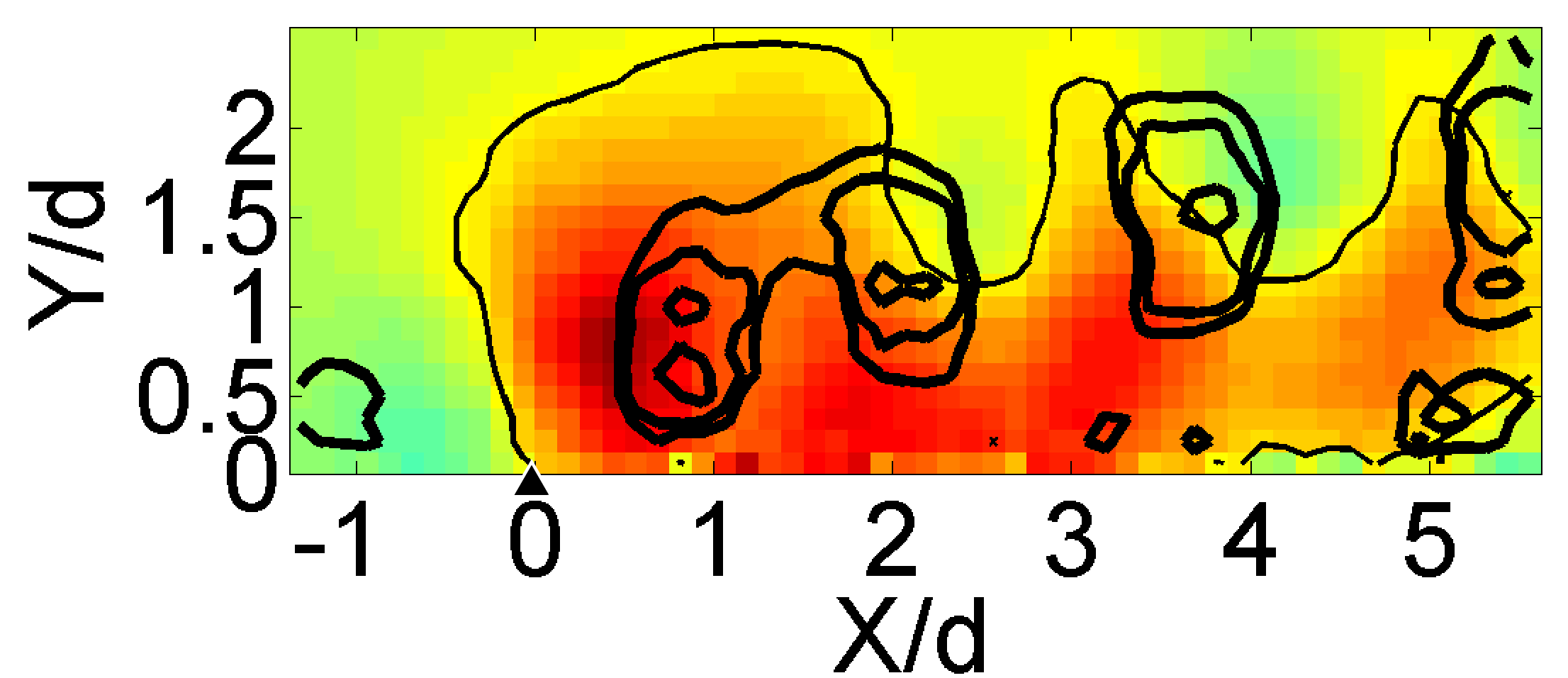}
    &    h)&\includegraphics[width=0.45\textwidth]{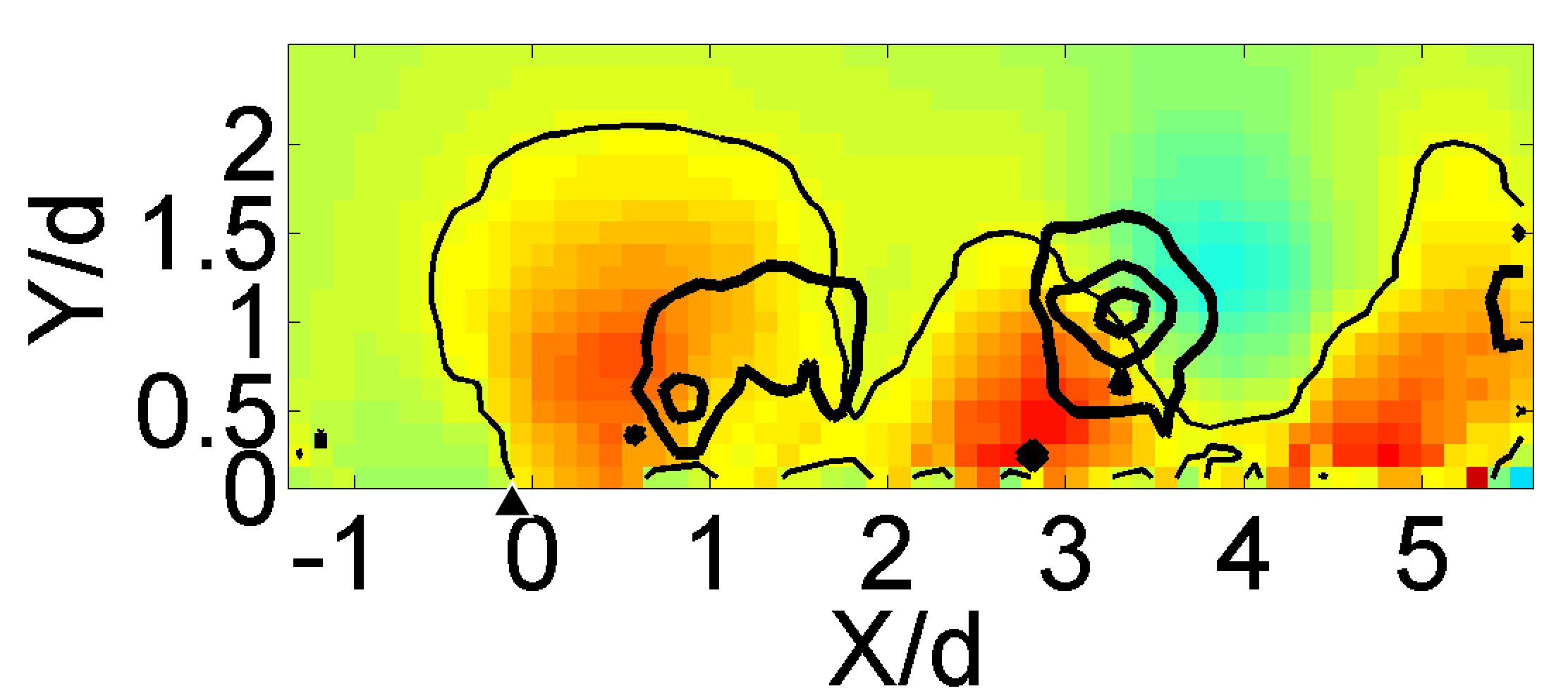}
    \\\hline
&R=0.34&&R=0.16\\
\Ghline
    \end{tabular}
    \caption{Time-series with $\Delta t\approx 7.1\cdot 10^{-2}$ s between each snapshot, for the cases R~$=0.34$ (a, b, c, d); R~$=0.16$ (e, f, g, h). Each image displays the vertical velocity in the symmetry plane, thick black isolines of $\lambda_{Ci\ Z}=\{0.5, 1, 2\}$ s$^{-1}$ and a thin black isoline of vertical velocity V~=~2~mm.s$^{-1}$.}
    \label{FigTopoJetRond:TrajCRVPinsta_highVR}
    \end{center}
 \end{figure}   
At the same time, the hairpin\rq{}s generation mechanism has changed. As can be seen in Fig. \ref{FigTopoJetRond:TrajCRVPinsta_XZ}d, it is no longer supported by the steady lateral shear-layers and the by pass of the very low velocity region. Indeed, the jet no longer is a steady obstacle for the crossflow.

In Fig. \ref{FigTopoJetRond:TrajCRVPinsta_highVR}, time-series snapshots show the vertical velocity in the symmetry plane, in the jet exit neighborhood. Thick isolines of $\lambda_{Ci\ Z}$ show the vortex positions and a fine isoline of vertical velocity V~=~2~mm.s$^{-1}$ show the onset of the jet vertical velocity. A triangular marker follows the location of this isoline at the wall. 

Figure~\ref{FigTopoJetRond:TrajCRVPinsta_highVR}a, b, c, d show the flow evolution for the R~$=0.34$ case, above the transition. The vertical velocity isolines starts near X/d~=~0, in the middle of the jet exit, illustrating the fact that the velocity profile is deformed by the crossflow velocity. Ineed, for a symmetric profile this isoline would begin near X/d~=~-0.5. Only a small oscillation of this position is perceptible around X/d~=~0 (see Fig.~\ref{FigTopoJetRond:TrajCRVPinsta_highVR}(a, b)).

Figure~\ref{FigTopoJetRond:TrajCRVPinsta_highVR}e, f, g, h show the flow evolution for the R~=~0.16 case, past the swept-jet transition.  A significant difference lays in the dynamic of the vertical velocity isoline. In Fig.~\ref{FigTopoJetRond:TrajCRVPinsta_highVR}e, f the iso-surface is completely disconnected from the jet exit. This absence of vertical velocity above the jet exit does not necessarily mean a total obstruction of the orifice by the crossflow. Indeed, if the obstruction is partial but important, i.e letting out only a thin stream of fluid, this stream would be bent inside the jet pipe and would not show any significant vertical velocity component. The jet momentum is then completely reoriented in the crossflow direction even before the jet fluid leaves the jet exit. This has an important effect on the hairpin\rq{}s vertical trajectory. They no longer are convected away by the jet exit. They only rise due to their own auto-induction and stay in the vicinity of the wall, embedded in the boundary layer. As a result, the swept jet matches a situation where the jet is attached to the wall.

In Fig.~\ref{FigTopoJetRond:TrajCRVPinsta_highVR}g the isoline seems to progressively reconnect with the jet exit region. Finally, in Fig.~\ref{FigTopoJetRond:TrajCRVPinsta_highVR}h the shape of the isoline matches the ones of a junction flow configuration, like the ones for R~=~0.34. 
At the same time, the formation of another hairpin is observed.

This observation is consistent with a two steps scenario where the jet is regularly swept by the crossflow. Figure~\ref{FigTopoJetRond:TrajCRVPinsta_highVR}e,f correspond to the obstruction phase, Fig.~\ref{FigTopoJetRond:TrajCRVPinsta_highVR}h to the ejection phase, and Fig.~\ref{FigTopoJetRond:TrajCRVPinsta_highVR}g to a transient step.
In the first step, the jet is swept by the crossflow. It results in an important obstruction of the pipe orifice by the crossflow fluid.  In the second step, an ejection phase occurs triggered by the overpressure created inside the pipe during the first phase: it allows for the jet momentum to be momentarily strong enough to push back the crossflow fluid. This ejection phase leads to the formation of another hairpin vortex.

\begin{figure}[htb!]
      \begin{center}
            \begin{tabular}{c}
\includegraphics[width=0.7\textwidth]{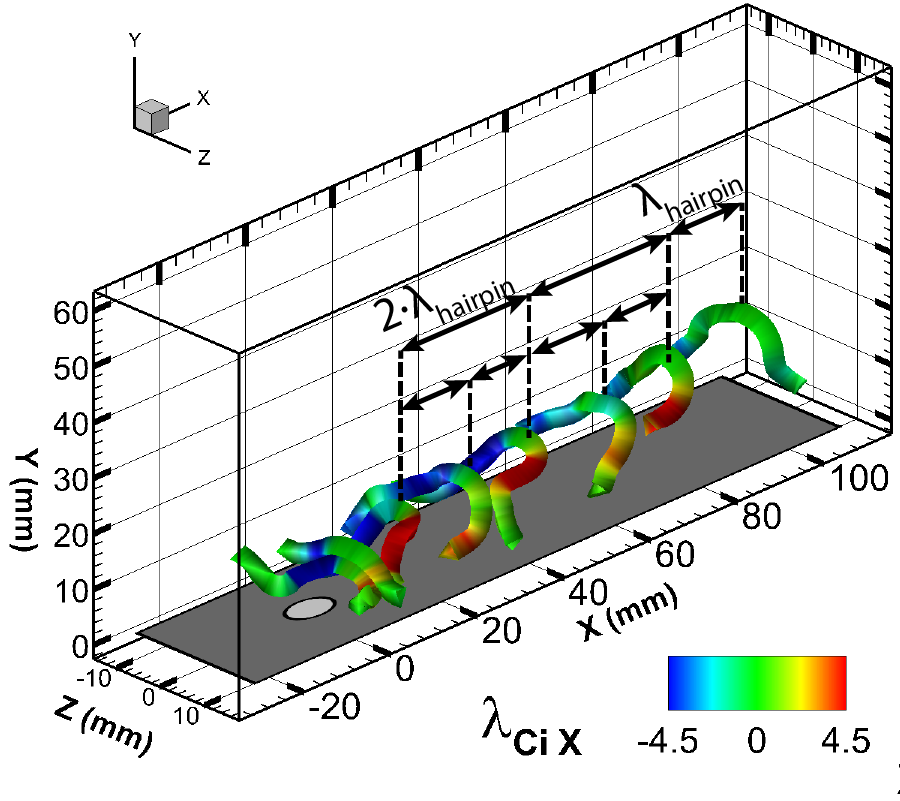}
    \end{tabular}
    \caption{ Visualization of vortex cores for the same instantaneous velocity field as Fig.~\ref{fig:isoLci_varVR_insta}f (R~=~0.16), colored with $\lambda_{Ci\ X}$. These vortex lines are defined as streamlines of the [$\lambda_{Ci\ X}$, $\lambda_{Ci\ Y}$, $\lambda_{Ci\ Z}$] field started from $\lambda_{Ci\ Z}$ maxima in the symmetry plane as in Fig. \ref{Fig:Topoinsta_vortexcores}.}
\label{fig:Isosurface_squelette_VR0v15}
    \end{center}
\end{figure}

The instantaneous 3D swirling structures corresponding to R~=~0.16 are visualized in Fig.~\ref{fig:Isosurface_squelette_VR0v15} using $\lambda_{Ci}$ vortex lines to define the vortex cores.  The vortex cores are colored using the $\lambda_{Ci\ X}$ to show the longitudinal swirl. As previously stated for Fig.~\ref{fig:isoLci_varVR_insta}f, an alley of hairpin vortices is very well visualized. 

Using the vortex cores instead of the isosurfaces allows the visualization of  more subtle characteristics. Indeed, one can clearly see that one pinched hairpin vortex with strong legs is followed by a vortex with more distant legs and weaker swirling intensity. It suggests a different origin for the vortex formation on each phase. A tentative explanation can be proposed to explain this observation. During the partial obstruction phase, the jet velocity is lowered as well as the velocity gradients leading to the vortex formation. The resulting hairpin should then be less intense, with weaker legs. On the other hand, during the ejection phase, the jet velocity $V_{J}$ increases leading to another hairpin with a greater swirling intensity. Due to the higher jet velocity, the hairpin\rq{}s head is convected higher by the jet velocity which stretches and pinches the hairpin, bringing its legs closer. Another explanation may lies in the interaction between the hairpin\rq{}s legs and wall side vortices. Indeed, 2 types of secondary side vortices with opposite streamwise vorticity have been found in the jet wake\citep{Bidan2013}. Depending on their swirling orientation, the induction between the hairpin leg and the side vortices tends to alternatively separate or bring together the hairpin\rq{}s legs. Both explanation scenario are non-exclusive

\subsection{Discussion about the transition scenario of the JICF at low velocity ratio proposed by \citet{Ilak2012}}\label{ssec:Ilakdiscussion}

\begin{figure}[htb!]
      \begin{center}
            \begin{tabular}{cc}
\multicolumn{2}{c}{ \includegraphics[width=1\textwidth]{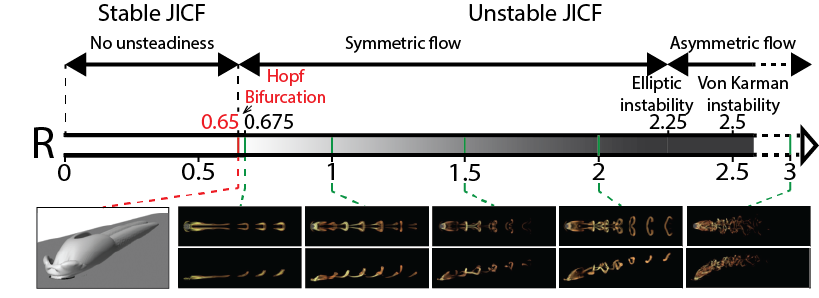}
}\\\multicolumn{2}{c}{a)}\\
    \includegraphics[width=0.55\textwidth]{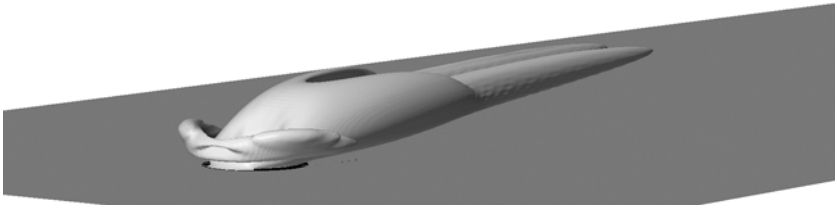}
    &\includegraphics[width=0.35\textwidth]{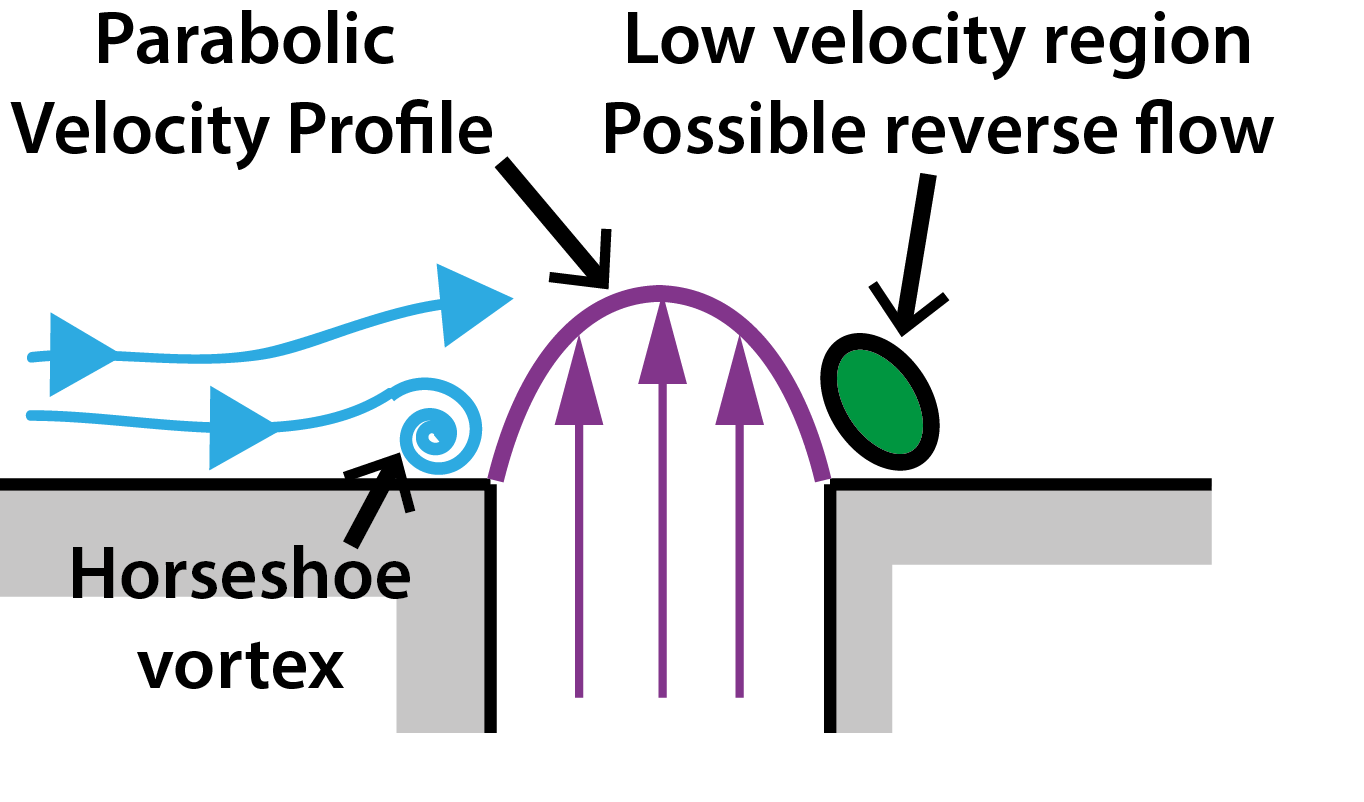}
    \\
    b)&c)
    \end{tabular}
    \caption{a) Overview of the results of \citet{Ilak2012}. b) $\lambda_2$ isosurface showing the steady state observed for R~=~0.65 (from \citet{Ilak2012}). c) Flow topology in the symmetry plane and parabolic velocity profile as imposed by \citet{Ilak2012}. The constant velocity profile used as the boundary condition of the direct numerical simulation is responsible of the transition toward a stable JICF state.}
\label{fig:Ilak}
    \end{center}
\end{figure}

The JICF topology observed in the present experiments at very low velocity ratios has to be compared with another transition scenario proposed recently by \citet{Ilak2012} using direct numerical simulations. Fig.~\ref{fig:Ilak}a sums up this scenario. It focuses on the apparition of different type of instabilities and on their influence on the stability of the JICF.
For the higher velocity ratios ($R>0.8$), their scenario compares well with ours.\\ 
At high R ($R>2.5$), the jet is very unstable and turbulent. The shear layer vortices (LEVs and TEVs) very quickly break down into smaller vortical structures. Due to the presence of an antisymmetric elliptic instability of the counter-rotating vortex pair (for $R>2.25$) and of a tornado looking type of instability near the wall, they qualify the flow as asymmetric. Below $R=2.25$ the flow is said symmetric. Of course such symmetry is only achievable numerically, and a slight asymmetry has been sometimes observed in our experiments as well as in  several other experimental and numerical studies \cite{KAMOTANI1972,Smith1998,Yuan1999,Marzouk2007,Kuzo1995}. 

Below $R=2.25$, the flow evolves then from a complex and a complicated quasi-periodic behavior of the classical JICF topology with both LEVs and TEVs ($1.2\lesssim R\lesssim2$), toward the deformed classical JICF topology with the progressive disappearance of LEVs ($0.8\lesssim R\lesssim1$). 

On the other hand, for the lower velocity ratios they observe a periodic vortex shedding which reaches a limit cycle of a Hopf bifurcation at R~=~0.675. Below this value, for R~=~0.65, their simulation exhibit a steady flow with a swirling structure composed of two counter-rotating branches very similar to a steady CRVP (Fig.~\ref{fig:Ilak}b). The first bifurcation occuring at $R=0.675$, and the observed shedding of hairpin vortices is related to the existence of a local absolute instability connected to the region of reversed flow immediately downstream of the jet.
 
No steady state could be observed in our experiments, even for the very low velocity ratios. This difference can be explained by some approximations and hypothesis made in this numerical study. Indeed, the key point lie in the choice of the boundary conditions used for the simulations. The jet inflow velocity profile they used is a Dirichlet boundary condition whose form corresponds to a constant laminar parabolic velocity profile (Fig.~\ref{fig:Ilak}c). With such a boundary condition, it is impossible to recover properly the jet physics at low R.  Indeed, the strong interactions between the pipe flow and the crossflow fluid leading to a strong deformation of the jet velocity profile at the pipe outlet can not be recovered. The authors are fully aware of this limitation as they acknowledge this fact, stating that "\emph{While at low values of R some backflow into the jet pipe is to be expected under realistic conditions,
it was demonstrated by Schlatter et al. (2010)\cite{Schlatter2011} that most of the relevant physics is still captured by the simulations, especially far away from the jet orifice (such as the CRVP). One feature that our simulations can not reproduce is the flow separation inside the pipe, which may play a significant role at low R. However, the qualitative similarity of our results to those of Ziefle (2007)\emph{[...]} indicates that even in the regime considered here the characteristic dynamics is reproduced.} "

Our results show that this assumption is false because the back-flow into the jet pipe cannot be overlooked, especially at very low velocity ratios. In this case, the main features and structures of the JICF as well as their dynamics are no longer located in the far field. For the high R cases, a parabolic profile can be considered as a good approximation of the jet velocity profile at the jet exit which explains that their simulations match the experiments. On the contrary, for the low R cases, the numerical simulations have to take into account the interaction between the jet and the crossflow inside the pipe, as demonstrated by \citet{Muppidi2005,Muppidi2007,Muppidi2008,Ziefle2009}. Forcing a parabolic velocity profile at low R forces  the flow onto a junction flow configuration with a blockage of the crossflow by the jet leading to the formation of a horseshoe vortex (Fig.~\ref{fig:Ilak}b) and the existence of a reverse flow region behind the jet (Fig.~\ref{fig:Ilak}c). These experiments proves that it is in reality no longer the case and no steady state exists in the very low velocity ratio limit.

\subsection{Discussion on the transition scenario of the JICF at low velocity ratio transition proposed by \citet{Bidan2013}}\label{ssec:Bidan}

\begin{figure}[htb!]
      \begin{center}
\begin{tabular}{c}
\includegraphics[width=\textwidth]{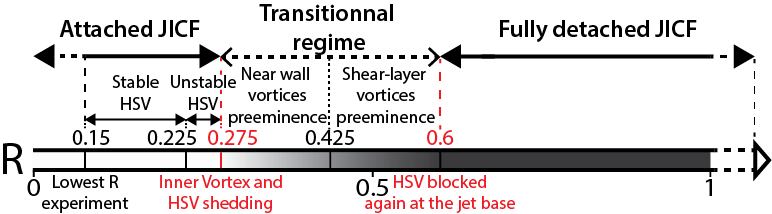}
\end{tabular}
    \caption{Overview of the results of \citet{Bidan2013}}
    \label{fig:Frise_Bidan}
    \end{center}
\end{figure}

The recent study of \citet{Bidan2013}, using both numerical simulations and experiments at very low velocity ratio, proposes a very interesting interpretation of the JICF topology consistent with our experiments. 
Figure~\ref{fig:Frise_Bidan} sums up their scenario.
Rather than the more classic velocity ratio, their main parameter is the jet to cross-flow mass-flux ratio or Blowing Ratio $BR=\rho_{j}V_{j}/\rho_{\infty}U_{\infty}$ which in their case equals the velocity ratio R since $\rho_{j}/\rho_{\infty}=1$.
It is worth noticing than \citet{Bidan2013} adopt a different approach than \citet{Ilak2012}, even if some descriptions of the JICF are very similar from both studies. While the scenario of \citet{Ilak2012} clearly adopts the point of view of the stability theory, the scenario of \citet{Bidan2013} is more clearly influenced by a film cooling approach, a community well-aware of the very low velocity ratio problematic and where an important distinction has to be made between an attached jet and a detached jet. 

To summarize their work, we think relevant to distinguish three different states for the JICF: attached, transitional or fully detached jet. 
In the "attached regime" the jet develops and stays embedded in the boundary layer with which it deeply interacts (Fig.~\ref{fig:isoLci_varVR_mean}f, \ref{fig:isoVormag_varVR_mean}f and \ref{FigTopoInstaJetRond:topoblob_varVR}h). In the "transitional regime", the jet still keeps a strong interaction with the boundary layer (Fig.~\ref{fig:isoLci_varVR_mean}d,e, \ref{fig:isoVormag_varVR_mean}d,e and \ref{FigTopoInstaJetRond:topoblob_varVR}f,g). As shown by the Fig.~\ref{fig:isoVormag_varVR_mean} and has already been noted by \citet{Gopalan2004}, the jet forms then a shell of shear layer which merges with the boundary layer.
In the "fully-detached jet regime", the interaction between the jet and the boundary layer is limited to the near-field of the jet exit, when the jet punctures the boundary layer (Fig. \ref{fig:isoVormag_varVR_mean}a,b and \ref{FigTopoInstaJetRond:topoblob_varVR}a,b).

In their study, the attached jet is observed at very low velocity ratio for $R<0.275$ when the jet momentum  is way weaker than the crossflow momentum. For \citet{Bidan2013}, the characteristic vortical structures are the HorseShoe Vortex (HSV), the Hovering Vortex (HoV) also called Inner Vortex, the Hairpin shear-layer Vortices (HpV), the inversed Hairpin Vortices (iHpV) and the quasi-streamwise secondary side vortices whose interaction forms X-shaped structures. 
The fully detached jet corresponds to the JICF topology at higher velocity ratio when the jet momentum is high enough to rapidly move away from the wall boundary layer.

The transition between the attached jet regime and the transitional regime is marked by the HSV and HoV behaviour. For $R<0.225$, a stable HSV is observed. It is stuck over the upstream part of the jet exit resulting in a pressure cap which blocks a part of the jet fluid and leads to the formation of a stable HoV inside the jet pipe. For $0.225<R<0.275$, the HSV starts to oscillate under the influence of the HoV and sporadically sheds. For $0.275<R<0.6$, the HSV as well as the HoV periodically shed.
The fully detached-jet regime occurs for $R>0.6$ when the pressure gradient in front of the jet fluid column stabilizes the HSV and stops its shedding. In-between, around $R\approx0.425$ a transition is found by wavelet analysis in the competition between the vortical structures. For $R<0.425$, the HSV and side-vortices are respectively preeminent above the hovering vortex and the hairpin vortices while for $R>0.425$, the hovering vortex grows stronger as well as the shear-layer vortices.\\

This scenario summarized on Fig. \ref{fig:Frise_Bidan} is in most part consistent with our experiments. In Fig. \ref{FigTopoInstaJetRond:topoblob_varVR}, the distribution of vortices in the symmetry plane complies with the scenario of \citet{Bidan2013}.
For $R<0.275$ (R = 0.16 and 0.18), a flat and attached-to-the-wall distribution of vortices corresponding to the hairpin\rq{}s head is observed (Fig.~\ref{FigTopoInstaJetRond:topoblob_varVR}h and Fig.~\ref{FigTopoJetRond:TrajCRVPinsta_highVR}e,f,g,h) while the jet starts to lift up for R=0.34 (Fig.~\ref{FigTopoInstaJetRond:topoblob_varVR}g and Fig. \ref{FigTopoJetRond:TrajCRVPinsta_highVR}a,b,c,d).
We also observe a disparition of the inner vortex in the jet pipe, with an intermittent shedding for R=0.34 (Fig. \ref{FigTopoInstaJetRond:topoblob_varVR}g), while the inner vortex is completely blocked inside the jet pipe for R = 0.16.
Most vortical structures associated to the attached and transitional regimes are also observed in our experiments: hairpin vortices, horseshoe vortices are systematically present. Due to our velocity field resolution, the smaller structures are more difficult to detect. Nevertheless, some instantaneous fields show vortices looking like inversed Hairpin Vortices and secondary side vortices. The influence of the latter can be seen in Fig.~\ref{fig:Isosurface_squelette_VR0v15} where it leads to the bi-periodicity of the spacing between the legs of the hairpins. Indeed, 2 types of secondary side vortices with opposite streamwise vorticity have been found. Depending on their swirling orientation, the induction between the hairpin leg and the side vortices tends to separate or bring together the hairpin\rq{}s legs.

The fully detached jet regime starts when the jet momentum allows for the HSV to stay blocked at the base of the jet. It is indeed an indubitable marker of the jet straightening, and as it will be discussed later it indeed can be seen as a transition value for the JICF topology. Nevertheless, it clearly is not enough to speak of a \lq\lq{}fully\rq\rq{} detached jet, with the classical JICF topology of the high velocity ratios. Main features of a fully detached jet regime should include a limited interaction between the jet and the boundary-layer as well as a similar swirling intensities between the upstream and downstream shear-layer vortices (LEV and TEV). According to the vortices distributions of Fig.~\ref{FigTopoInstaJetRond:topoblob_varVR}, the fully detached jet regime transition occurs between  $1.16<R<1.39$.

\subsection{Swept-jet transition parameter and topological transition scenario of the JICF at low velocity ratios}

As shown before, the spatial distribution of the vortices in the symmetry plane is a good way to evaluate the state of the flow.  At very low R, the disappearance of the hovering vortex in the jet pipe without shedding can be directly related to the triggering of the swept-jet state where the jet is attached to the wall. It can therefore be used to detect the transition. Since this transition results from the competition between the jet and the crossflow momenta, a more relevant transition parameter is introduced: $S_J = R(d/\delta)$.  It depends on the physical parameter of the problem: the velocity ratio $R$, the diameter of the jet $d$ and the boundary layer thickness $\delta$. Indeed, a greater velocity ratio or diameter should both increase the jet momentum and should therefore delay the swept-jet transition.

The vortex count $n_{LEV}$ applied to the LEV's distribution along the curvi-linear abscissa S of the upstream shear layer trajectory ($3<S/d<8$) in the symmetry plane is used to compute $N_{LEV}=n_{LEV}/n$ (n=1000). 

\begin{figure}[htb!]
      \begin{center}\begin{tabular}{cc}
\includegraphics[width=0.5\textwidth]{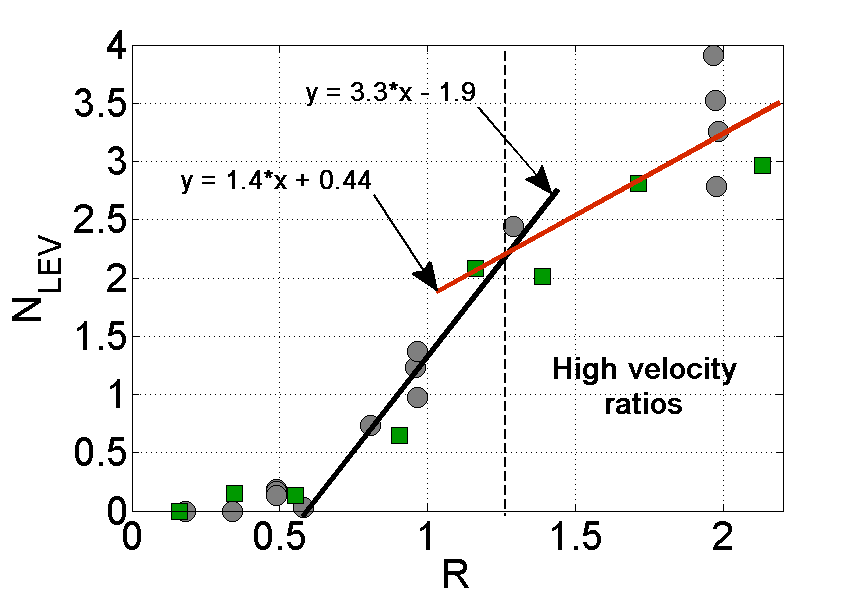}
&\includegraphics[width=0.5\textwidth]{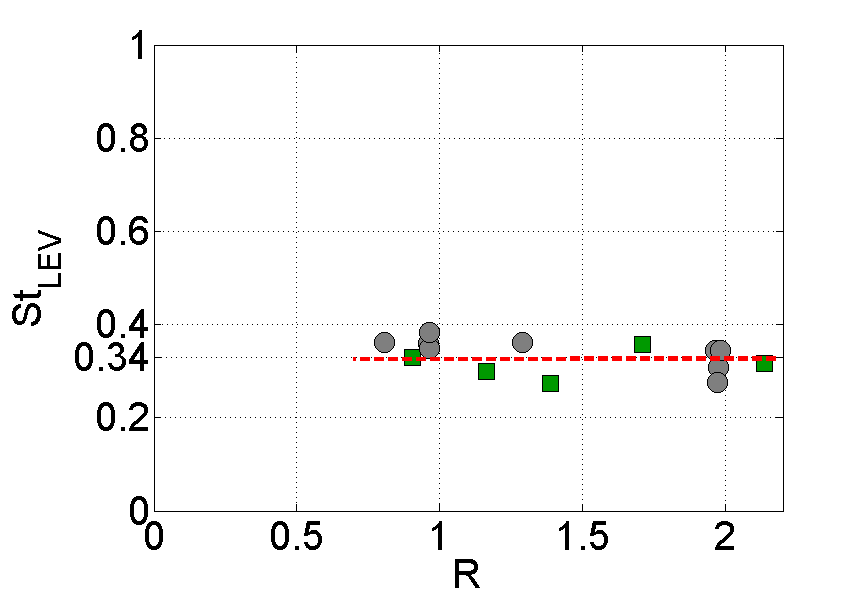}
\\
a)&b)
\end{tabular}
    \caption{a) Number of Leading-Edge Vortex by instantaneous velocity field between $S/d=3$ and $S/d=8$ $N_{HSV}$ with respect to the velocity ratio R, where  $S/d$ is the curvi-linear abscissa along the upstream shear layer trajectory non-dimensionalized by the jet exit diameter. b) LEV\rq{}s Strouhal numbers as a function of the velocity ratio $R$. The particular experiments shown in this article in the Fig. \ref{fig:isoLci_varVR_mean}, \ref{fig:isoVormag_varVR_mean}, \ref{fig:isoLci_varVR_insta}, \ref{FigTopoInstaJetRond:topoblob_varVR}, \ref{FigTopoJetRond:TrajCRVPinsta_backgrounds}, \ref{FigTopoJetRond:TrajCRVPinsta_XZ}, \ref{FigTopoJetRond:TrajCRVPinsta_highVR} are marked with green square markers.}
    \label{fig:NbLEV-StLEV}
    \end{center}
\end{figure}

Fig.~\ref{fig:NbLEV-StLEV}a shows $N_{LEV}$  as a function of the velocity ratio. One should first  notice that for the LEV the relevant parameter is no longer $S_j$ but R. For the high velocity ratio cases (R~$>1.25$), $N_{LEV}$ regularly increases which results from the more frequent destabilization of the upstream shear layer when the velocity ratio is increased. At least 2 to 4 LEVs are present on each instantaneous velocity field.  For  $0.55<R<1.25$, the upstream shear-layer weakening strongly decreases the number of produced leading-edge vortices. It allows for a less intuitive definition of the high velocity ratios and of the fully detached jet state, as the intersection of the linear tendencies associated with both regime.  This value is consistent with the one obtained with the boundary-layer vortex distribution.

One can see that there is also a clear transition for the LEV: for $R<0.55$, less than 0.2 vortex can be found per time steps and mostly corresponds to sporadic events or unfiltered noise vortical structures. The LEVs appears for $R>0.6$ with a growing number of occurrence when increasing the velocity ratio. Another related phenomenon occurs near this critical $R=0.55$ value: the end of the HSV shedding\cite{Bidan2013}. For $R>0.6$, the HSV stay blocked at the base of the jet. The jet is then detached enough for the LEVs to grow on the unperturbed upstream shear layer.

The LEVs growth has also been characterized by their Strouhal number. This latter has been computed as follow. 
Firstly, the LEV distribution is used to measure the spatial periodicity of the LEVs along the curvi-linear abscissa $S$. For each time-step, the distance between two consecutive LEVs is measured, non-dimensionalized by the jet exit diameter and associated with the position S along the curvi-linear abscissa of the first LEV. Cumulating this information over 1000 time-steps, the time-averaged distance $\lambda_{LEV}$ between two consecutive LEVs is retrieved for each curvi-linear position S. 
Each LEV has also been tracked between consecutive time-steps allowing for an estimation of its individual transport velocity, which is then associated with the mean position S along the curvi-linear abscissa of this structure between the two time-steps. Cumulating this information over 1000 time-steps, it becomes possible to define the mean transport velocity $V_{LEV}$ along the curvilinear abscissa.  The mean transport velocity $V_{LEV}$ along the curvilinear abscissa $S$ and the mean distance $\lambda_{LEV}$ between two consecutive LEVs along this abscissa are used to precisely retrieve the frequency between consecutive LEVs along S. The Strouhal number of leading-edge vortices $St_{LEV}$ is then defined as:

$$St_{LEV}=\frac{d}{V_{J}}\int_{S/d=3}^{S/d=8} V_{LEV}(S/d)\cdot\lambda_{LEV}(S/d)\frac{dS}{d}$$

Figure~\ref{fig:NbLEV-StLEV}b shows the LEV Strouhal number $St_{LEV}$ of each configuration as a function of the velocity ratio. For the velocity ratios $R>0.8$ where the definition of a spatial periodicity is relevant (Fig.~\ref{fig:NbLEV-StLEV}a),  a mean Strouhal value can be estimated, $\overline{St_{LEV}}=0.337$ with a standard deviation $\sigma_{St_{LEV}}=0.039$. This result is in good agreement with the Strouhal number at high velocity ratios found in the numerical study of \citet{Ilak2012}.\\

\begin{figure}[htbp!]
      \begin{center}
\hspace*{-1cm}\begin{tabular}{c}
\includegraphics[width=1.1\textwidth]{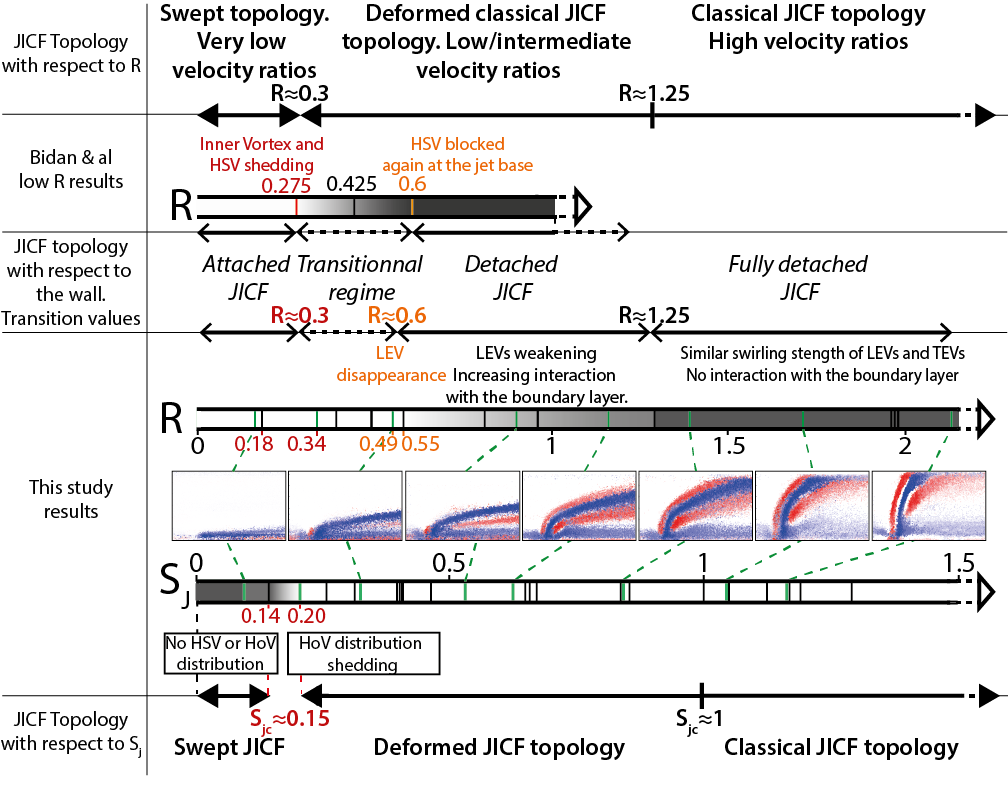}
\end{tabular}
    \caption{Evolution of the low velocity ratio JICF topology as a function of $S_J$ and $R$ illustrated by the spatial distributions of swirling structures in the symmetry plane. Each vertical bar corresponds to an experiments of table \ref{Tbl:VRdtable}.  Like in table \ref{Tbl:VRdtable}, the particular experiments shown in this article are located using green bars. The swept JICF transition (in red) occurs in the interval $S_J=[0.14\ 0.20]$. The low R results of \citet{Bidan2013} emphasizes the consistency of their observations  with ours.}
    \label{fig:Frise}
    \end{center}
\end{figure}

Figure~\ref{fig:Frise} summarizes the global transition scenario of the JICF topology as a function of the  $S_J$ and as a function of R based on our experiments. Both variables are close but different. Most of the variations of $S_J$ in our experiments are due to the velocity ratio variation, but by definition $S_J$ takes into account the diameter and boundary layer thickness influences and is therefore best suited to describe the swept jet topological transition from a detached to an attached-to-the-wall state, and all phenomena occurring at very low velocity ratio. On the other hand, the progressive LEVs weakening is only governed by the velocity ratio and does not depend on the diameter and  the boundary layer thickness.
Figure.~\ref{fig:Frise} also includes all the latest valid results of the litterature\cite{Bidan2013} to provide the most complete scenario possible of the JICF topological transition for velocity ratios $R<2$. 
They observe transitions of the JICF topology at R values very close from our results. To simplify this overview, we adopt here rounded transition values :
\begin{itemize}
\item $R\approx0.3$ for the swept jet transition.
\item $R\approx0.6$ for the detached JICF transition characterized the stabilization of the HSV position and the formation of LEV on the upstream shear-layer.
\item $R\approx1.25$ for the fully detached JICF which corresponds to the beginning of the classical topology associated with the high velocity ratios.
\end{itemize}
 
\section{Conclusion}
For the first-time, a thorough experimental study of low-velocity ratio round jet in cross-flow has been presented. Volumetric velocimetry measurements were used to characterize the main 3D swirling structures present in both the instantaneous and time-averaged velocity fields. A qualitative analysis was first proposed using visualizations of the swirling structures through $\lambda_{Ci}$ isosurfaces for increasing velocity ratios. To go further and deeper into the analysis, a statistical analysis of the time-series of instantaneous 3D velocity fields was introduced. Using statistical occurrence of the main swirling structures, it was possible to identify the main transitions of the flow in the low velocity ratio regime. As a result, a global evolution scenario of the round JICF topology has been proposed.

For $R>1.25$ (approximately $S_J\gtrsim1$), the classical JICF topology for the high velocity ratios is recovered (Fig. \ref{fig:JICF_Topologies}). The jet is fully detached and the interactions with the boundary layers are negligible. The jet velocity is large enough with respect to the crossflow velocity to form upstream and downstream shear-layer (respectively LEV and TEV)  of similar shear intensity (respectively swirling intensity). The jet  shortly crosses the boundary layer without significant interactions with this latter. The resulting interaction with the boundary layer is weak.Below $R=1.25$, the velocity ratios can be considered as low.

For $0.6< R\lesssim1.25$, a progressive deformation of this classical topology occurs while $R$ is decreased. In this regime, the jet is still partially detached. The jet trajectory is progressively bent, which results in a stronger interaction with the boundary layer. In the meantime, the progressive weakening, and followed by disappearance of the upstream shear layer leads to a similar dynamic of the LEVs. The lower jet trajectory increases the interaction between the jet and the boundary layer, between the Recirculation Vortices (RcV) and the upper Boundary Layer transverse Vortices (BLV) whose both swirling intensities significantly increase.

For $0.3<R < 0.6$, even if the JICF still follows a junction flow topology, the LEV distribution has totally disappeared. The jet is barely detached and strongly interacts with the boundary layer with which it forms a vorticity shell. The crossflow momentum becomes strong enough to push back the jet, and the crossflow starts invading the jet pipe. The horseshoe vortex advances above the jet exit. The hivering/inner vortex is progressively pushed back in the jet pipe by the HSV.

Finally below $R<0.3$, a topological transition occurs. The jet momentum is not strong enough to sustain a steady obstacle for the crossflow and the JICF is periodically by the crossflow.
As a result, this new flow topology is called swept-jet topology. It is made of alternating phases of obstruction and ejection.
Each phase may lead to a vortex formation whose characteristics depends on the phase. This double periodicity is observed on the instantaneous velocity fields, with pinched hairpin vortices with strong legs followed by hairpin vortices with more distant legs and weaker swirling intensity.

This evolution scenario has been confronted with two recent transition scenarios. The one proposed by \citet{Ilak2012} was proven to be wrong especially for the very low velocity ratios. The present study proves the necessity to take into account the interactions of the crossflow fluid with the jet pipe for the low velocity ratio JICF simulations. In these cases, the pipe flow has to be simulated with great care using at least a 10 diameters long pipe\cite{Muppidi2005} in order to get physically relevant jet inflow velocity profile at the jet exit. In particular, a parabolic velocity profile has to be excluded since it forces the JICF onto a junction flow configuration which no longer exists.
On the opposite, the transition scenario proposed by \citet{Bidan2013} is consistent with our experiments. They observe three different states for the JICF: attached, transitional and fully detached jet and obtain the same transition value of R for each regime.
A global sketch including all the latest valid results of the literature (Fig.~\ref{fig:Frise}) provides  an up-to-date complete overview of the transition scenario of the JICF topology for velocity ratios $R<2$. This transition scenario is expressed as a function of R and as a function of the  $S_J$, a transition parameter we think best suited to describe the phenomena occurring at very low velocity ratio.

\bibliographystyle{plainnat}
\bibliography{LowVelocityRatioTopology2}

\begin{thebibliography}{51}
\providecommand{\natexlab}[1]{#1}
\providecommand{\url}[1]{\texttt{#1}}
\expandafter\ifx\csname urlstyle\endcsname\relax
  \providecommand{\doi}[1]{doi: #1}\else
  \providecommand{\doi}{doi: \begingroup \urlstyle{rm}\Url}\fi

\bibitem[Andreopoulos and Rodi({1984})]{ANDREOPOULOS1984}
J.~Andreopoulos and W.~Rodi.
\newblock {Experimental investigation of jets in a cross-flow}.
\newblock \emph{J. Fluid Mech.}, {138}:\penalty0 {93--127}, {Jan} {1984}.

\bibitem[Bagheri et~al.({2009})Bagheri, Schlatter, Schmid, and
  Henningson]{Bagheri2009}
S.~Bagheri, P.~Schlatter, P.J. Schmid, and D.S. Henningson.
\newblock {Global stability of a jet in crossflow}.
\newblock \emph{Journal of Fluid Mechanics}, {624}:\penalty0 {33--44}, {Apr 10}
  {2009}.
\newblock ISSN {0022-1120}.

\bibitem[Baker({1979})]{BAKER1979}
C.J. Baker.
\newblock {Laminar Horseshoe Vortex}.
\newblock \emph{{J. Fluid Mech.}}, {95}\penalty0 ({Nov}), {1979}.

\bibitem[Baker(1980)]{BAKER1980}
C.J. Baker.
\newblock The turbulent horseshoe vortex.
\newblock \emph{J. Wind Eng. Ind. Aero.}, 6\penalty0 (1-2):\penalty0 9--23,
  1980.

\bibitem[Bidan and Nikitopoulos(2013)]{Bidan2013}
G~Bidan and DE~Nikitopoulos.
\newblock On steady and pulsed low-blowing-ratio transverse jets.
\newblock \emph{J. Fluid Mech.}, 714:\penalty0 393--433, 2013.

\bibitem[Blanchard et~al.({1999})Blanchard, Brunet, and Merlen]{Blanchard1999}
J.N. Blanchard, Y.~Brunet, and A.~Merlen.
\newblock {Influence of a counter rotating vortex pair on the stability of a
  jet in a cross flow: an experimental study by flow visualizations}.
\newblock \emph{Exp. Fluids}, {26}\penalty0 ({1-2}):\penalty0 {63--74}, {Jan}
  {1999}.

\bibitem[Broadwell and Breidenthal({1984})]{BROADWELL1984}
J.E. Broadwell and R.E. Breidenthal.
\newblock {Structure and mixing of a transverse jet in incompressible flow}.
\newblock \emph{J. Fluid Mech.}, {148}\penalty0 ({NOV}):\penalty0 {405--412},
  {1984}.
\newblock ISSN {0022-1120}.

\bibitem[Cambonie(2012)]{CAMBONIEthesis}
T.~Cambonie.
\newblock \emph{\'Etude par v\'elocim\'etrie volumique d\rq{}un jet dans un
  \'ecoulement transverse \`a faibles ratios de vitesses. Experimental study of
  a Jet in Crossflow at low velocity ratios using volumetric velocimetry}.
\newblock PhD thesis, Pierre and Marie Curie University (UPMC Paris 6), 2012.

\bibitem[Cambonie and Aider(2012)]{CaAi12}
T.~Cambonie and J.L. Aider.
\newblock Optimal seeding for high spatial resolution instantaneous volumetric
  measurements: application to low velocity ratios jets in cross-flow.
\newblock In \emph{16th International Symposium on Laser Techniques to Fluid
  Mechanics}, 2012.

\bibitem[Cambonie and Aider(2014)]{CaAi14}
T.~Cambonie and J.L. Aider.
\newblock Seeding optimization for instantaneous volumetric velocimetry.
  application to a jet in crossflow.
\newblock \emph{Optics and Laser Engineering}, 56:\penalty0 99 -- 112, 2014.

\bibitem[Cambonie et~al.(2013)Cambonie, Gautier, and Aider]{Cambonie2013}
T.~Cambonie, N.~Gautier, and J.L. Aider.
\newblock Experimental study of counter-rotating vortex pair trajectories
  induced by a round jet in cross-flow at low velocity ratios.
\newblock \emph{Exp. Fluids}, 54\penalty0 (3):\penalty0 1--13, 2013.

\bibitem[Camussi et~al.({2002})Camussi, Guj, and Stella]{Camussi2002}
R.~Camussi, G.~Guj, and A.~Stella.
\newblock {Experimental study of a jet in a crossflow at very low Reynolds
  number}.
\newblock \emph{J. Fluid Mech.}, {454}:\penalty0 {113--144}, {Mar 10} {2002}.

\bibitem[Chakraborty et~al.(2005)Chakraborty, Balachandar, and
  Adrian]{Chakraborty2005}
P.~Chakraborty, S.~Balachandar, and R.J. Adrian.
\newblock On the relationships between local vortex identification schemes.
\newblock \emph{J. Fluid Mech.}, 535:\penalty0 189--214, July 2005.

\bibitem[Chakraborty et~al.(2007)Chakraborty, Balachandar, and
  Adrian]{Chakraborty2007}
P.~Chakraborty, S.~Balachandar, and R.J. Adrian.
\newblock Kinematics of local vortex identification criteria.
\newblock \emph{J. Vis.}, 10\penalty0 (2):\penalty0 137--140, April 2007.

\bibitem[Christensen and Adrian(2002)]{Christensen2002}
K.T. Christensen and R.J. Adrian.
\newblock The velocity and acceleration signatures of small-scale vortices in
  turbulent channel flow.
\newblock \emph{J. Turb.}, 3:\penalty0 023, April 2002.

\bibitem[Coelho and Hunt({1989})]{COELHO1989}
S.L.V. Coelho and J.C.R. Hunt.
\newblock {The dynamics of the near-field of strong jets in crossflows}.
\newblock \emph{J. Fluid Mech.}, {200}:\penalty0 {95--120}, {MAR} {1989}.
\newblock ISSN {0022-1120}.

\bibitem[Cortelezzi and Karagozian({2001})]{Cortelezzi2001}
L.~Cortelezzi and A.R. Karagozian.
\newblock {On the formation of the counter-rotating vortex pair in transverse
  jets}.
\newblock \emph{J. Fluid Mech.}, {446}:\penalty0 {347--373}, {Nov 10} {2001}.

\bibitem[Davitian et~al.({2010})Davitian, Hendrickson, Getsinger, M'Closkey,
  and Karagozian]{Davitian2010}
J.~Davitian, C.~Hendrickson, D.~Getsinger, R.~T. M'Closkey, and A.~R.
  Karagozian.
\newblock {Strategic Control of Transverse Jet Shear Layer Instabilities}.
\newblock \emph{AIAA Journal}, {48}\penalty0 ({9}):\penalty0 {2145--2156},
  {Sep} {2010}.
\newblock ISSN {0001-1452}.
\newblock {AIAA 46th Aerospace Sciences Meeting and Exhibit, Reno, NV, Jan
  07-10, 2008}.

\bibitem[Fric and Roshko({1994})]{FRIC1994}
T.F. Fric and A.~Roshko.
\newblock {Vortical structure in the wake of a transverse jet}.
\newblock \emph{J. Fluid Mech.}, {279}:\penalty0 {1--47}, {Nov} {1994}.

\bibitem[Gopalan et~al.(2004)Gopalan, Abraham, and Katz]{Gopalan2004}
S.~Gopalan, B.M. Abraham, and J.~Katz.
\newblock The structure of a jet in cross flow at low velocity ratios.
\newblock \emph{Phys. Fluids}, 16\penalty0 (6):\penalty0 2067--2087, June 2004.

\bibitem[Hasselbrink and Mungal(2001)]{Hasselbrink2001}
E.~F. Hasselbrink and M.G. Mungal.
\newblock Transverse jets and jet flames. part 1. scaling laws for strong
  transverse jets.
\newblock \emph{J. Fluid Mech.}, 443:\penalty0 1--25, 2001.

\bibitem[Ilak et~al.(2012)Ilak, Schlatter, Bagheri, and Henningson]{Ilak2012}
M.~Ilak, P.~Schlatter, S.~Bagheri, and D.S. Henningson.
\newblock Bifurcation and stability analysis of a jet in cross-flow: onset of
  global instability at a low velocity ratio.
\newblock \emph{J. Fluid Mech.}, 696:\penalty0 94--121, 2012.

\bibitem[Kamotani and Greber({1972})]{KAMOTANI1972}
Y~Kamotani and I~Greber.
\newblock {Experiments on a turbulent jet in a cross flow}.
\newblock \emph{AIAA Journal}, {10}\penalty0 ({11}):\penalty0 {1425--\&},
  {1972}.
\newblock ISSN {0001-1452}.

\bibitem[Karagozian({2010})]{Karagozian2010}
A.R. Karagozian.
\newblock {Transverse jets and their control}.
\newblock \emph{Prog. Energy Combust. Sci.}, {36}\penalty0 ({5}):\penalty0
  {531--553}, {Oct} {2010}.

\bibitem[Keffer and Baines(1963)]{KEFFER1963}
J.F. Keffer and W.D. Baines.
\newblock The round turbulent jet in a cross-wind.
\newblock \emph{J. Fluid Mech.}, 15\penalty0 (4):\penalty0 481--\&, 1963.

\bibitem[Kelso and Smits({1995})]{KELSO1995}
R.M. Kelso and A.J. Smits.
\newblock {Horseshoe vortex systems resulting from the interaction between a
  laminar boundary-layer and a transverse jet}.
\newblock \emph{Phys. Fluids}, {7}\penalty0 ({1}):\penalty0 {153--158}, {Jan}
  {1995}.

\bibitem[Kelso et~al.({1996})Kelso, Lim, and Perry]{Kelso1996}
R.M. Kelso, T.T. Lim, and A.E. Perry.
\newblock {An experimental study of round jets in cross-flow}.
\newblock \emph{J. Fluid Mech.}, {306}:\penalty0 {111--144}, {Jan 10} {1996}.

\bibitem[Krothapalli et~al.(1990)Krothapalli, Lourenco, and
  Buchlin]{KROTHAPALLI1990}
A.~Krothapalli, L.~Lourenco, and J.~M. Buchlin.
\newblock Separated flow upstream of a jet in a crossflow.
\newblock \emph{AIAA Journal}, 28\penalty0 (3):\penalty0 414--420, March 1990.
\newblock ISSN 0001-1452.

\bibitem[Kuzo(1995)]{Kuzo1995}
D.M. Kuzo.
\newblock \emph{Jet to Freestream Velocity Ratio Computations for a Jet in a
  Crossflow}.
\newblock Ph. D. Thesis of the California Institute of Technology, 1995.

\bibitem[Lim et~al.({2001})Lim, New, and Luo]{Lim2001}
T.T. Lim, T.H. New, and S.C. Luo.
\newblock {On the development of large-scale structures of a jet normal to a
  cross flow}.
\newblock \emph{Phys. Fluids}, {13}\penalty0 ({3}):\penalty0 {770--775}, {Mar}
  {2001}.

\bibitem[Margason and Tso(1993)]{Margason1993}
R.J. Margason and J.~Tso.
\newblock \emph{Jet to Freestream Velocity Ratio Computations for a Jet in a
  Crossflow}.
\newblock American Institute of Aeronautics and Astronautics, 1993.

\bibitem[Marzouk and Ghoniem({2007})]{Marzouk2007}
Y.M. Marzouk and A.F. Ghoniem.
\newblock {Vorticity structure and evolution in a transverse jet}.
\newblock \emph{J. Fluid Mech.}, {575}:\penalty0 {267--305}, {Mar} {2007}.

\bibitem[Megerian et~al.({2007})Megerian, Davitian, Alves, and
  Karagozian]{Megerian2007a}
S.~Megerian, J.~Davitian, L.~S. De~B. Alves, and A.~R. Karagozian.
\newblock {Transverse-jet shear-layer instabilities. Part 1. Experimental
  studies}.
\newblock \emph{J. Fluid Mech.}, {593}:\penalty0 {93--129}, {Dec 25} {2007}.
\newblock ISSN {0022-1120}.

\bibitem[Moussa et~al.({1977})Moussa, Trischka, and Eskinazi]{MOUSSA1977}
Z.~M. Moussa, J.~W. Trischka, and S~Eskinazi.
\newblock {Near-field mixing of a round jet with a cross-stream}.
\newblock \emph{{J. Fluid Mech.}}, {80}\penalty0 ({APR4}):\penalty0 {49--\&},
  {1977}.
\newblock ISSN {0022-1120}.

\bibitem[Muppidi and Mahesh({2005})]{Muppidi2005}
S.~Muppidi and K.~Mahesh.
\newblock {Study of trajectories of jets in crossflow using direct numerical
  simulations}.
\newblock \emph{J. Fluid Mech.}, {530}:\penalty0 {81--100}, {May} {2005}.

\bibitem[Muppidi and Mahesh({2007})]{Muppidi2007}
S.~Muppidi and K.~Mahesh.
\newblock {Direct numerical simulation of round turbulent jets in crossflow}.
\newblock \emph{J. Fluid Mech.}, {574}:\penalty0 {59--84}, {Mar} {2007}.

\bibitem[Muppidi and Mahesh({2008})]{Muppidi2008}
S.~Muppidi and K.~Mahesh.
\newblock {Direct numerical simulation of passive scalar transport in
  transverse jets}.
\newblock \emph{J. Fluid Mech.}, {598}:\penalty0 {335--360}, {Mar} {2008}.

\bibitem[Muppidi and Mahesh({2006})]{Muppidi2006}
Suman Muppidi and Krishnan Mahesh.
\newblock {Two-dimensional model problem to explain counter-rotating vortex
  pair formation in a transverse jet}.
\newblock \emph{Phys. Fluids}, {18}\penalty0 ({8}), {AUG} {2006}.
\newblock ISSN {1070-6631}.

\bibitem[New et~al.(2006)New, Lim, and Luo]{New2006}
T.~H. New, T.~T. Lim, and S.~C. Luo.
\newblock Effects of jet velocity profiles on a round jet in cross-flow.
\newblock \emph{Exp. Fluids}, 40\penalty0 (6):\penalty0 859--875, June 2006.

\bibitem[Pereira and Gharib({2002})]{Pereira2002}
F.~Pereira and M.~Gharib.
\newblock {Defocusing digital particle image velocimetry and the
  three-dimensional characterization of two-phase flows}.
\newblock \emph{Meas. Sci. Technol.}, {13}\penalty0 ({5}):\penalty0 {683--694},
  {May} {2002}.

\bibitem[Pereira et~al.({2000})Pereira, Gharib, Dabiri, and
  Modarress]{Pereira2000}
F.~Pereira, M.~Gharib, D.~Dabiri, and D.~Modarress.
\newblock {Defocusing digital particle image velocimetry: a 3-component
  3-dimensional DPIV measurement technique. Application to bubbly flows}.
\newblock \emph{Exp. Fluids}, {29}\penalty0 ({Suppl. S}):\penalty0 {S78--S84},
  {Dec} {2000}.

\bibitem[Pereira et~al.({2006})Pereira, Stuer, Graff, and Gharib]{Pereira2006}
F.~Pereira, H.~Stuer, E.C. Graff, and M.~Gharib.
\newblock {Two-frame 3D particle tracking}.
\newblock \emph{Meas. Sci. Technol.}, {17}\penalty0 ({7}):\penalty0
  {1680--1692}, {Jul} {2006}.

\bibitem[Peterson and Plesniak({2004})]{Peterson2004}
S.D. Peterson and M.W. Plesniak.
\newblock {Evolution of jets emanating from short holes into crossflow}.
\newblock \emph{J. Fluid Mech.}, {503}:\penalty0 {57--91}, {Mar} {2004}.

\bibitem[Salewski et~al.(2008)Salewski, Stankovic, and Fuchs]{Salewski2008}
M.~Salewski, D.~Stankovic, and L.~Fuchs.
\newblock Mixing in circular and non-circular jets in crossflow.
\newblock \emph{Flow Turb. Combust.}, 80\penalty0 (2):\penalty0 255--283, March
  2008.

\bibitem[Schlatter and Henningson({2011})]{Schlatter2011}
S.~Schlatter, P.and~Bagheri and D.S. Henningson.
\newblock {Self-sustained global oscillations in a jet in crossflow}.
\newblock \emph{{Theor. Comput. Fluid Dyn.}}, {25}\penalty0 ({1-4,
  SI}):\penalty0 {129--146}, {Jun} {2011}.

\bibitem[Simpson(2001)]{Simpson2001}
R.~L. Simpson.
\newblock Junction flows.
\newblock \emph{Annu. Rev. Fluid Mech.}, 33:\penalty0 415--443, 2001.

\bibitem[Smith and Mungal({1998})]{Smith1998}
SH~Smith and MG~Mungal.
\newblock {Mixing, structure and scaling of the jet in crossflow}.
\newblock \emph{J. Fluid Mech.}, {357}:\penalty0 {83--122}, {Feb 25} {1998}.
\newblock ISSN {0022-1120}.

\bibitem[SU and MUNGAL(2004)]{SU2004}
L.~K. SU and M.~G. MUNGAL.
\newblock Simultaneous measurements of scalar and velocity field evolution in
  turbulent crossflowing jets.
\newblock \emph{J. Fluid Mech.}, null:\penalty0 1--45, 8 2004.
\newblock ISSN 1469-7645.

\bibitem[Yuan et~al.(1999)Yuan, Street, and Ferziger]{Yuan1999}
L.~L. Yuan, R.~L. Street, and J.~H. Ferziger.
\newblock Large-eddy simulations of a round jet in crossflow.
\newblock \emph{J. Fluid Mech.}, 379:\penalty0 71--104, January 1999.

\bibitem[Zhou et~al.(1999)Zhou, Adrian, Balachandar, and Kendall]{Zhou1999}
J.~Zhou, R.~J. Adrian, S.~Balachandar, and T.~M. Kendall.
\newblock Mechanisms for generating coherent packets of hairpin vortices in
  channel flow.
\newblock \emph{J. Fluid Mech.}, 387:\penalty0 353--396, May 1999.

\bibitem[Ziefle and Kleiser({2009})]{Ziefle2009}
J.~Ziefle and L.~Kleiser.
\newblock {Large-Eddy Simulation of a Round Jet in Crossflow}.
\newblock \emph{AIAA Journal}, {47}\penalty0 ({5}):\penalty0 {1158--1172},
  {2009}.

\end{thebibliography}

\end{document}